\documentclass{article}
\usepackage{graphicx} 
\usepackage{biblatex} 
\usepackage{appendix}
\usepackage{float}
\usepackage{amsmath}
\usepackage{amsfonts}
\usepackage{amsthm}
\theoremstyle{definition}
\newtheorem{definition}{Definition}[section]
\title{Accessibility of Modified $NK$ Fitness Landscapes}
\author{Henry Nunns\\
\texttt{Rose-Hulman Institute of Technology}
\and
Manda Riehl\\
\texttt{Rose-Hulman Institute of Technology}\\
\texttt{riehl@rose-hulman.edu}
}
\date{June 2025}

\begin{document}

\maketitle
\begin{abstract}
In this paper we present two modifications of traditional $NK$ fitness landscapes, the $\theta NK$ and $HNK$ models, and explore these modifications via accessibility and ruggedness. The $\theta NK$ model introduces a parameter $\theta$ to integrate local Rough Mount Fuji-type correlations in subgenotype contributions, simulating more biologically realistic correlated fitness effects. The $HNK$ model incorporates gene regulation effects by introducing a masking mechanism where certain loci modulate the expression of other loci, simulating effects observed in gene regulatory networks without modeling the full network. Through extensive simulations across a wide range of parameters ($N$, $K$, $\theta$, and $H$), we analyze the impact of these modifications on landscape accessibility and the number of local optima. We find that increasing $\theta$ or the number of masking loci ($H$) generally enhances accessibility, even in landscapes with many local optima, showing that ruggedness doesn't necessarily hinder evolutionary pathways. Additionally, distinct interaction patterns (blocked, adjacent, random) lead to different observations in accessibility and optimum structure. While more complex than traditional $NK$, we believe each model provides a new biologically relevant facet to fitness landscapes and provides insight into how genetic and regulatory structures influence the evolutionary potential of populations. 
\end{abstract}
\section{Background}
The question of how to model relationships between an organism's genotype and its evolutionary fitness has been heavily studied in recent decades by both biologists and biomathematicians. The interest in this topic lies in its ability to explain how evolution (as induced by random mutation and natural selection) can find optimal or near-optimal genotypes by traversing rough fitness landscapes where there may be many local optima.

A fitness landscape is most commonly considered to be an $N$-dimensional Boolean lattice with directed edges such that it is acyclic \cite{KL}, as this allows the directedness of an edge to stand in for a relationship of increasing fitness when moving along that edge. From this starting point, lower and upper bounds on descriptors of ``roughness" like the number of local optima can be derived from other features like reciprocal sign epistasis \cite{RSE} using techniques from graph theory, providing a model-agnostic way to analyze fitness landscapes \cite{RSE2, ROSI, walsh}. However, most research in this field instead uses numerical scores of fitness assigned according to a specific ``model" as a function of the organism's genotype. Deriving the direction of edges from numerical scores naturally guarantees that the lattice is acyclic.

Several models for assigning numerical fitness exist, but of interest to us in this paper primarily are three: House-of-Cards, $NK$, and Rough Mount Fuji (RMF). 

House-of-Cards is the simplest, and assigns a fitness determined at random from a non-specific distribution, such that changing any part of the genotype will result in a completely new fitness score, as though knocking down a house of cards \cite{access}. While mathematically approachable, this model is biologically unrealistic, as fitness functions found in nature are correlated \cite{KL}.

Modeling correlation requires more complex models, and generally requires the notion of tunable ruggedness (some parameter that controls how many local optima to expect for a fixed $N$) in order to make the model useful. 

In the $NK$ model first proposed by Kauffman and Levin in 1986, the fitness of an $N$-length binary genotype was determined by assigning each locus of the genotype $K$ interactors \cite{KL}. This creates structure in the resultant fitness landscape, and is tunably rugged in the sense that $K=1$, having no interactions, would be exhibit additive fitness and $K=N$, having every possible interaction, would revert to House-of-Cards. The structure of the interactions between loci can be patterned or random, leading to different emergent landscape behaviors \cite{adaptive, NK2, NK3}. 

In the RMF model proposed by Aita et. al., a global optimum genotype is chosen at random, and all other genotypes are assigned fitness scores based on their Hamming distance from the global optimum (the ``Mount Fuji" shape) plus an element of noise (the ``Rough"-ness) \cite{aita}. 

Many emergent properties of a landscape can be studied using either of these models. Commonly treated properties like the mean distance between local optima and the correlation in height of local optima \cite{estimatingheights, NK2, interactioneffects} as analogues to the ruggedness of the landscape. To accommodate the additional complexity we will be introducing to the $NK$ structure, we will instead focus instead on the number of local optima and the proportion of the total number of genotypes that have an evolutionarily accessible path to the global optimum. These measures give direct insight into the roughness of the landscape and the degree of difficulty associated with evolution to high fitnesses within systems that resemble the model conditions \cite{tunably}, and are more straightforward to interpret from simulation results.

The question of accessibility is of particular interest to us in this investigation. Both $NK$ and RMF create the implication that a more rugged landscape would be less accessible, yet it is possible for landscapes examined under laboratory conditions to be both rugged (having many local optima) and easily navigable by evolution (high accessibility) \cite{ruggedyet, biophys, RMFadapt}. This leaves open the question of whether a more detailed model with more parameters could expose new behaviors that reflect biological realities more closely.

Our goal is to propose and simulate alterations to the $NK$ model that expand its domain. In particular, we aim to allow the structure of the contributions to fitness by interacting groups of loci to be correlated, and to model the gene activation/regulation effects observed in biological systems without the complexity of a Gene Regulatory Network model \cite{GRN}. 

\section{Definitions}
We start by defining many commonly used terms in this field so that there will be no ambiguity as to our use of them.

We define a \textit{genotype} $\sigma$ to be an $N$-length string of binary bits (1s or 0s) where each position in the string is a locus and the bit values correspond to two possible states for that locus' gene. Unlike in many other papers, we will not necessarily imply that one of these is the wild type and another is the mutant type. 

We define $\sigma_\ell$ to be the $\ell^{th}$ bit of $\sigma$. We will allow the subscript to be a set, extracting multiple loci from $\sigma$ rather than just one, and we will always assume that extracted loci retain their order in the genotype relative to each other -- that is, $\sigma_{1, 4, 9, 7}=\sigma_1 \sigma_4 \sigma_7 \sigma_9$. 

We specify that different genotypes of the same length will be distinguished by superscripts. We define the fitness of an $N$-length genotype $f(\sigma)$ to be a real number, output by an injective function $f:\{0, 1\}^N\rightarrow \mathbb{R}$, which we will call a fitness function. We define $f(\sigma^1)>f(\sigma^2)$ to mean $\sigma^1$ is more fit (better adapted) than $\sigma^2$.

We define the \textit{Hamming distance} between $\sigma^1$ and $\sigma^2$ to be \[d(\sigma^1, \sigma^2)=\displaystyle \sum_{\sigma^1_i\neq\sigma^2_i}1,\] the count of the number of loci in which $\sigma^1$ and $\sigma^2$ differ. 

We define an $N$-dimensional \textit{fitness landscape} based on a fitness function $f$ to be the directed $N$-dimensional hypercube with the $2^N$ possible $N$-locus genotypes at its vertices and with any two vertices connected if and only if the Hamming distance between them is 1. Each edge is directed such that it points to the more fit of the two vertices it connects.

We define an \textit{adaptive walk} of length $M$ to be a sequence of unique genotypes $\sigma^1, \sigma^2, ..., \sigma^M$ where $f(\sigma^1), f(\sigma^2), ..., f(\sigma^M)$ is a monotonically increasing sequence, and where $d(\sigma^i, \sigma^{i+1})=1$ for all $i=1, 2, ..., M-1$.

We define a genotype $\sigma$ to be a \textit{local optimum} if $f(\sigma) \geq f(\tau)$ for all $\tau$ such that $d(\sigma, \tau) = 1$. 

We define a genotype $\sigma$ to be the \textit{global optimum} if $f(\sigma) \geq f(\tau)$ for all $\tau$.

We define the global optimum of an $N$-dimensional fitness landscape to be \textit{accessible} from a genotype $\sigma$ if there exists an adaptive walk from $\sigma$ to the global optimum.

We define $p_1$ for an $N$-dimensional fitness landscape to be the probability that the global optimum is accessible from a uniformly randomly chosen initial genotype. The subscript of 1 is included because some works in the literature will allow adaptive walks to take steps larger than a Hamming distance of 1, and thus accessibility defined based on these alternate forms would be different. Here, we have restricted adaptive walks to a step length of 1, so we will use only $p_1$.

\subsection{The $NK$ Model}
We define an $NK$ fitness landscape to be an $N$-dimensional fitness landscape where the associated fitness function $f$ is constructed as follows.
\begin{enumerate}
    \item For each locus $\ell$, there exists $B_\ell$, the set of interacting loci.
    \item For all $B_\ell$, $\left| B_\ell \right|=K$, and $\ell \in B_\ell$.
    \item For each locus $\ell$, there exists $f_\ell(\sigma, B_\ell)$, a real-valued injective function.
    \item For each $\sigma$ in the landscape, let $\sigma_{B_\ell}$ be the ``subgenotype" of $\sigma$ restricted only to the loci in $B_\ell$, and let $f(\sigma)=\displaystyle\sum_{\ell=1}^{N} f_\ell(\sigma_{B_\ell})$, which must also be an injection.
\end{enumerate}

It is constrained that $1 \leq K \leq N$. 

We define a \textit{classic} $NK$ fitness landscape to be one in which each $f_\ell$ has a range comprised of observations of a standard normal -- that is, $N(0, 1)$ -- random variable. The observations comprising this range are only taken once -- that is to say, the outputs of $f_\ell$ themselves are not random variables. We remark that this is analogous to each $f_\ell$ being a House-of-Cards fitness landscape of its own over a $K$-dimensional neighborhood of loci.

The pattern of interactions between loci influences the structure of an $NK$ landscape heavily \cite{interactioneffects} even though the specific locus positions are arbitrary. As such, we define the following common neighborhood interaction patterns of interest to our investigation.

We define an $NK$ fitness landscape with \textit{blocked} neighborhoods to be an $NK$ fitness landscape where $B_\ell = \{i \in \mathbb{N} : \lfloor \frac{i}{K} \rfloor = \lfloor \frac{\ell}{K} \rfloor\}$. That is, the first $K$ loci interact with each other, the second $K$ loci interact with each other, and so on until the final group of $N \mod K$ remaining loci form a last, smaller interaction group. This is called a ``blocked" neighborhood style because the interactions, if written as a matrix, would appear to form square blocks.

We define an $NK$ fitness landscape with \textit{adjacent} neighborhoods to be an $NK$ fitness landscape where $B_\ell = \{i \in \mathbb{N} : (\ell - \lceil \frac{K}{2} \rceil) \mod N < i \leq (\ell + \lfloor\frac{K}{2} \rfloor) \mod N\}$. That is, the $K$ loci centered on $\ell$ interact with $\ell$. For even $K$, the extra interacting locus is added to the higher-indexed portion. We will use the interpretation that a negative integer $x$ modulo $N$ should be congruent to the positive integer $N-x$.

We define an $NK$ fitness landscape with \textit{random} neighborhoods to be an $NK$ fitness landscape where each $B_\ell$ contains $\ell$ and $K-1$ uniformly randomly selected other loci.

These neighborhood definitions are consistent with those given in \cite{adaptive}.

\subsubsection{The $\theta NK$ Model}
An $NK$ fitness landscape imposes some restrictions onto the shape of the landscape itself. This is most easily seen in its classic variety, where the subgenotype fitness functions $f_\ell$ are required to be House-of-Cards. A real biological system need not conform to this regimen. We propose instead that any particular subgenotype could be expected to have an ideal configuration, corresponding with the best performance of whatever phenotypical trait locus $\ell$ controlled, and every mutation of the genotype away from that ideal configuration could be deleterious to fitness, within some level of noise.

This idea on a landscape-wide scale motivated Aita et. al. in the development of the Rough Mount Fuji model \cite{aita}, but we propose instead that it can be applied to subgenotypes.

\begin{definition}[$\theta NK$]
    A $\theta NK$ fitness landscape is an $NK$ fitness landscape where each $f_\ell = -\theta d(\delta_\ell, \sigma_{B_\ell}) + N(0, 1)$, where $\delta_\ell$ is a uniformly randomly chosen ``peak" subgenotype of length $K$ and $\theta$ is a parameter of the landscape such that $\theta \geq 0$. 
\end{definition}

This is the fitness function of a Rough Mount Fuji landscape \cite{aita} applied to the contributions to fitness rather than the overall fitness function. Note that the neighborhood styles previously defined for $NK$ landscapes are all directly applicable to $\theta NK$ landscapes.

\subsubsection{The $HNK$ Model}
Another potential modification to $NK$ reflects the ability of genes in the genotype of real organisms to regulate activity. Unlike in gene regulatory networks, we will not attempt to model stimulus and activation \cite{GRN}, but rather to examine only the effect on accessibility under a regimen where certain genes ``mask" the identity of others.

The biological motivation for this masking behavior is that there exist Hox genes capable of activating or turning off other genes' expression in biological systems \cite{hox}. Activation and inhibition are complementary processes, so we considered it sufficient to only model one.

\begin{definition}[$HN$ genotypes]
    An $HN$ genotype is a genotype of length $H+N$ divided into two segments, the first of length $H$ and the second of length $N$, indexed sequentially with the $H$-length segment coming first.
\end{definition}

For convenience, we will refer to these as the $H$ segment and $N$ segment going forward. We require that $H, N \in \mathbb{N}$ and $H < N$. We will denote the $H$ segment of a genotype $\sigma$ as $\sigma_H$ and the $N$ segment as $\sigma_N$.

\begin{definition}[Mask function]
    The mask function $m$ for a given $H$ and $N$, taking as input an $HN$ genotype $\sigma$, a set $M$ of loci between $H+1$ and $N+H$, and a locus index $\ell$ between $H+1$ and $N+H$, is $m(\sigma, M, \ell) = \sigma_\ell$ if $\ell \in M$ and $m(\sigma, M, \ell) = 1$ otherwise.
\end{definition}

This allows us to establish our notion of gene activation for formalizing our new landscape:

\begin{definition}[$HNK$]
An $HNK$ fitness landscape is an $(N+H)$-dimensional fitness landscape where the associated fitness function $f$ formed as follows:
\begin{enumerate}
    \item For each locus $h$ in the $H$ segment, there exists $G_h$, the set of ``governed" loci, which are entirely contained within the $N$ segment. We require $|G_h| = K$.
    \item For each genotype $\sigma$ in the landscape, there exists $\gamma$ to be a genotype of length $N$ where $\gamma_\ell = \displaystyle \prod_{h = 1}^H m(\sigma, G_h, \ell+H) \sigma_{\ell+H}$.  We will refer to forming $\gamma$ as ``masking."
    \item We define an $NK$ fitness landscape with fitness function $f'$, and require $f$ to be $f(\sigma) = f'(\gamma)$. We will refer to this underlying $NK$ landscape as the $\gamma$ landscape.
\end{enumerate}
\end{definition}

We observe that this is not a fitness function, or indeed a fitness landscape at all in the sense defined previously, because it contains tied fitnesses. Therefore, we define an $HNK$ fitness function to be to be a real-valued function where $f(\sigma^1)>f(\sigma^2)$ implies $\sigma^1$ is more fit (better adapted) than $\sigma^2$, and where $f(\sigma^1) = f(\sigma^2)$ if and only if $f'(\gamma^1) = f'(\gamma^2)$. 

Similarly, we will allow a fitness landscape in an $HNK$ context to be a $2^{H+N}$-dimensional hypercube with edges of Hamming length 1 oriented to point toward increasing fitness, and with adjacent genotypes of tied fitness possessing a bidirectional connection.

We will refer to a genotype $\sigma$ where its masked genotype $\gamma$ satisfies $\sigma_N = \gamma$ as being \textit{fully expressed}.

Similarly, we will refer to a local optimum in an $HNK$ context to be a fully expressed genotype $\sigma$ where $f(\sigma) > f(\sigma^i)$ for all $\sigma^i$ such that $d(\sigma^i, \sigma^j) = 1$ and $\sigma^i$ is also fully expressed. We comment that this will necessarily mean that a genotype $\sigma$ is a local optimum on an $HNK$ landscape if and only if $\sigma_H = 11\dots1$ and $\sigma_N$ is a local optimum on the $\gamma$ landscape.

We note that masking loci in this way is not truly analogous to inhibition of expression, as in real systems the behaviors of allele 1 and allele 0 would both be different from the lack of gene expression whatsoever, but to preserve the biallelic landscapes structure that allows fitness landscapes to be represented as Boolean lattices, we have chosen not to model gene inhibition as a tristate process. We maintain that masking is sufficient to represent inhibition, as it still allows for the most important characteristic -- that is, the ability for the phenotype (and therefore fitness) to go unchanged while the genotype is modified -- to be modeled.

\subsection{Focuses of Investigation}
The common denominator between these two modifications of the $NK$ model is that they aim to introduce a greater degree of biological reality. The goal of making any model like these two is that its behaviors -- both in the limit for large $N$ and in the patterns formed by changing $N$ and $K$ -- will provide an insight into the structures present in biological systems.

The best analogue for ``structure" in this sense is a measure of how well evolution performs in landscapes like these. Accessibility and the number of local optima both provide insight into this property, since accessibility is the probability the global optimum can be reached by an evolutionary process from a random starting genotype and the optimum count provides an analogue to ruggedness. As in \cite{ruggedyet}, these are not necessarily the same thing -- many local optima can exist with accessibility still being high.

Our goal, therefore, is to identify how these modifications impact accessibility and the optimum count. We also relate these impacts to real biological processes that they either correspond with or contradict to examine if the new models better agree with these processes. A secondary aim is to identify patterns and structure in the new models that will allow us to make predictions about their end behaviors and the accessibility and optimum counts of arbitrary landscapes.

It should be noted that other measures of roughness, such as the performance of greedy adaptive walks (those that are only permitted to make \emph{the steepest} uphill step, rather than simply any improving step as in our definition) or the size of an optimum's basin of attraction have been used in other works \cite{tunably}, but we have chosen not to focus on these. Our primary reasoning is that the two we have selected are sufficient, and that behaviors like greedy walks will usually make evolution seem to under-perform, as the implied ultra-selective environment that makes greedy adaptation sensible (ie. the steepest step will only out-compete all others every time if the environment is extremely harsh) goes beyond the typical strong selection, weak mutation assumptions laid out by Kauffman and Levin \cite{KL}.

A final topic we have chosen to exclude, though one of considerable value to potential future investigations, is that of time-dynamic landscapes. Some work has been done in the literature on how the variation in a fitness landscape over time can be modeled, and on how that might impact accessibility to the global optimum \cite{deloc, terraced}. One might imagine that introducing an element of time to the fitness function $f$ and some mutational rate bound to time could yield different results in all manner of landscape structures. A potential angle is suggested in the Future Directions.

\section{Methods}
In this investigation, the primary mode of analysis on the $\theta NK$ and $HNK$ models was simulation. Code written in C implemented the two types of landscape and was used to repeatedly generate independent instances of each type of landscape with varying parameters, such that the influence of $H$, $N$, $K$, and $\theta$ could each be understood in relation to each other.

The code used an exhaustive approach to calculating accessibility. Since $p_1$ can be calculated as the proportion of the genotypes in the landscape that have an adaptive path to the global optimum, it is possible to calculate it by starting from the global optimum and performing a recursive depth-first search that requires each step be between genotypes of Hamming distance 1 and be monotonically decreasing in fitness. It should be noted that in $HNK$ landscapes, where different genotypes can have tied fitness, traversal along any non-increasing path was allowed. This led to an extreme degree of recursion that demanded the call stack be unrolled.

Counting the number of local optima is simpler, and requires only iteration over the landscape to look at every genotype in relation to its Hamming distance 1 neighbors.

All fitness calculations in a landscape were performed once when it was generated, and then stored to memory so that they could be looked up based on the genotype later. This was done for efficiency, to avoid redundant computation.

In the final implementation of the code, it was possible to sweep from $N=1$ to $N=20$ (and from $K=1$ to $K=N$ for each $N$, and from $\theta=0$ to $\theta=3$ in steps of size 1 for each $K$) in approximately 38 hours. Each $\theta, N, K$ combination (or $H, N, K$ combination, depending on the model) was used to generate 3000 different landscapes, and the final value reported for accessibility and the number of optima was the average across the accessibilities and numbers of optima on those 3000 trial landscapes. For more information about the formation of these trial landscapes, see Appendix A.

The source for the code written in this investigation is available at \url{https://github.com/rose-nunnshe/ma_capstone}.

\section{Results}
The simulation results obtained in this investigation showed that both increasing $\theta$ and increasing $H$ led to significant increases in the average accessibility of landscapes for a fixed $N$ and $K$, and that unlike in the case where $\theta=0$ and the end behavior as $(N=K) \rightarrow \infty$ (where $p_1$ tends to a nonzero value less than 1) all cases where $\theta > 0$ and $(N=K) \rightarrow \infty$ led to $p_1 \rightarrow 1$. 

We also found that the number of optima decreased as $\theta$ increased, but that for large values of $\theta$ these quantities did not follow simple monotonic or proportional/inverse-proportional relationships, as when $\theta \rightarrow \infty$, whether $\theta$ is even or odd introduces a bias that dominates the overall correlation of the landscape. 

For $HNK$, it should be noted that our definition of a local optimum applies only to the underlying $\gamma$-landscape, so the average optima count in $HNK$ necessarily does not change with $H$. 

\subsection{Accessibility in $\theta NK$}
\begin{figure}[H]
    \begin{center}
        \includegraphics[width=0.7\linewidth]{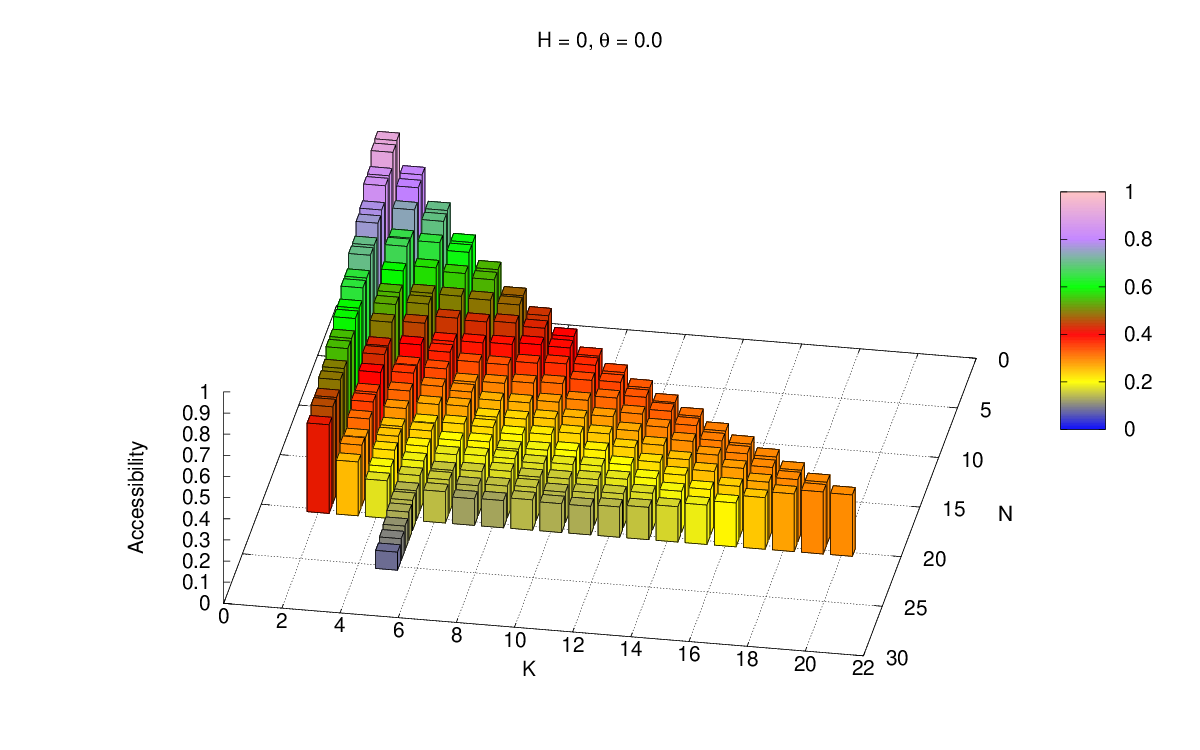}
        \caption{$\theta=0$ sweep of $N,K$ with blocked neighborhoods}
    \end{center}
    \label{TNK_longer}
\end{figure}

We investigated the accessibility of $\theta NK$ landscapes in terms of plots comparing accessibility (represented by the height of a bar chart) with both $N$ and $K$, with plots being made separately for different values of $\theta$. The base case in which $\theta NK$ reverts to the classic $NK$ landscape is $\theta=0$, graphed in Figure \ref{TNK_longer} for a blocked neighborhood $\theta NK$ landscape.

The broad bowl shape of this accessibility graph is expected, and the monotonically decreasing accessibility along rows of constant $K$ for decreasing $N$ is consistent with the fact that accessibility in a blocked neighborhood $NK$ landscape can be represented as the product of the accessibilities within each of the blocks. 

In terms of the number of optima, we found it most useful to draw comparisons between the number of optima and the accessibility in the form of scatterplot graphs in two or three dimensions. If we defined the optima density as the ratio of the number of optima in a landscape to the number of genotypes total in a landscape, we can find that there are smooth, predictive bands relating this quantity to accessibility and $K$ across different values of $N$.

\begin{figure}[H]
    \begin{center}
        \includegraphics[width=0.7\linewidth]{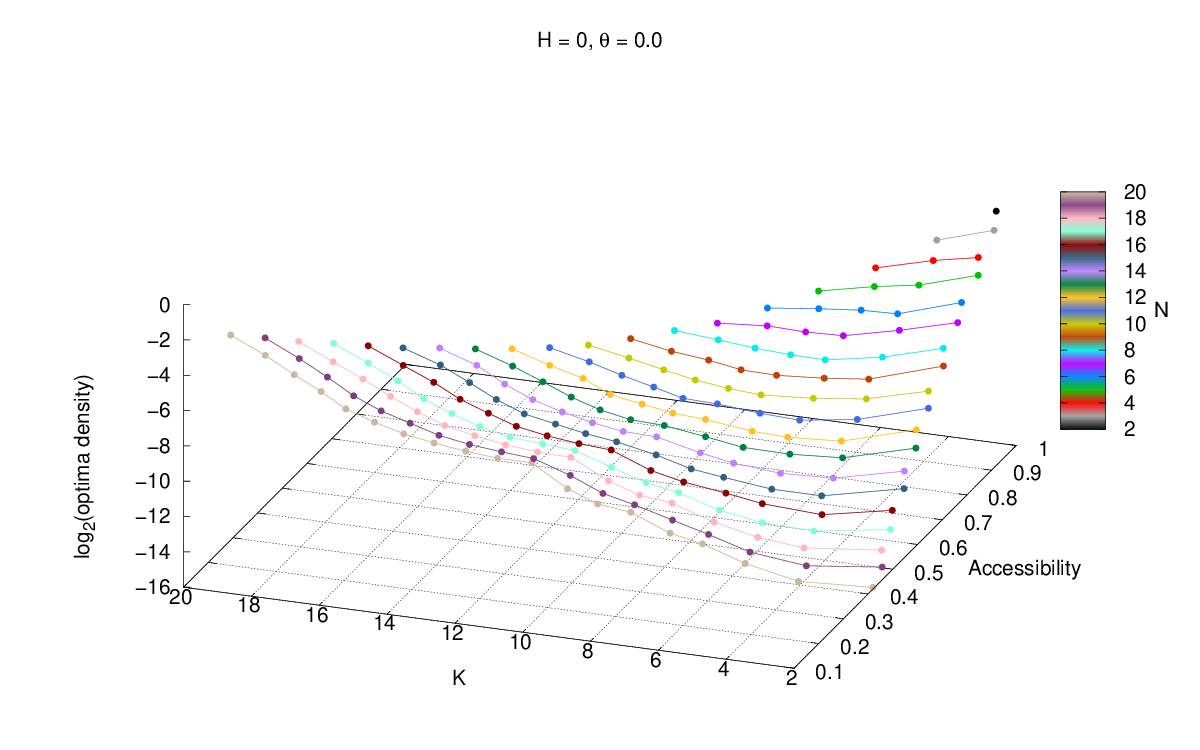}
        \caption{Scatterplot of $\theta=0$ $N,K$ sweep}
    \end{center}
\end{figure}

In the specific case of blocked neighborhood landscapes, increasing $\theta$ introduces a new shape to the $N=K$ ``ridge" of accessibility graphs, where accessibility falls for increasing $N$ between $N=1$ and another point, usually between $N=5$ and $N=8$, and then begins to rise again, eventually tending toward $p_1 = 1$.

\begin{figure}[H]
    \begin{center}
    \begin{tabular}{c}
    \includegraphics[width=0.4\linewidth]{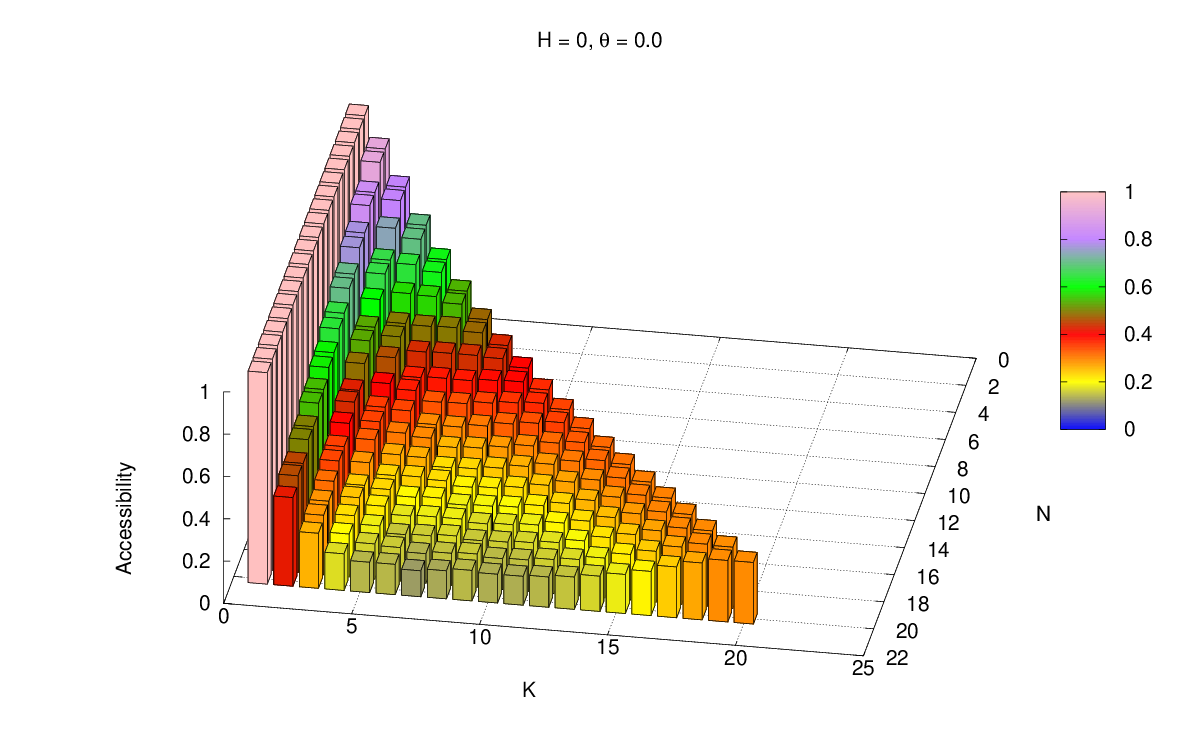}
    \\
    \includegraphics[width=0.4\linewidth]{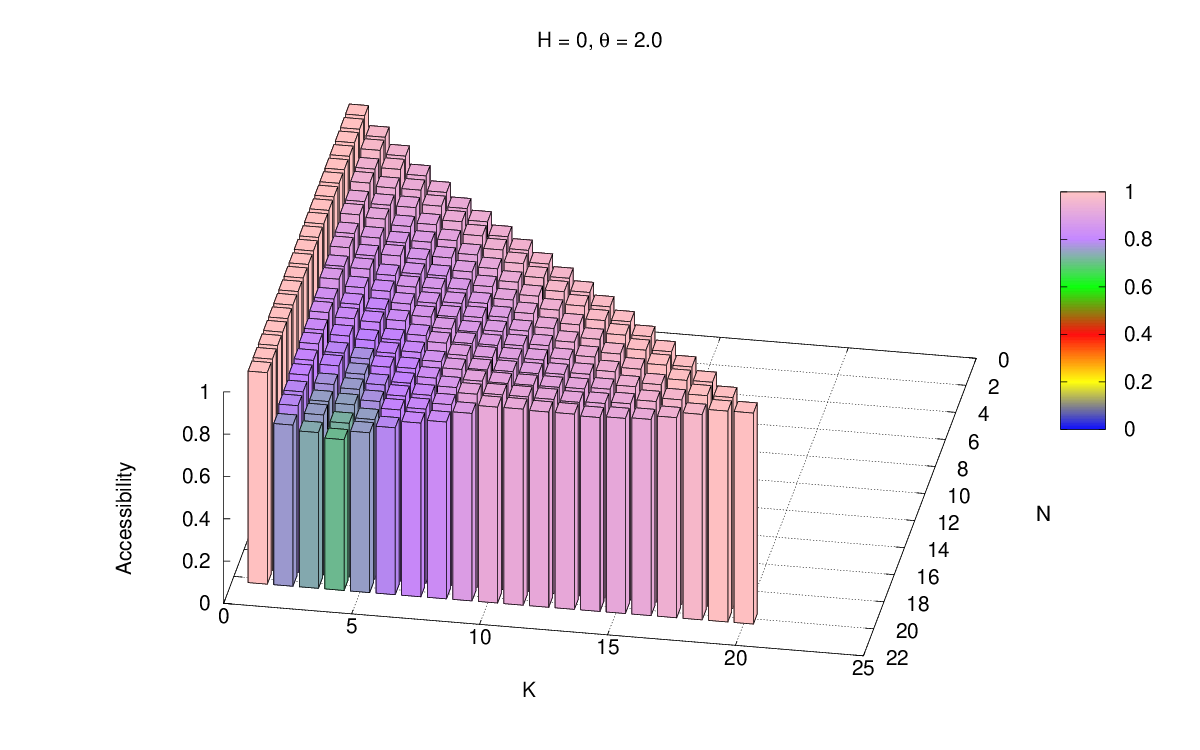}
    \end{tabular}
    \begin{tabular}{c}
    \includegraphics[width=0.4\linewidth]{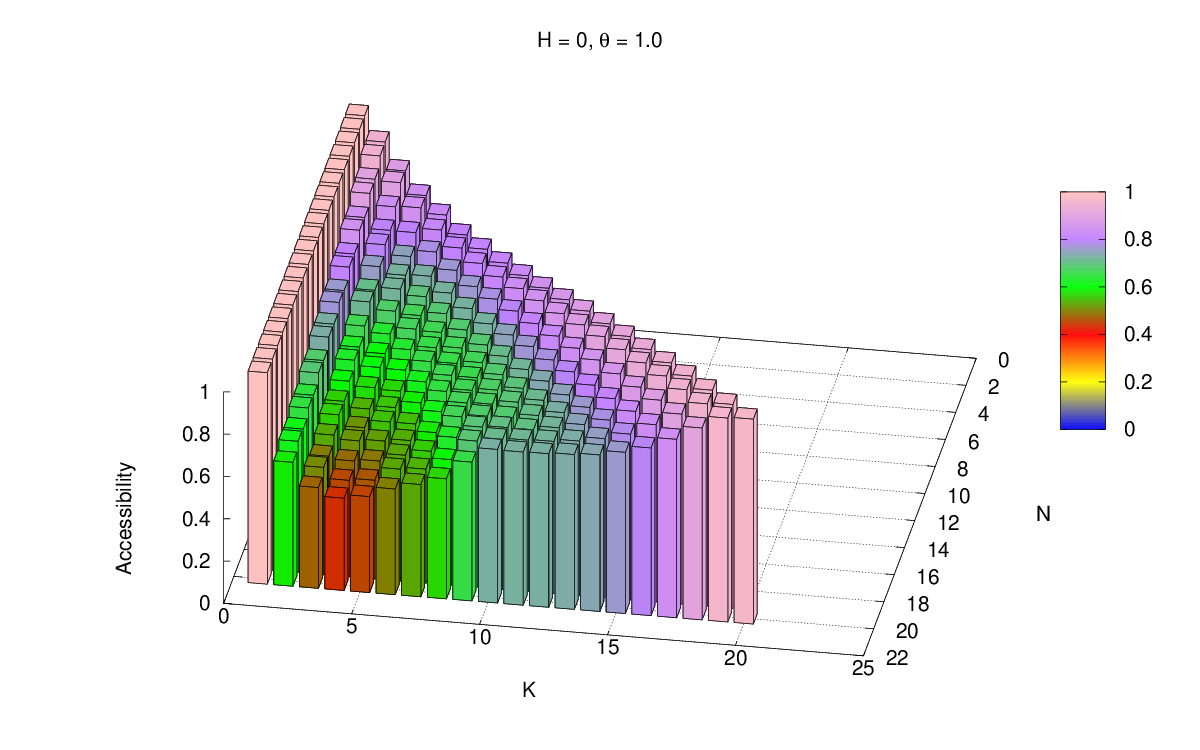}
    \\
    \includegraphics[width=0.4\linewidth]{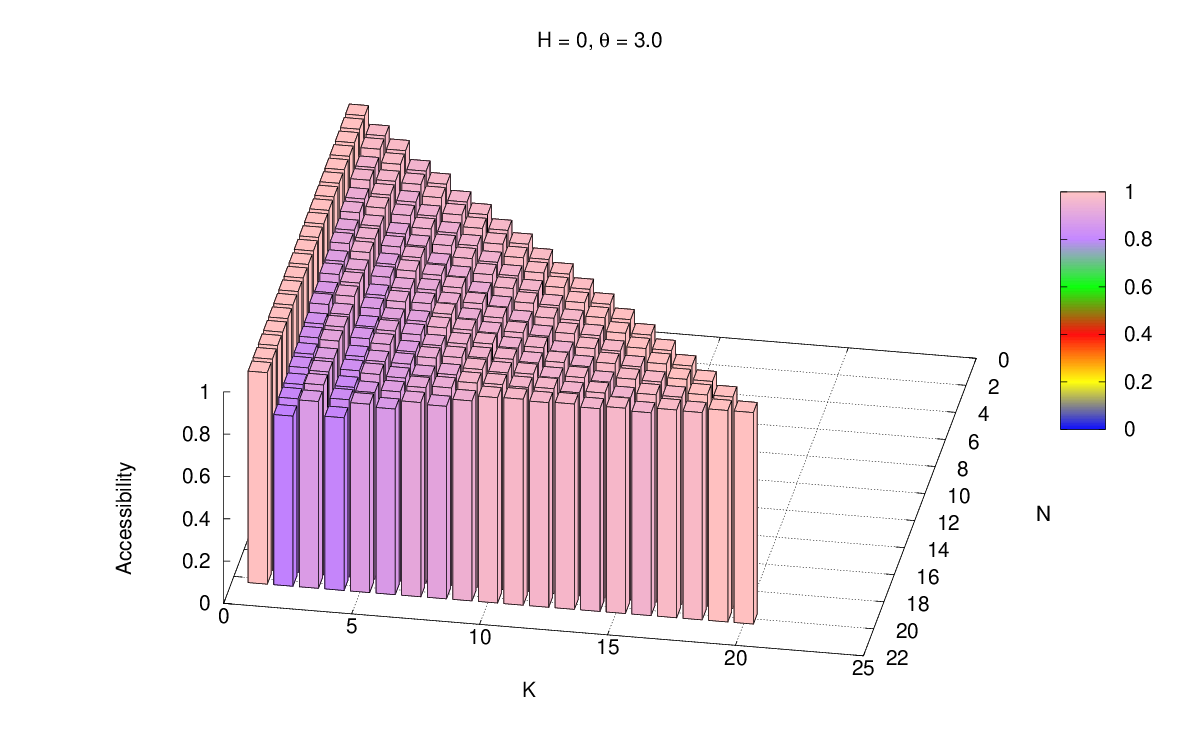}
    \end{tabular}
    \end{center}
    \caption{$\theta NK$ sweep across $\theta=0,1,2,3$, $N,K$ from 1 to 20, with blocked neighborhoods}
\end{figure}

Because blocked neighborhood landscapes admit a prediction of the accessibility of any $N,K$ parameter pair according to the products of the accessibilities of the $N_2=K$ sub-landscapes their blocks represent, this feature allows prediction of the shape of the $K<N$ region of the graphs.

In comparing optima count and accessibility, increasing $\theta$ primarily demonstrates a different effect. On the $N=K$ ridge, as $\theta$ becomes large compared to the $N(0, 1)$ noise element added into fitness, the overall fitness of a genotype becomes dominated by whether each of its loci either agree or disagree with the majority of $\delta_\ell$ peak genotypes in said locus. If $N$ is odd, then then for all loci $\ell$ there exists a genotype that is exhibited in locus $\ell$ by more $\delta$ peaks than not, but when $N$ is even, ties are possible, allowing the random noise to exhibit a persistent effect.

This odd-even alternation dominated for $\theta>2$, creating a jagged shape:

\begin{figure}[H]
    \begin{center}
    \begin{tabular}{c}
    \includegraphics[width=0.4\linewidth]{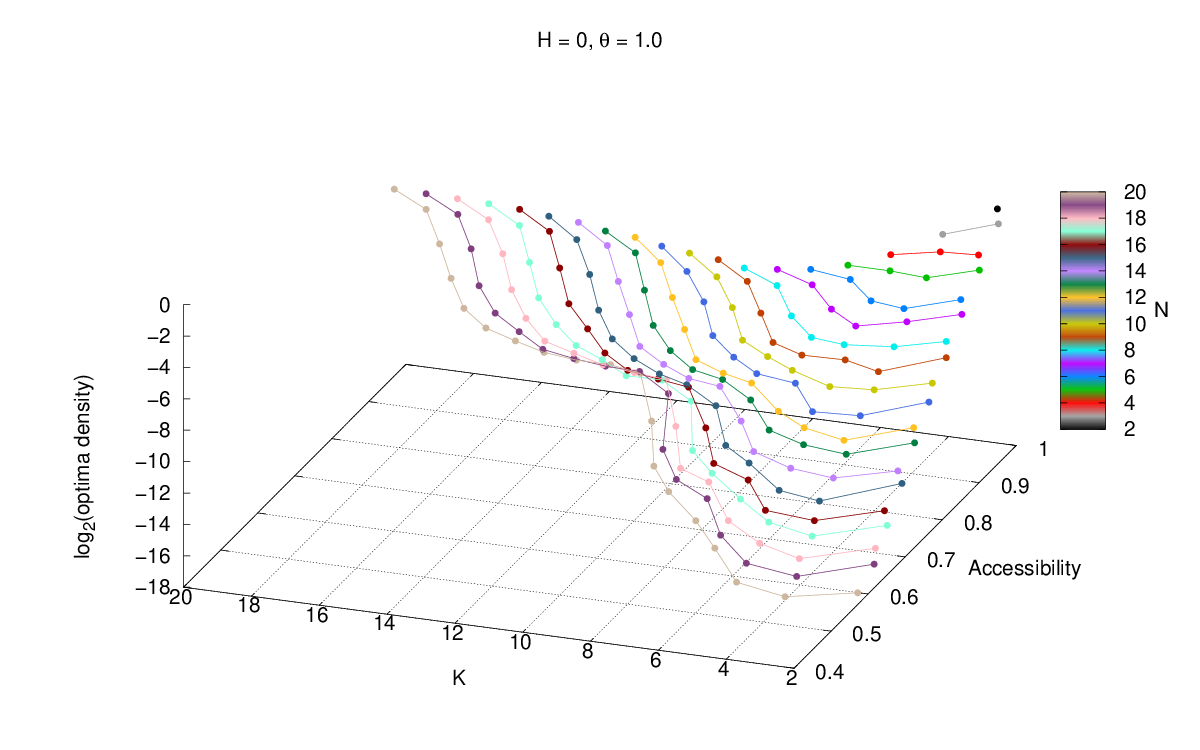}
    \\
    \includegraphics[width=0.4\linewidth]{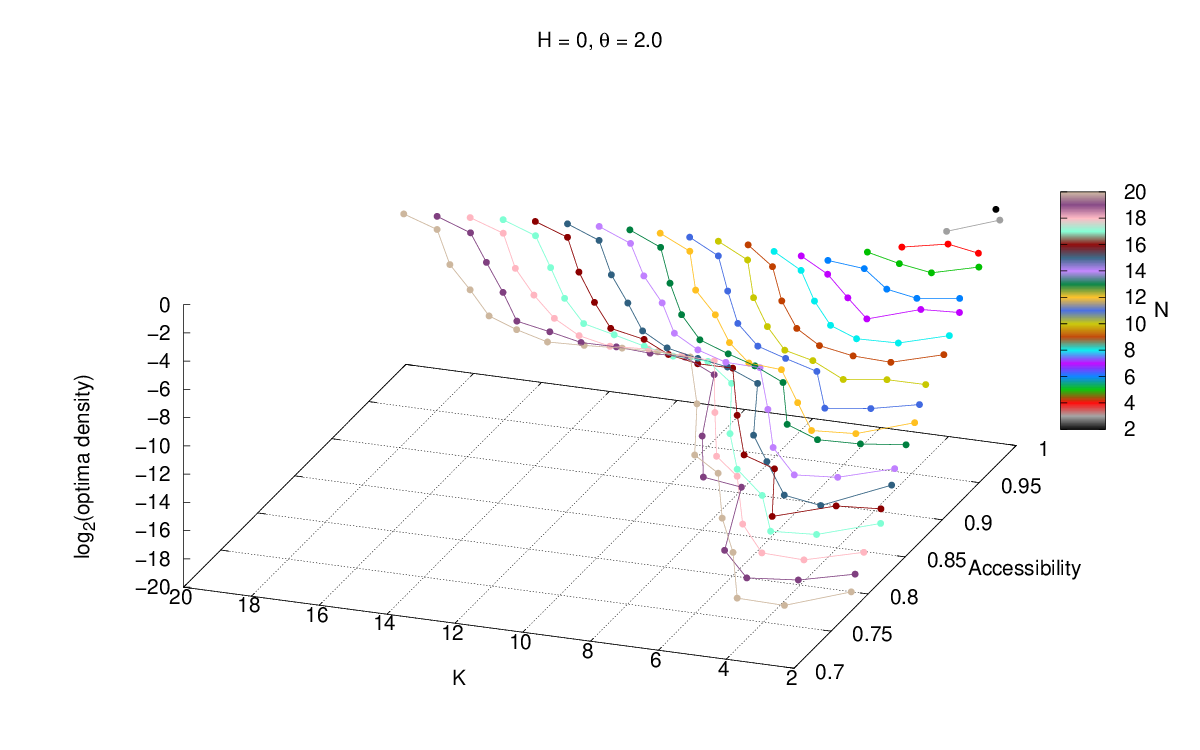}
    \end{tabular}
    \begin{tabular}{c}
    \includegraphics[width=0.4\linewidth]{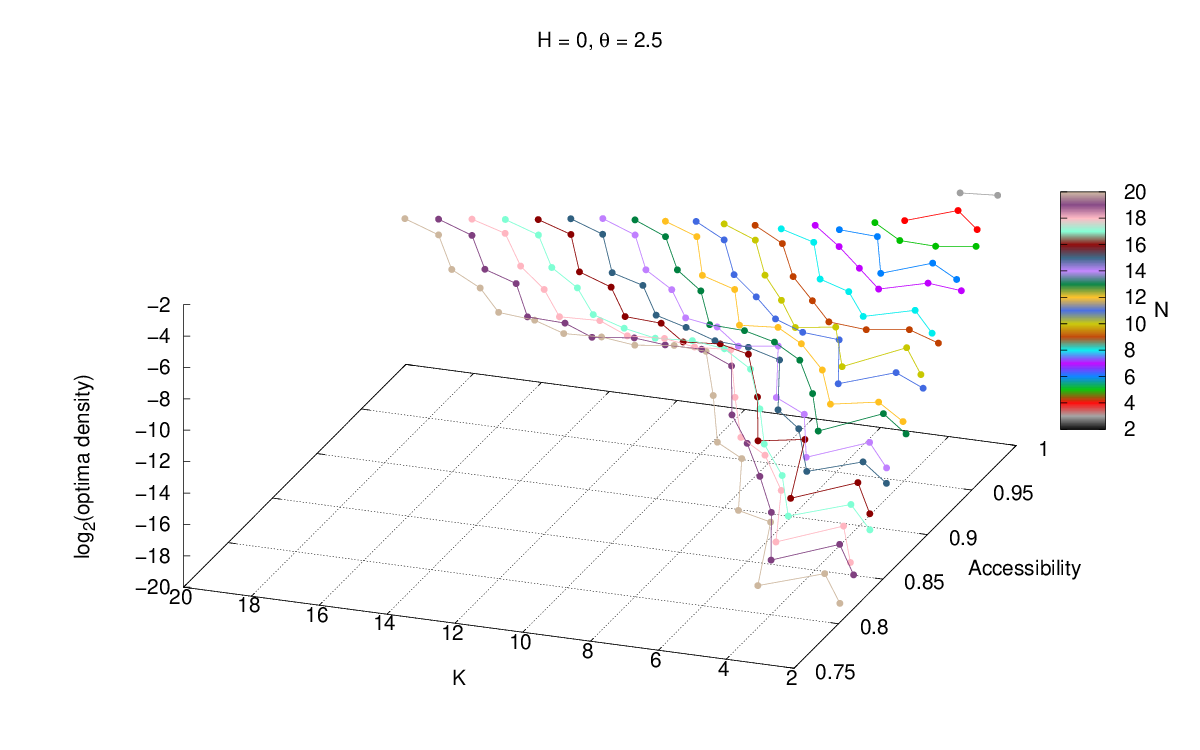}
    \\
    \includegraphics[width=0.4\linewidth]{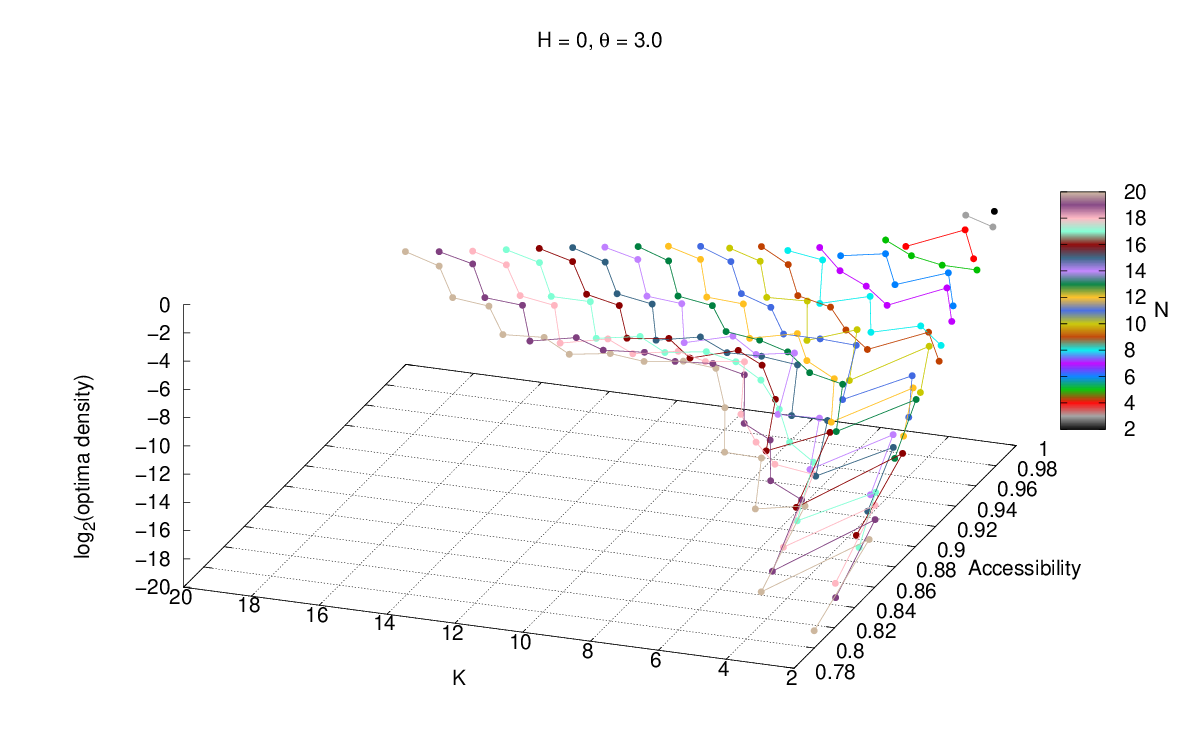}
    \end{tabular}
    \end{center}
    \caption{$\theta NK$ sweep across $\theta=0,2,2.5,3$ scatterplots}
\end{figure}

\begin{figure}[H]
    \begin{center}
    \begin{tabular}{c}
    \includegraphics[width=0.4\linewidth]{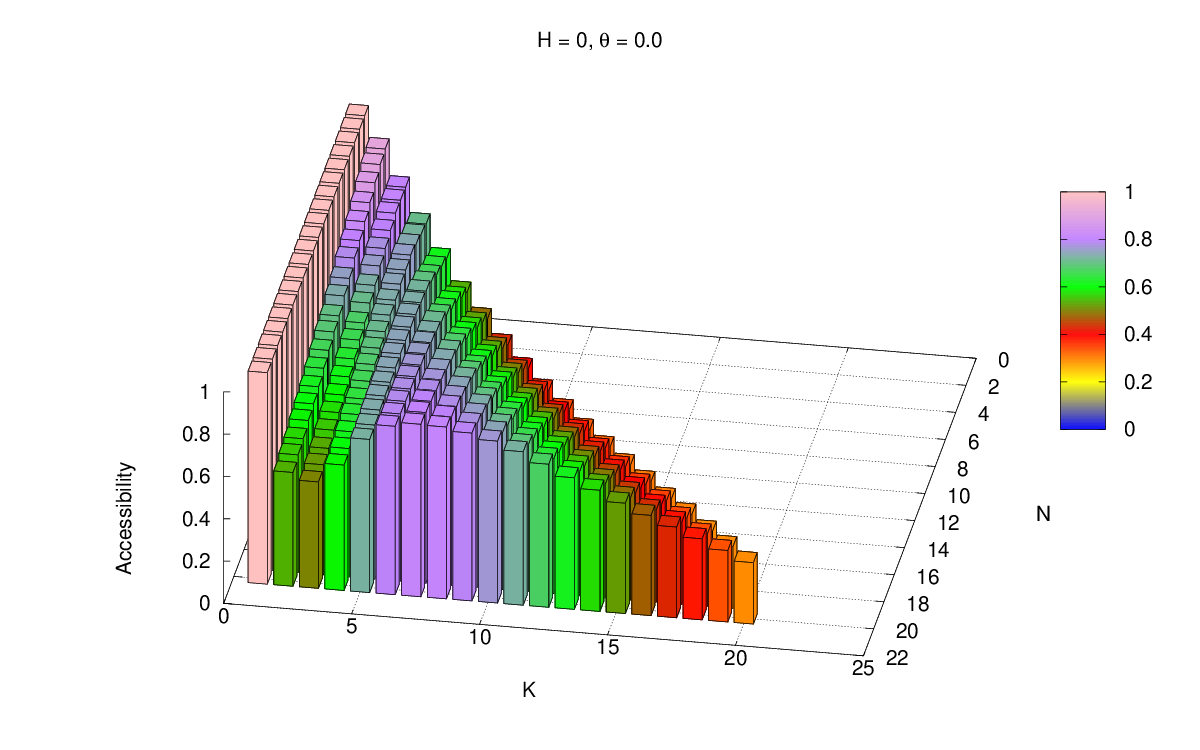}
    \\
    \includegraphics[width=0.4\linewidth]{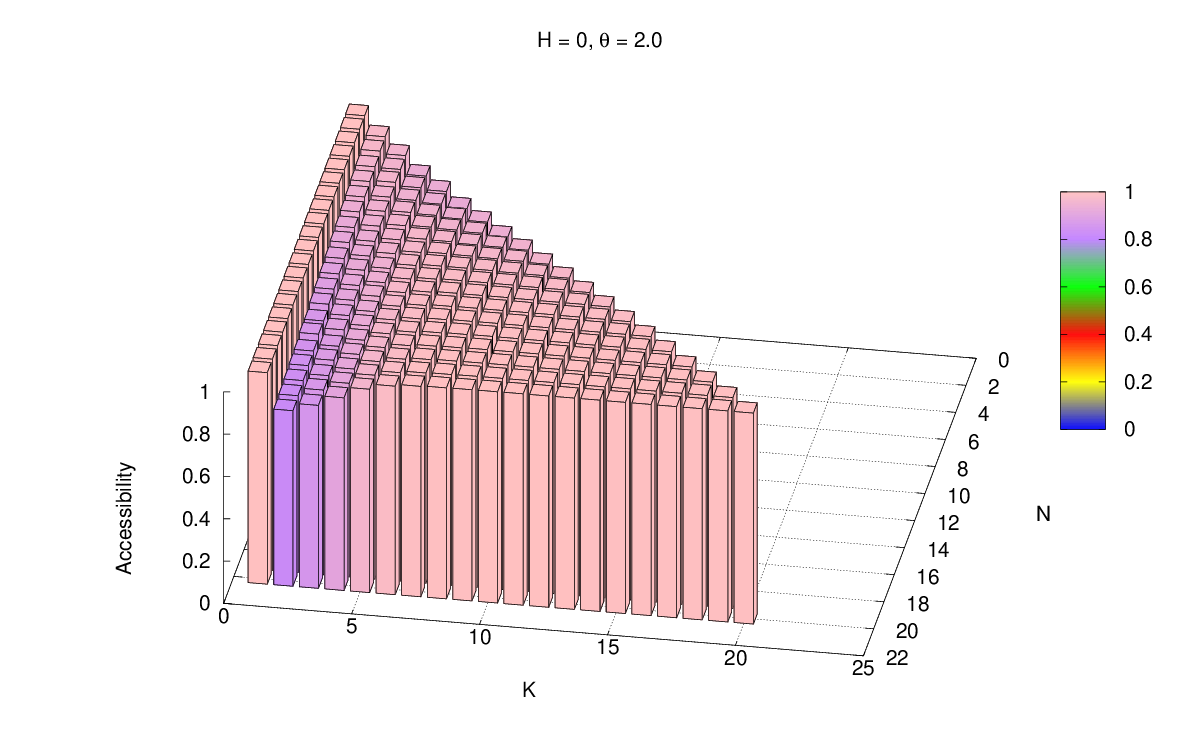}
    \end{tabular}
    \begin{tabular}{c}
    \includegraphics[width=0.4\linewidth]{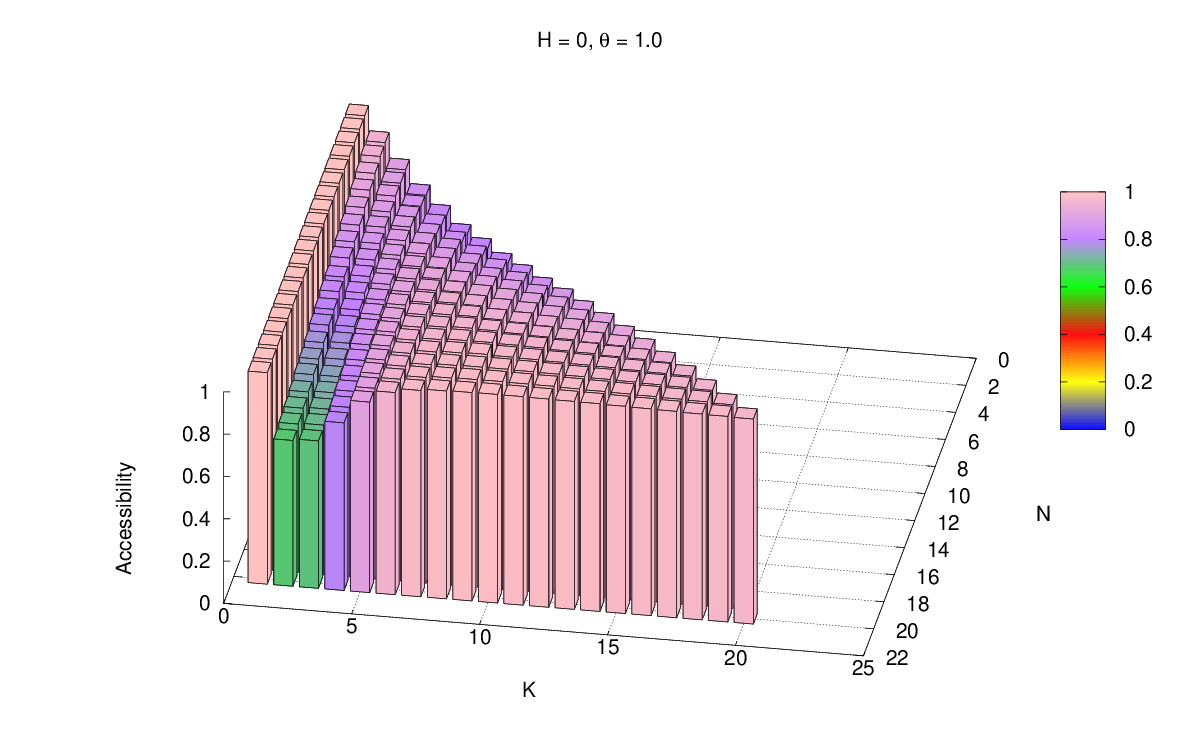}
    \\
    \includegraphics[width=0.4\linewidth]{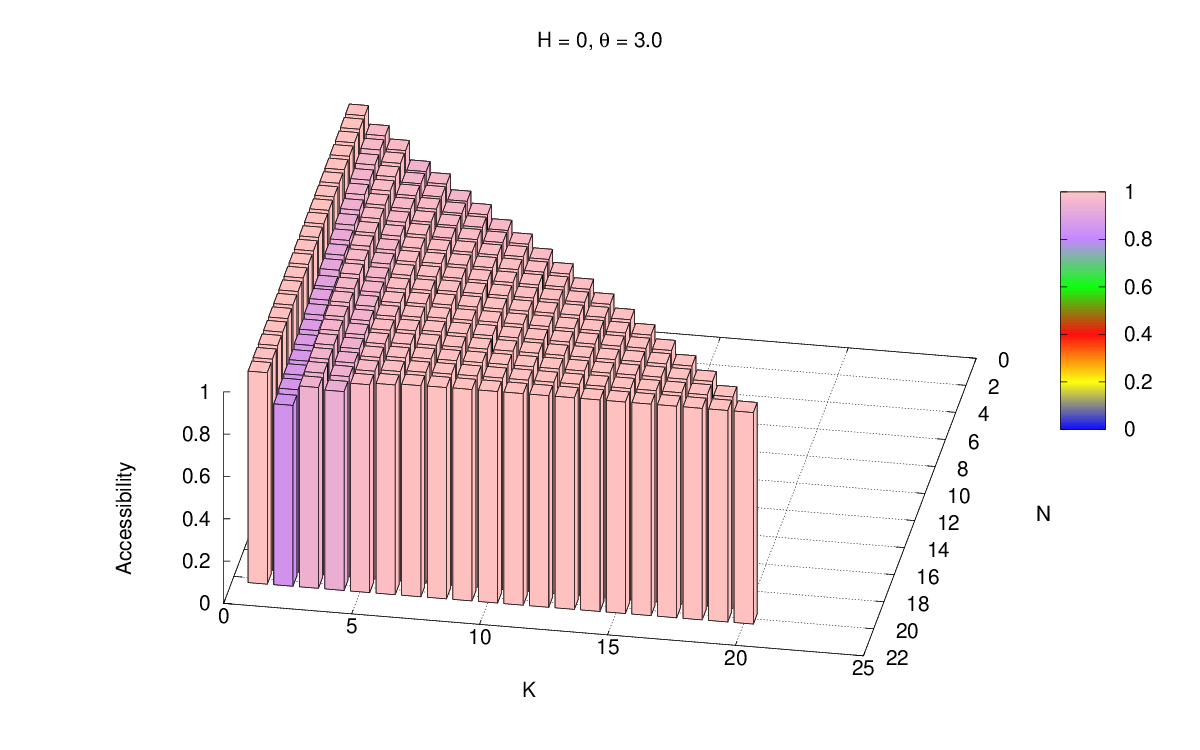}
    \end{tabular}\\
    \end{center}
    \caption{$\theta NK$ sweep across $\theta=0,1,2,3$, $N,K$ from 1 to 20, with adjacent neighborhoods}
    \label{adj}
\end{figure}

\begin{figure}[H]
    \begin{center}
    \begin{tabular}{c}
    \includegraphics[width=0.4\linewidth]{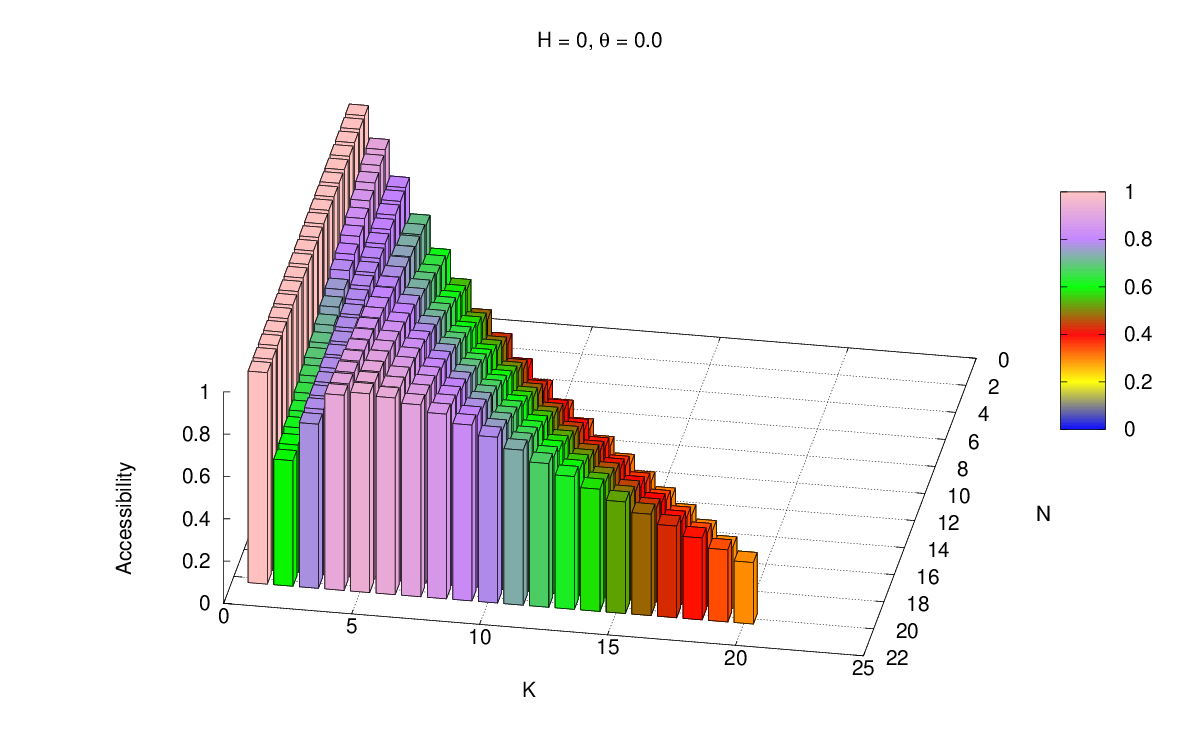}
    \\
    \includegraphics[width=0.4\linewidth]{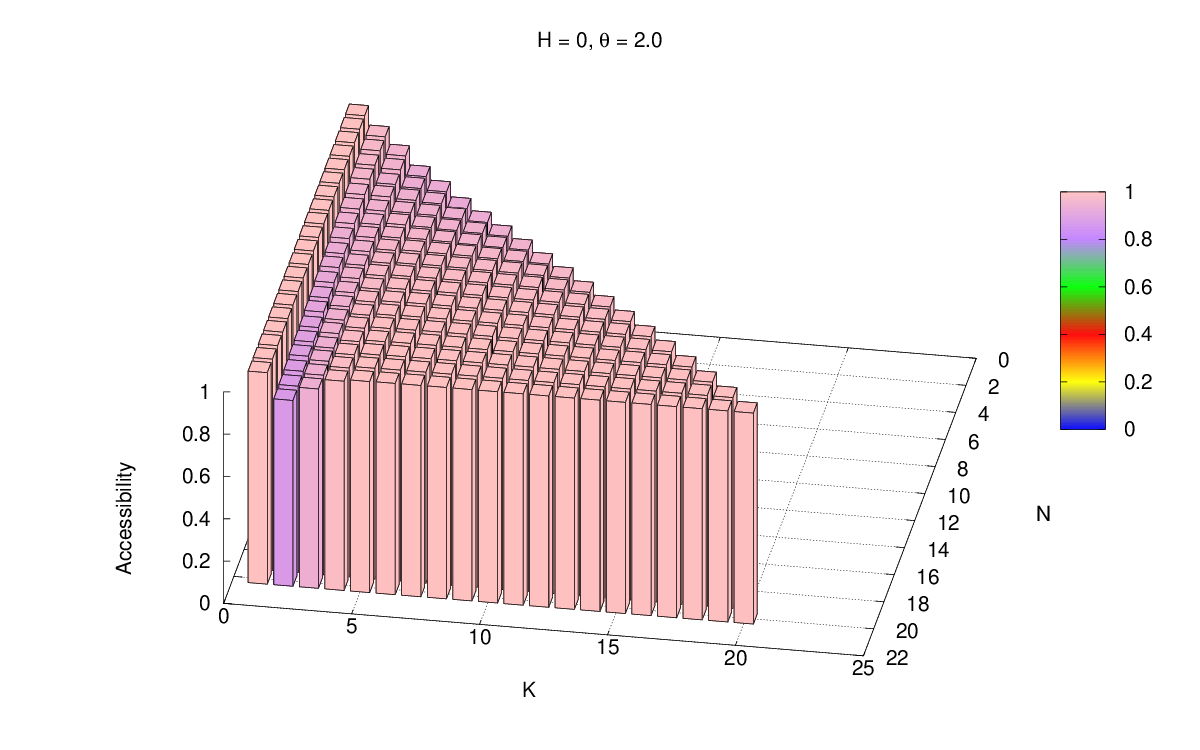}
    \end{tabular}
    \begin{tabular}{c}
    \includegraphics[width=0.4\linewidth]{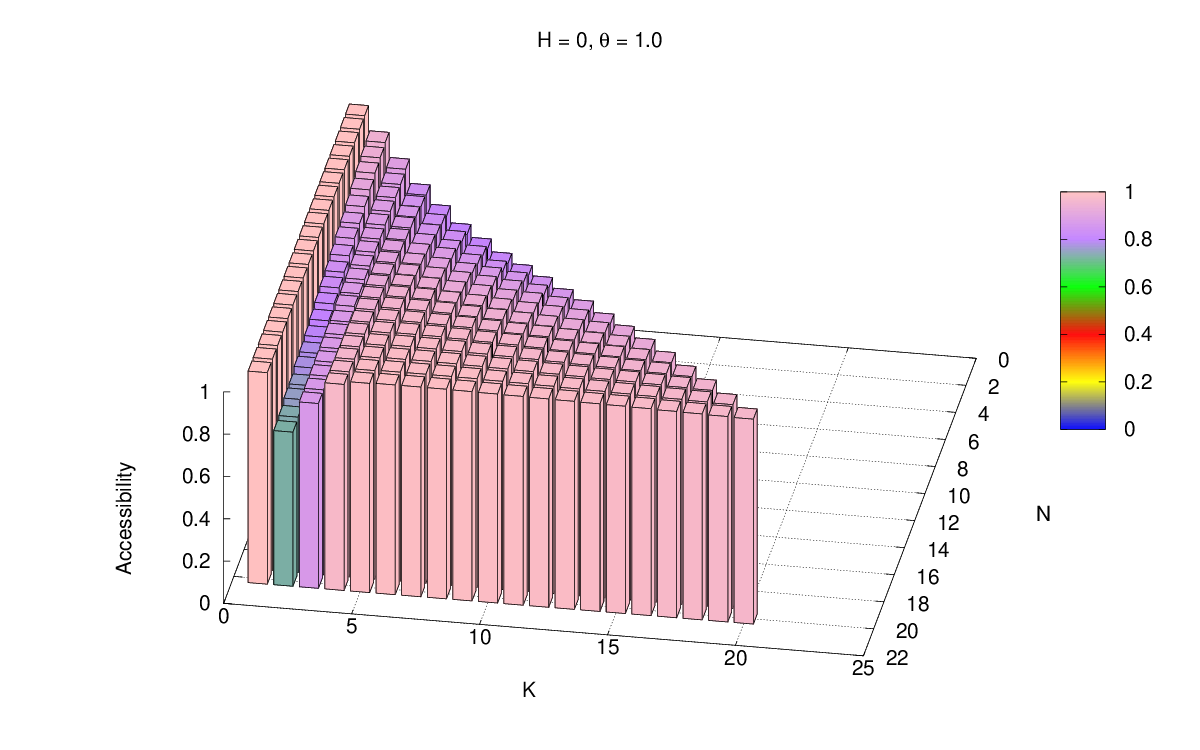}
    \\
    \includegraphics[width=0.4\linewidth]{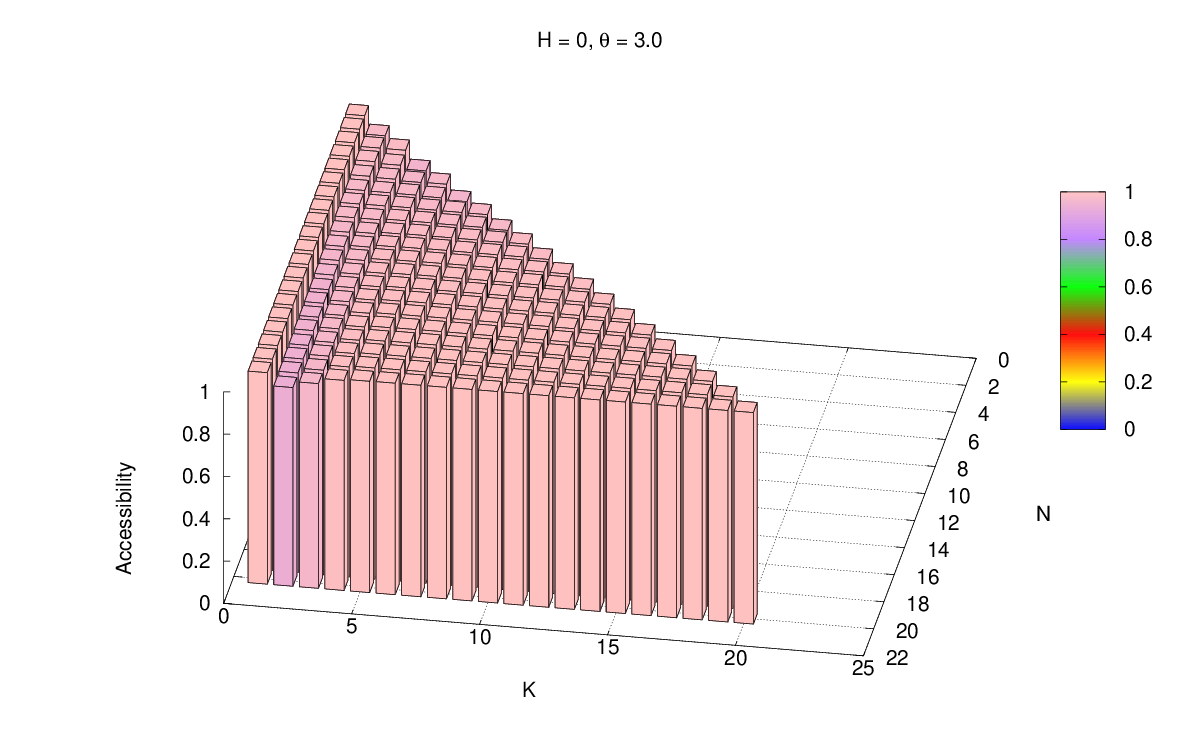}
    \end{tabular}\\
    \caption{$\theta NK$ sweep across $\theta=0,1,2,3$, $N,K$ from 1 to 20, with random neighborhoods}
    \label{rdm}
    \end{center}
\end{figure}

Because they do not contain cloistered blocks, neither adjacent neighborhood landscapes (Figure \ref{adj}) nor random neighborhood landscapes (Figure \ref{rdm}) show these same structures.

Our focus primarily became the blocked neighborhood landscapes because of their ability to be decomposed into smaller landscape sizes where $N=K$ always. Fitting a function to the accessibility data we obtained in simulation offered a method to predict $p_1$ for arbitrary $\theta NK$ landscapes, and so became of interest to our investigation. This topic is explored more in Appendix B.

\subsection{Accessibility in $HNK$}
\begin{figure}[H]
    \begin{center}
        \includegraphics[width=0.6\linewidth]{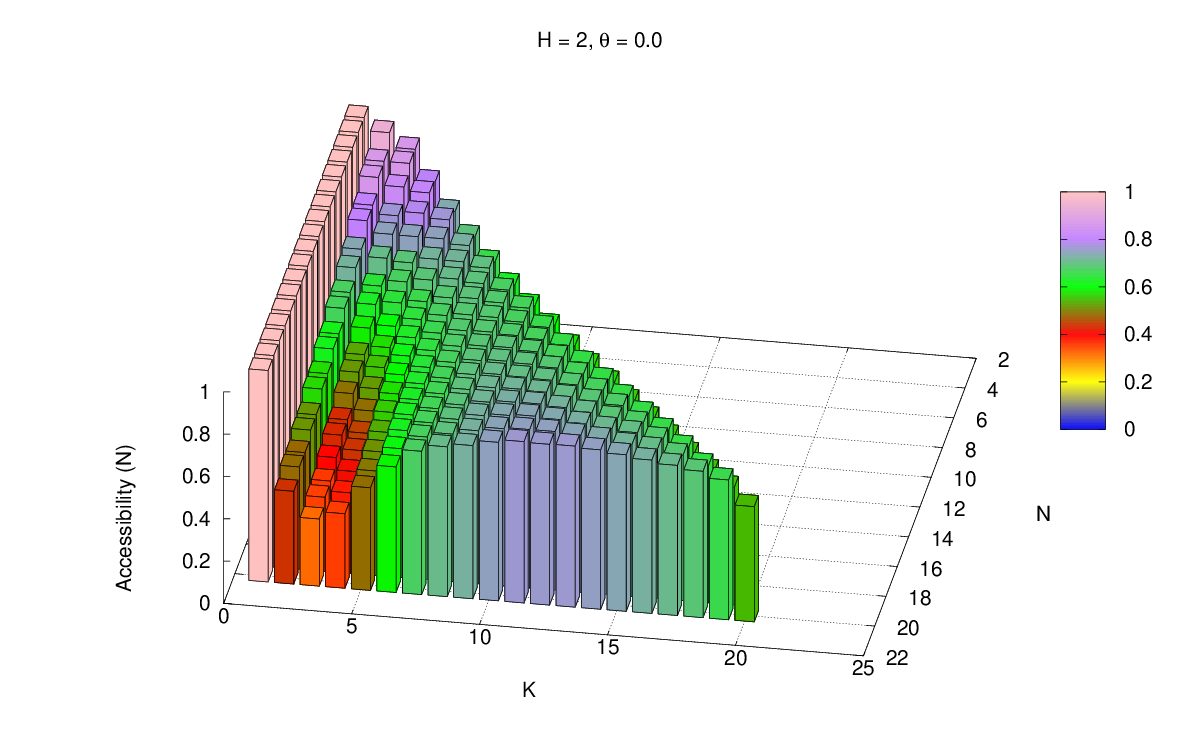}
        \includegraphics[width=0.6\linewidth]{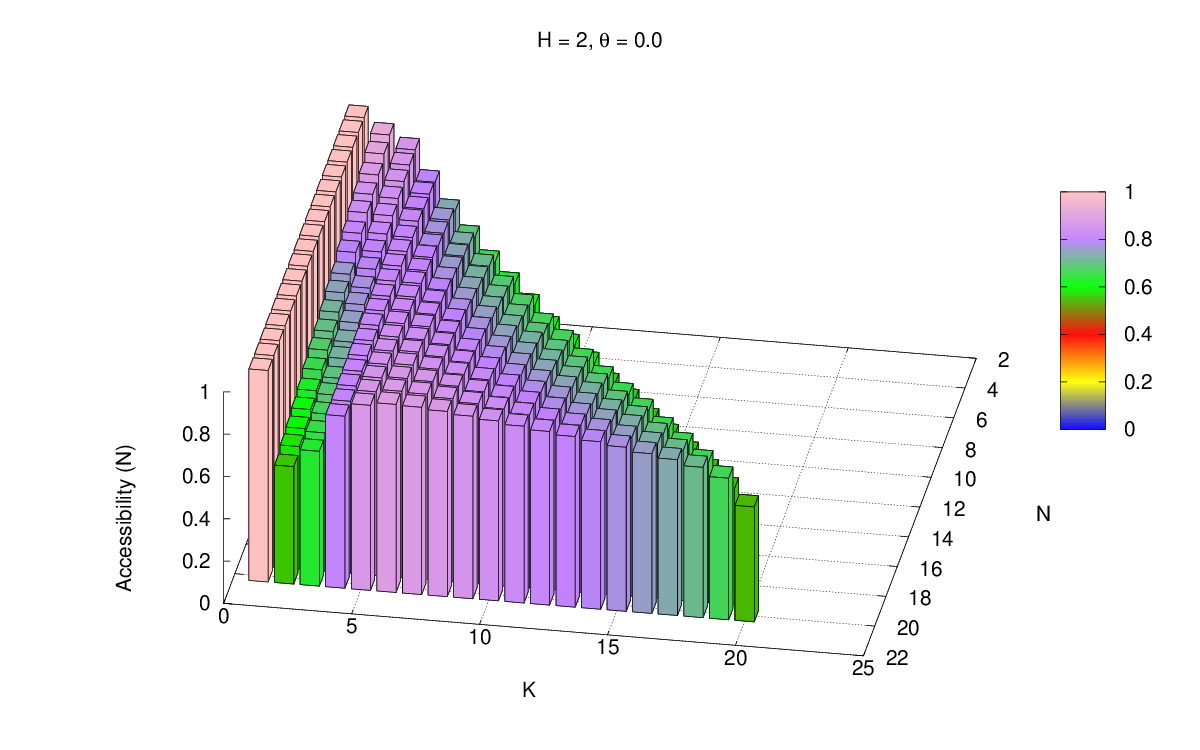}
        \includegraphics[width=0.6\linewidth]{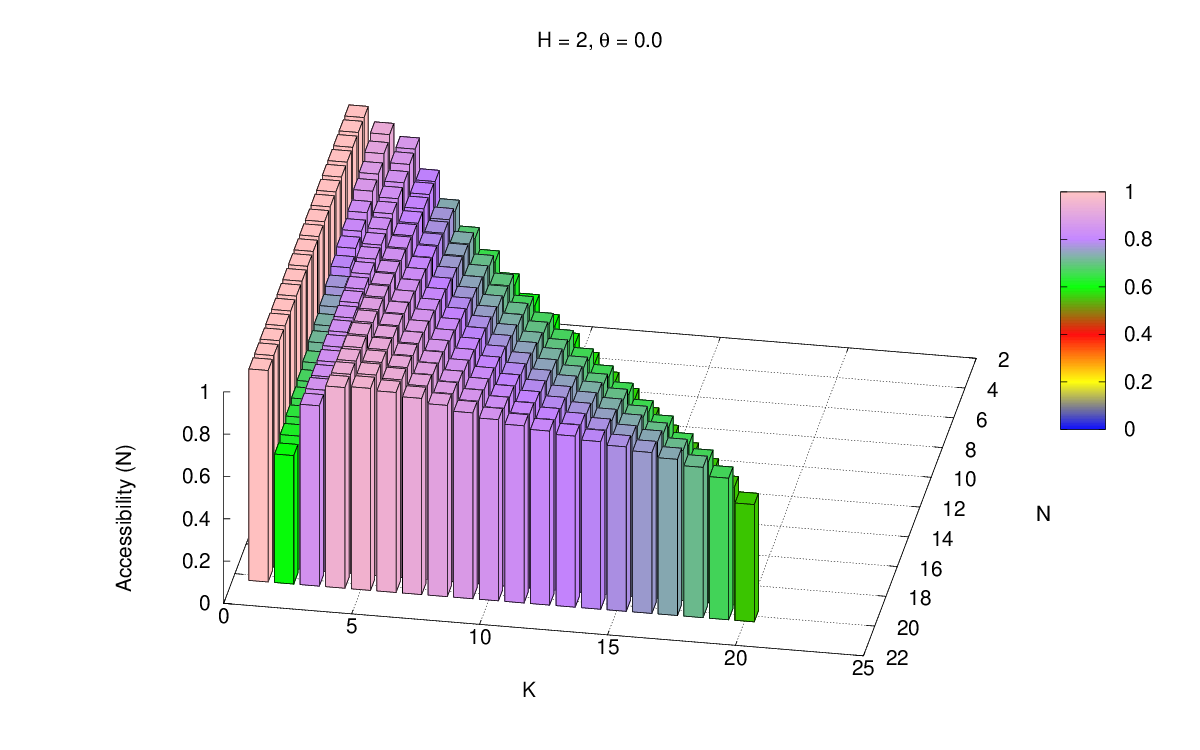}
    \end{center}
    \caption{$HNK$ accessibility plots for $H=2$, $N,K$ from 1 to 20, across the three primary neighborhood types; blocked on top, adjacent in the middle, and random on the bottom}
    \label{HNK}
\end{figure}

The simulation results in $HNK$ showed that introducing $H$-segment loci naturally broke the multiplicative decomposition pattern seen in blocked neighborhood $NK$ landscapes and created a similar shape among all three-dimensional accessibility graphs. Figure \ref{HNK} shows the graphs for blocked neighborhoods, adjacent neighborhoods, and random interactions, all for the case of $H=2$ starting at $N=3$ (respectively, from left to right). 

The common shape where accessibility is greater in the region where $K$ is close to $\frac{N}{2}$ and decreases near the edges (except $K=1$, of course, since this always causes a landscape to revert to additive fitness across the loci) is created by the fact that our formulation of $HNK$ required each $H$-segment locus to interact with precisely $K$ $N$-segment loci. This makes $H$ more potent as $K$ increases, but causes a sharp reversion toward the $H=0$ accessibility when $K$ approaches $N$ due to the fact that each $H$-segment locus masks all but 1 or 2 loci, requiring that $\sigma_H=11\dots1$ at the global optimum unless $\sigma_N=00\dots0$ is the optimum.

Accessibility does not reach 1 even at its greatest extent for $K>1$ and $H>0$ on any of these landscapes. From a biological perspective, this suggests that even landscapes where many genes can be masked at a time inevitably struggle to reach full accessibility because performing that masking requires the phenotype used for calculating fitness ($\gamma$) to assume some particular pattern (in our case, many loci being allele 0) that may be significantly less fit than the fully-expressed starting genotypes that do not have adaptive walks to the global optimum. As such, the genotypes that lack accessible pathways would have to first decrease in fitness to reach a region of genotypes where an $H$-segment locus could be switched to 0, and then other loci in the $N$-segment could change to eventually emerge on the path to the global optimum.

Distinction was drawn in simulating $HNK$ between cases where the global optimum was fully expressed and those in which it was not, and the accessibility of non-fully expressed global optima was higher than that of fully expressed optima. The number of fully expressed optima changes sharply with $K$, from almost all optima being non-fully expressed when $K=1$ to almost all optima being fully expressed as $K$ approached $N$. 

\begin{figure}[H]
    \begin{center}
        \begin{tabular}{c c c c}
        \includegraphics[width=0.2\linewidth]{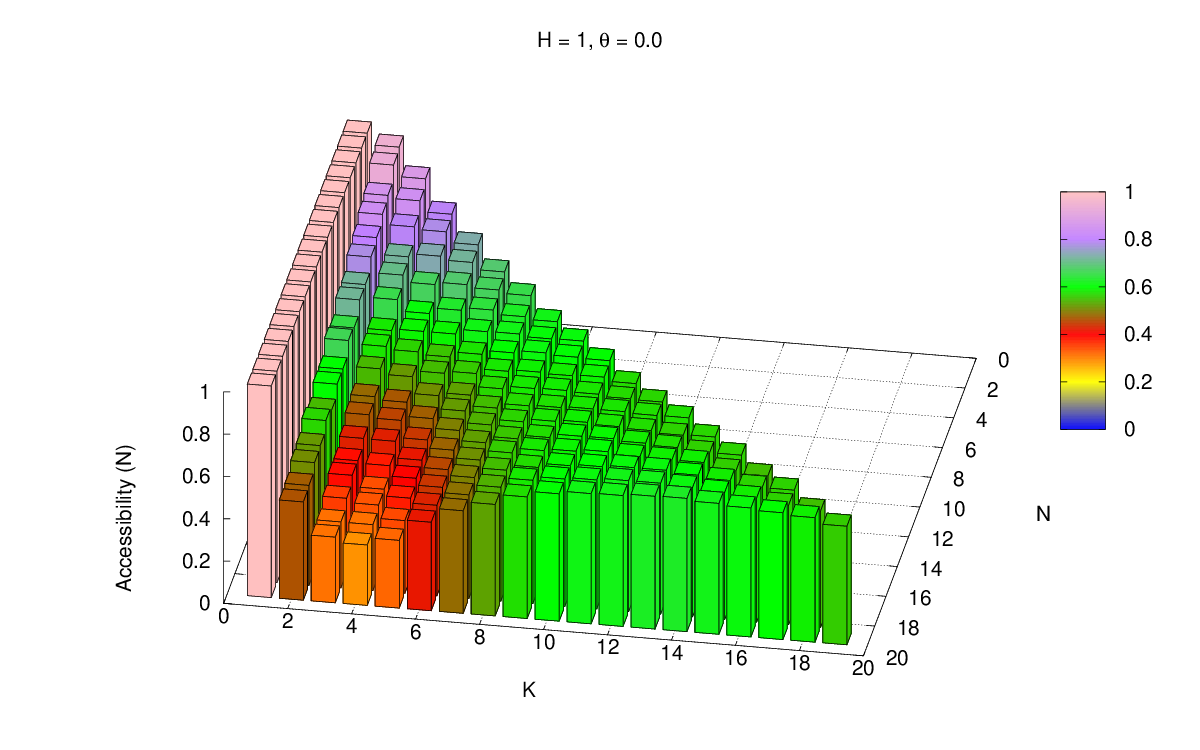}
        &
        \includegraphics[width=0.2\linewidth]{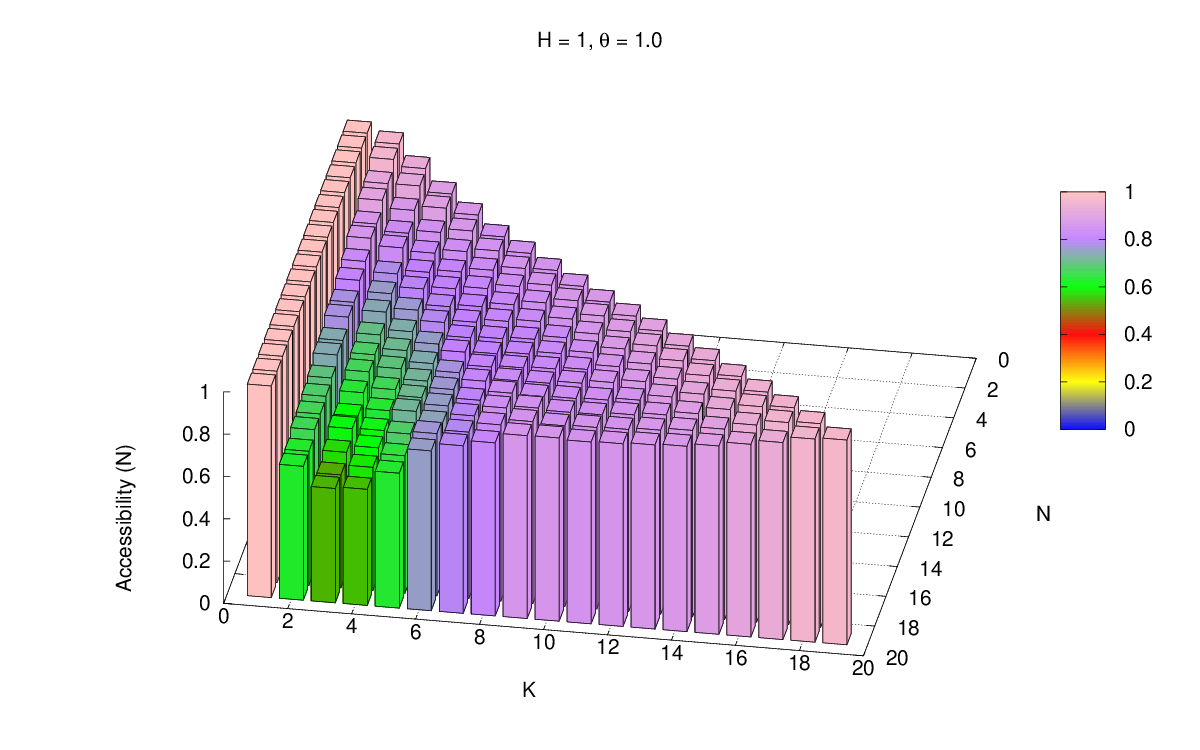}
        &
        \includegraphics[width=0.2\linewidth]{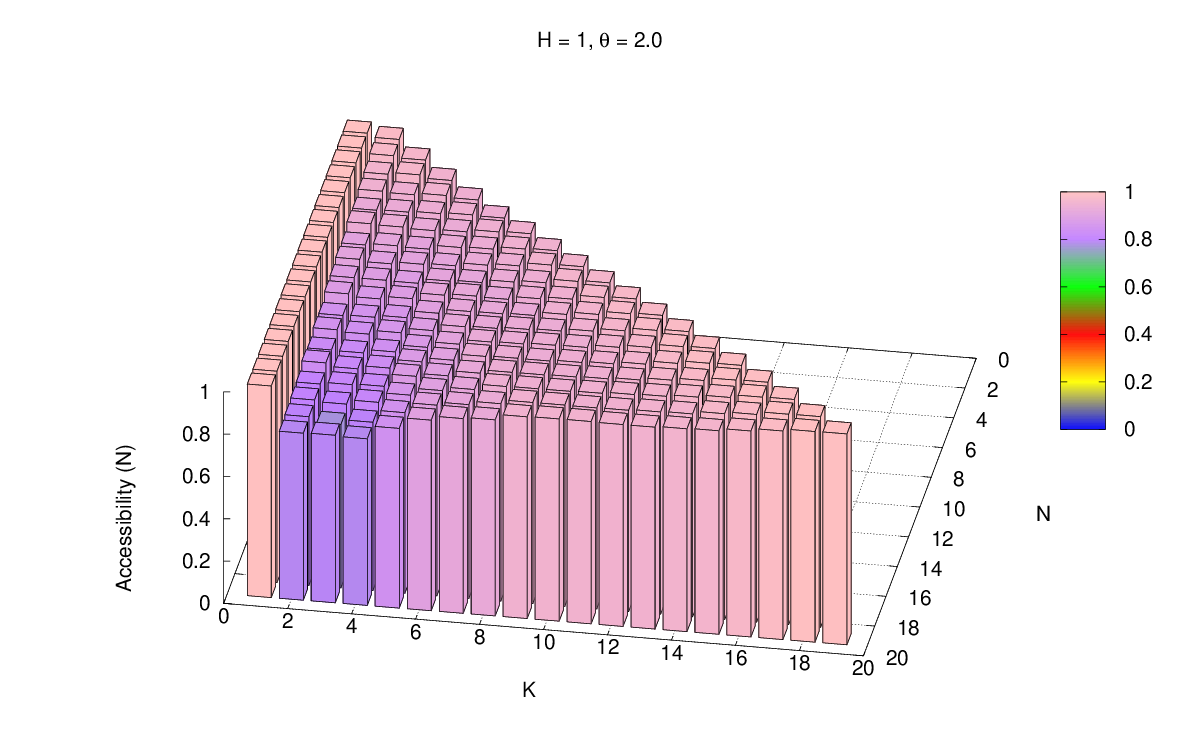}
        &
        \includegraphics[width=0.2\linewidth]{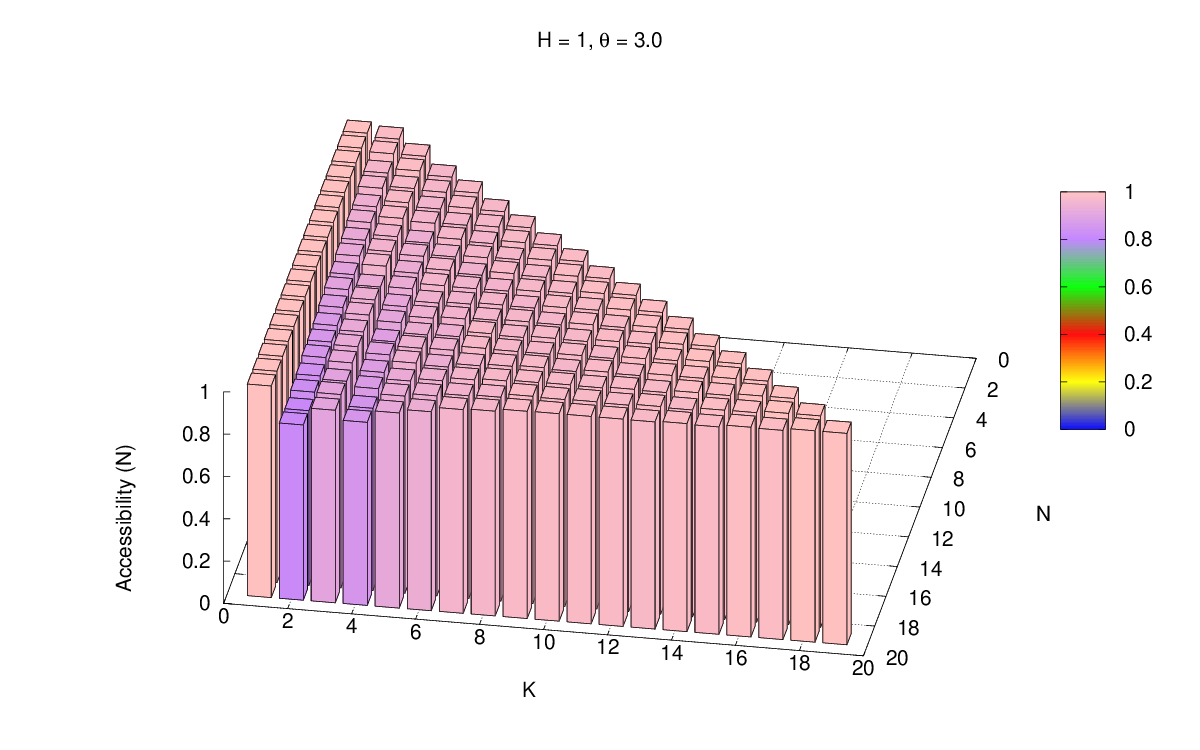}
        \\
        \includegraphics[width=0.2\linewidth]{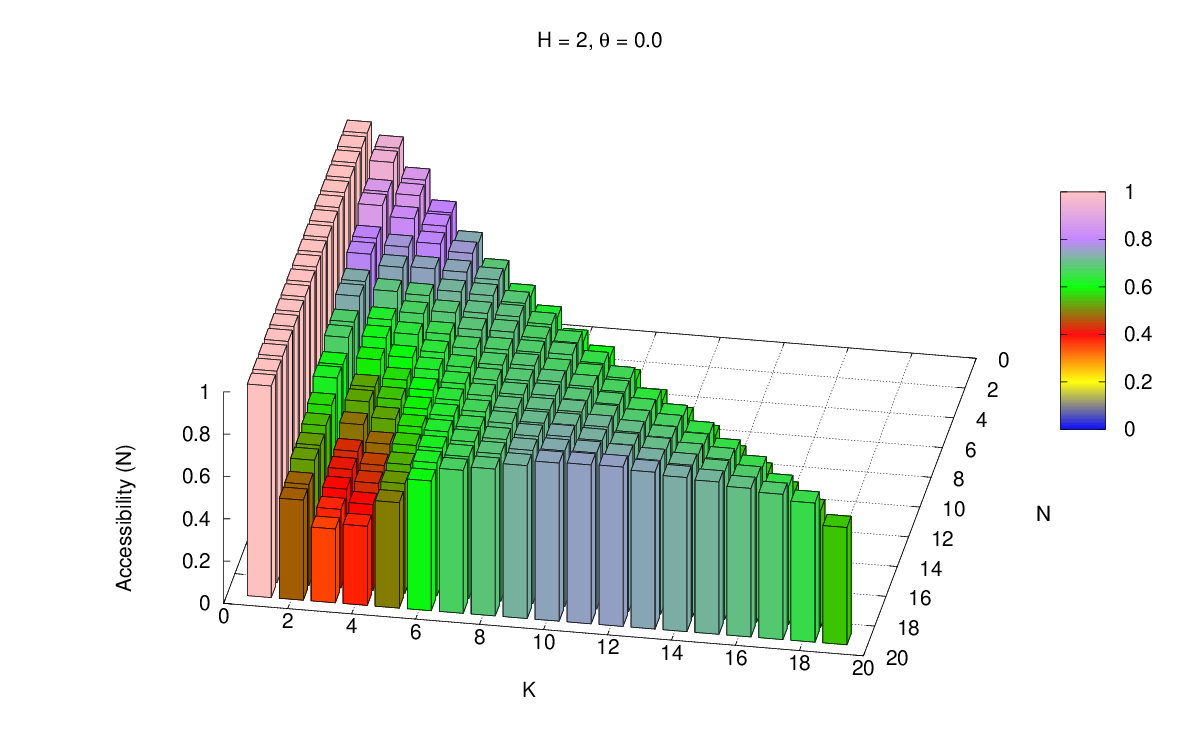}
        &
        \includegraphics[width=0.2\linewidth]{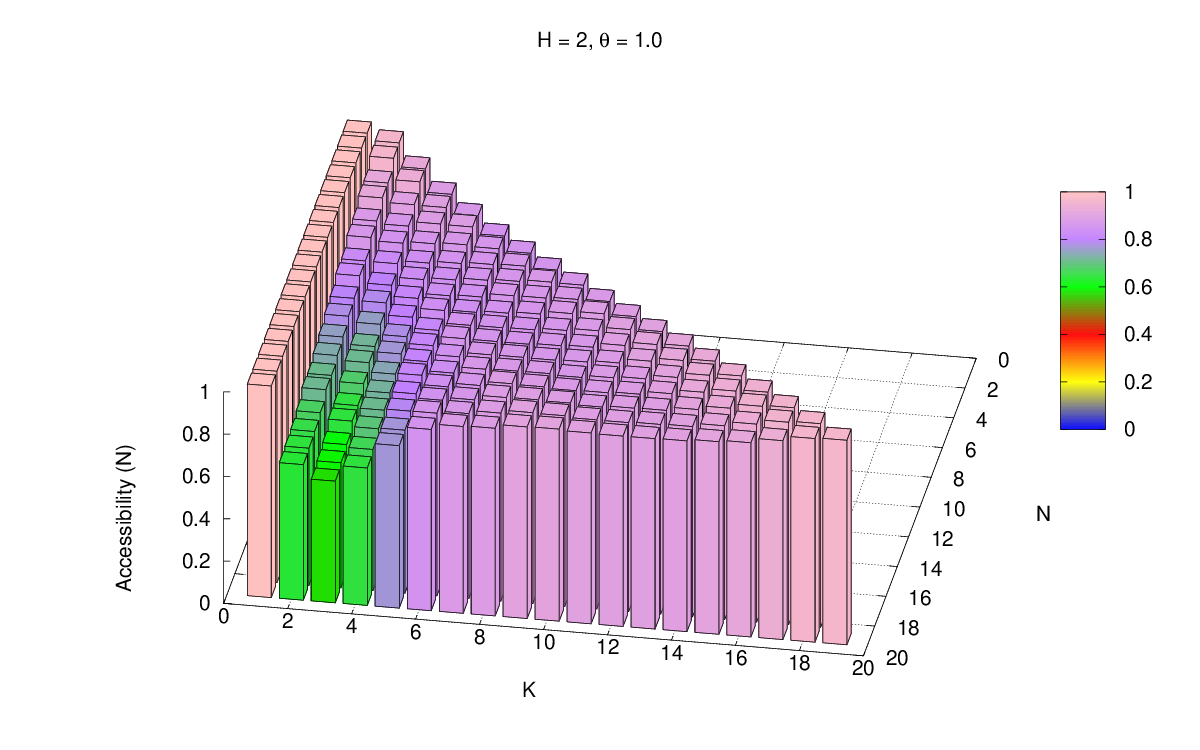}
        &
        \includegraphics[width=0.2\linewidth]{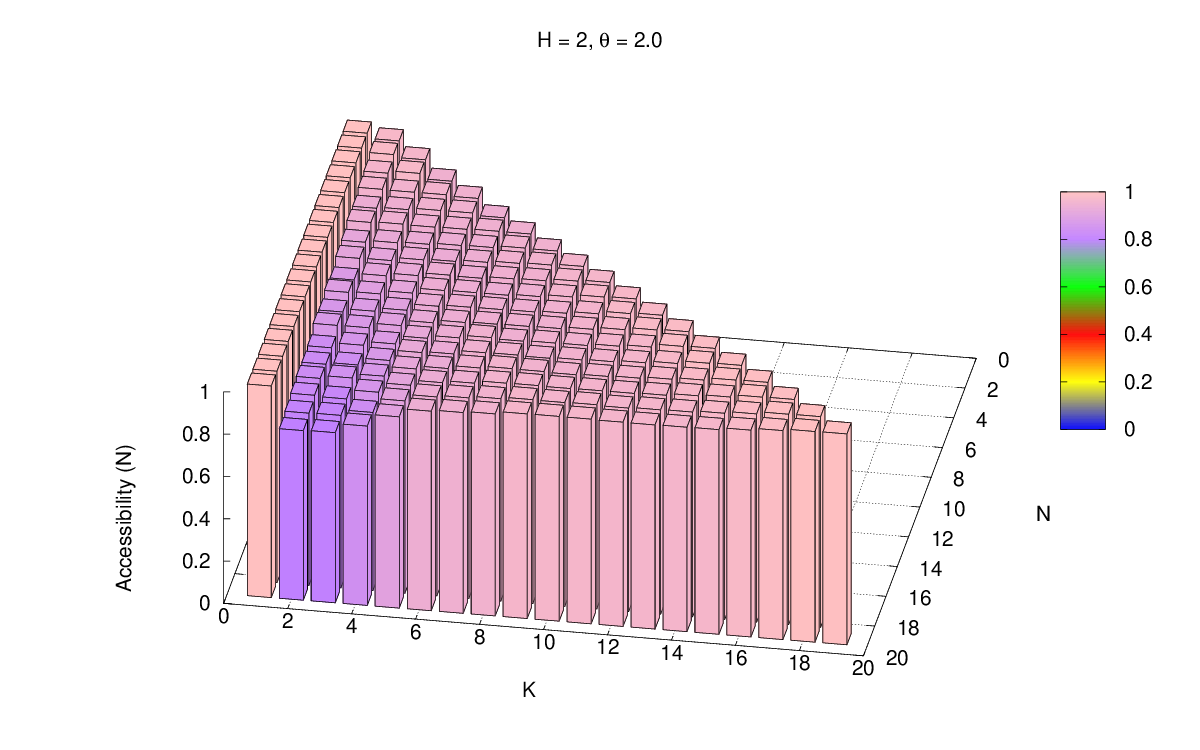}
        &
        \includegraphics[width=0.2\linewidth]{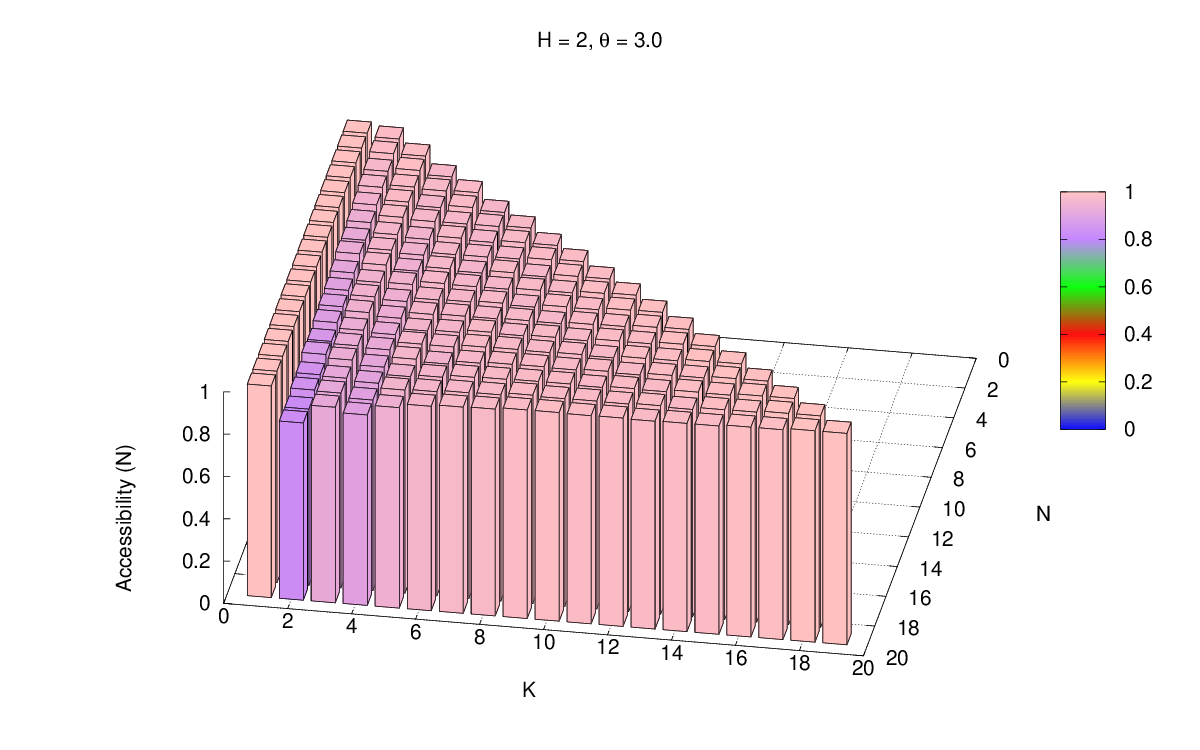}
        \\
        \includegraphics[width=0.2\linewidth]{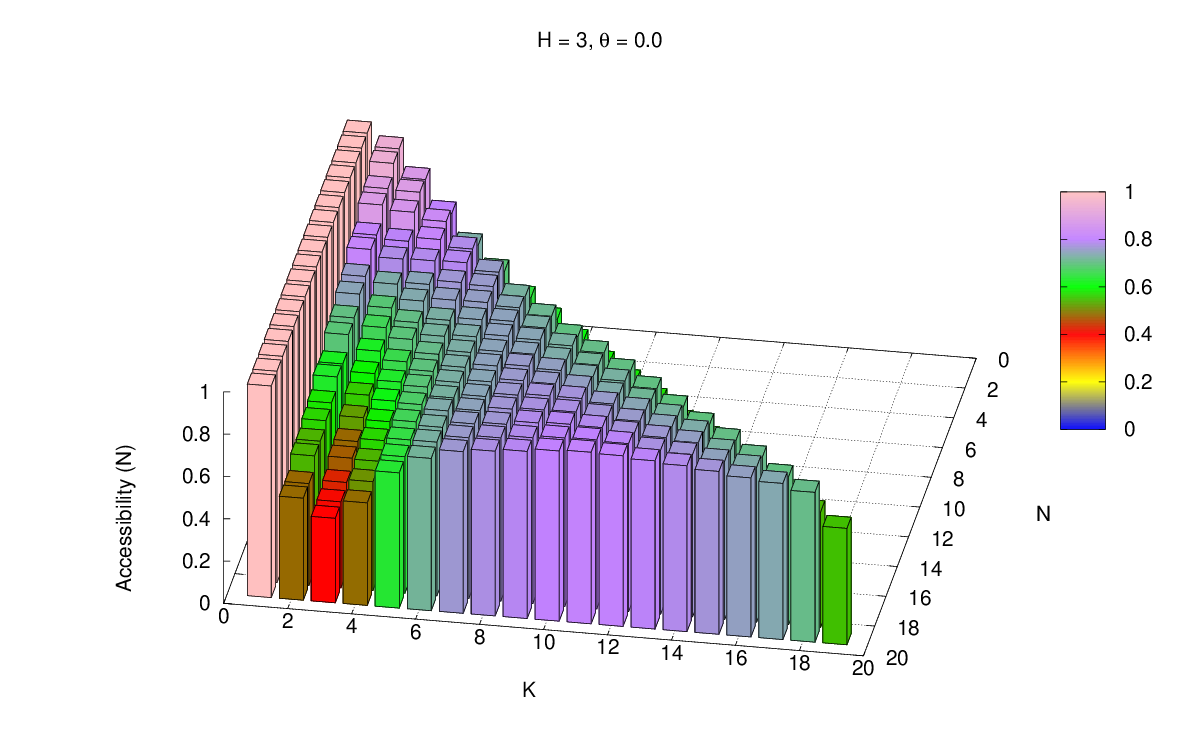}
        &
        \includegraphics[width=0.2\linewidth]{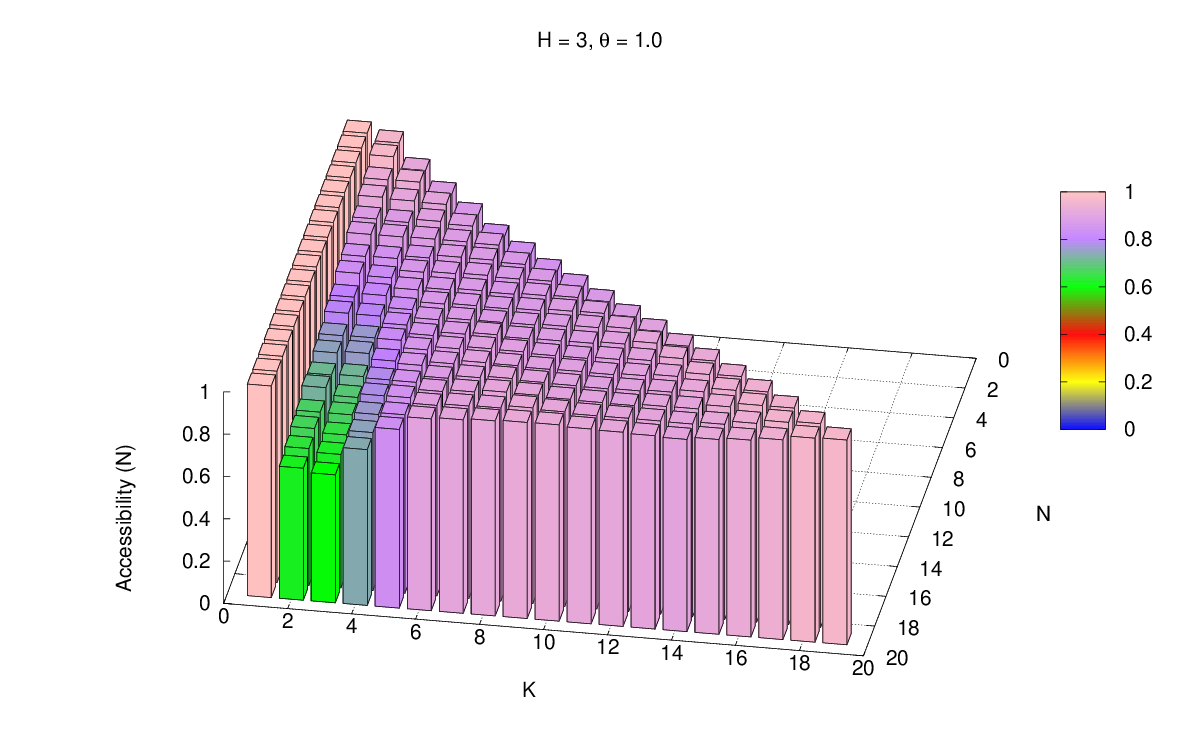}
        &
        \includegraphics[width=0.2\linewidth]{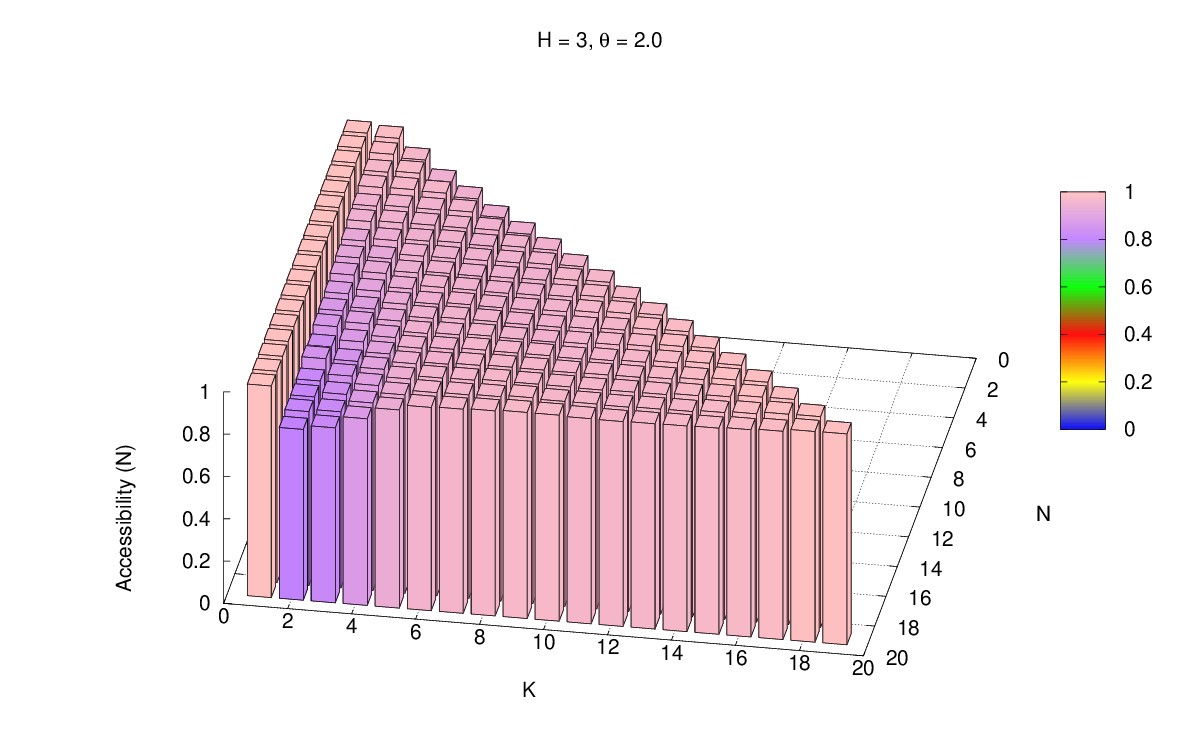}
        &
        \includegraphics[width=0.2\linewidth]{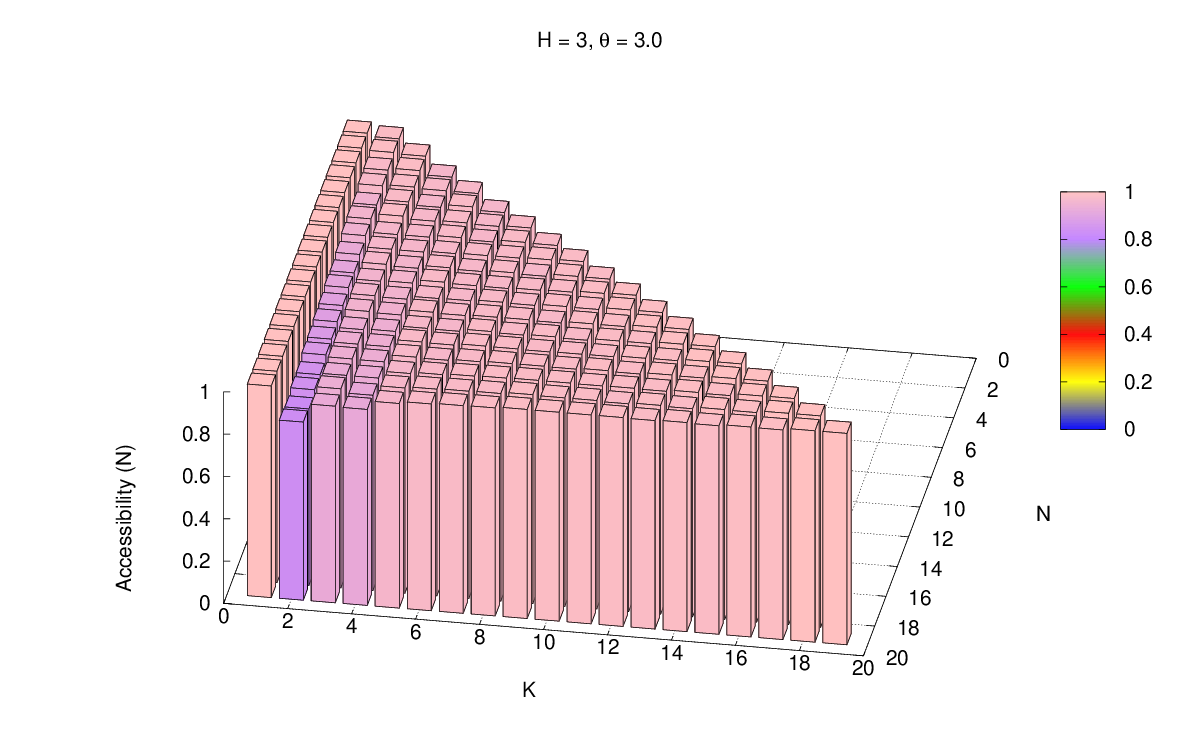}
        \\
        \includegraphics[width=0.2\linewidth]{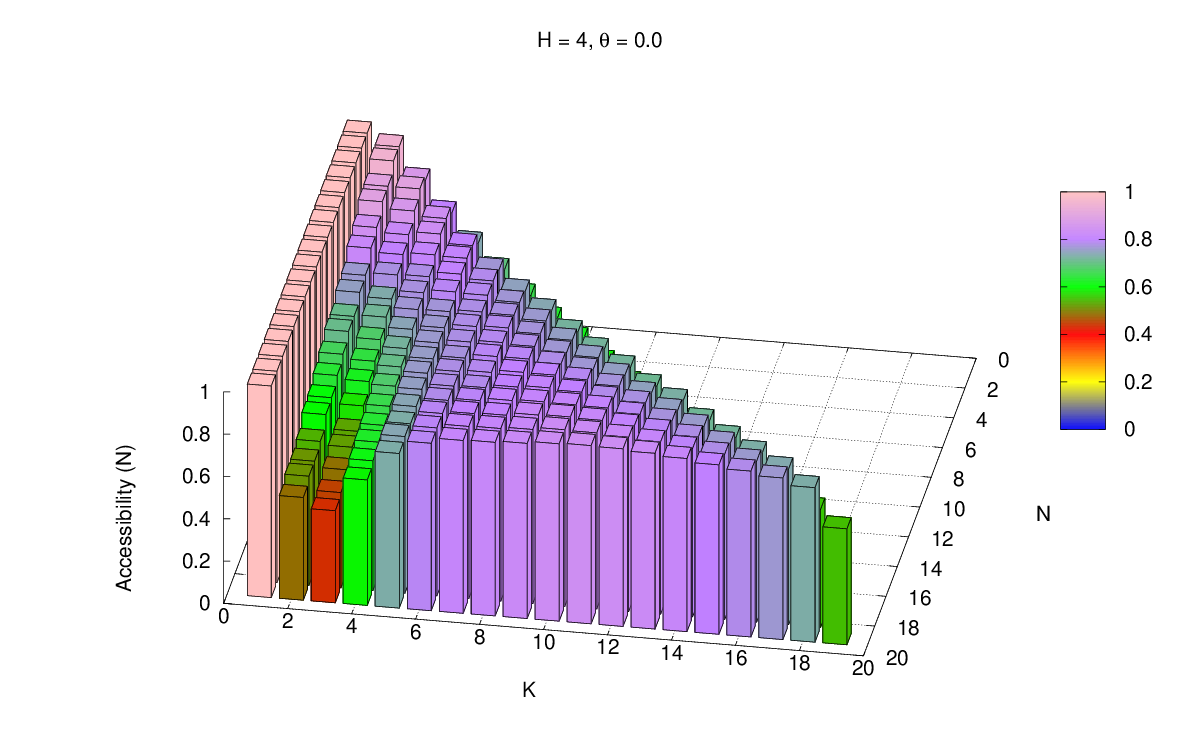}
        &
        \includegraphics[width=0.2\linewidth]{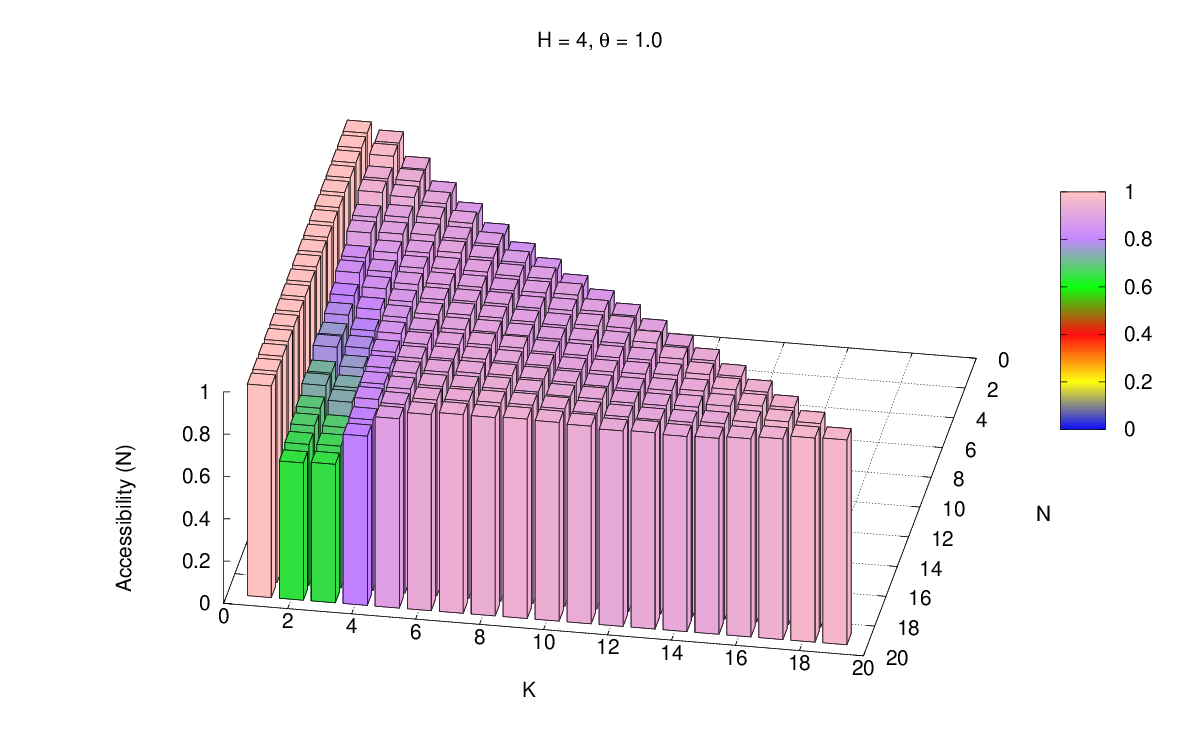}
        &
        \includegraphics[width=0.2\linewidth]{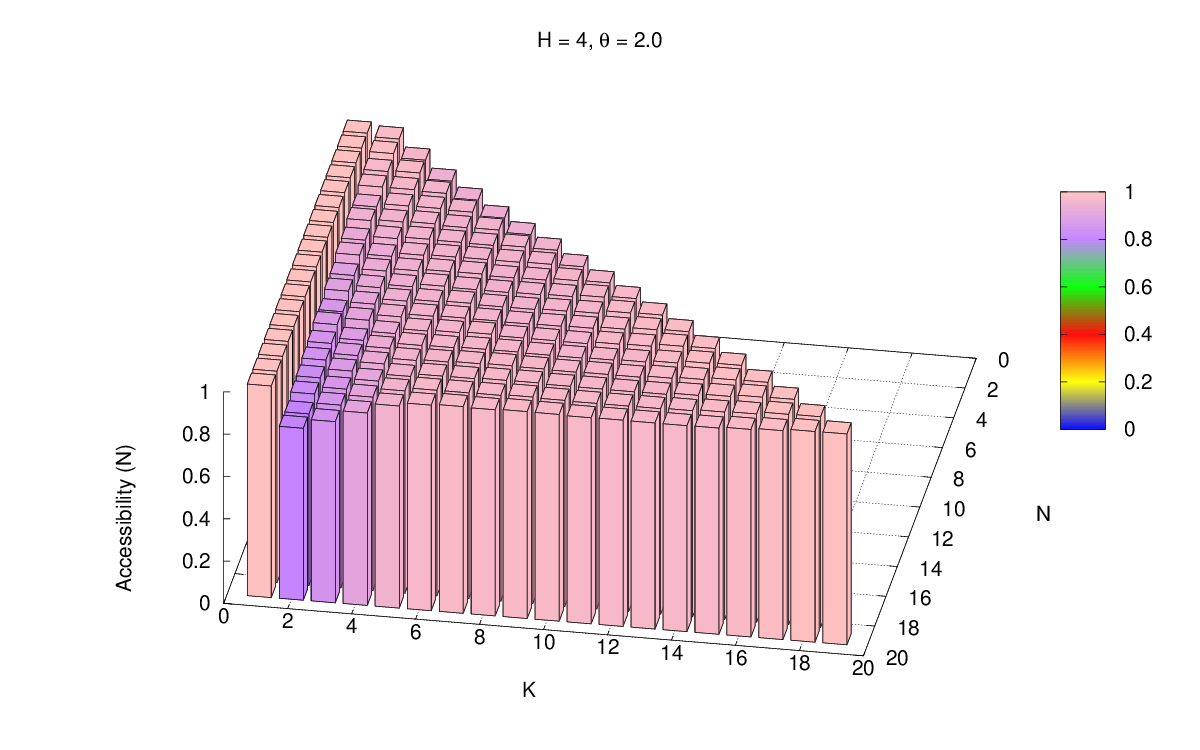}
        &
        \includegraphics[width=0.2\linewidth]{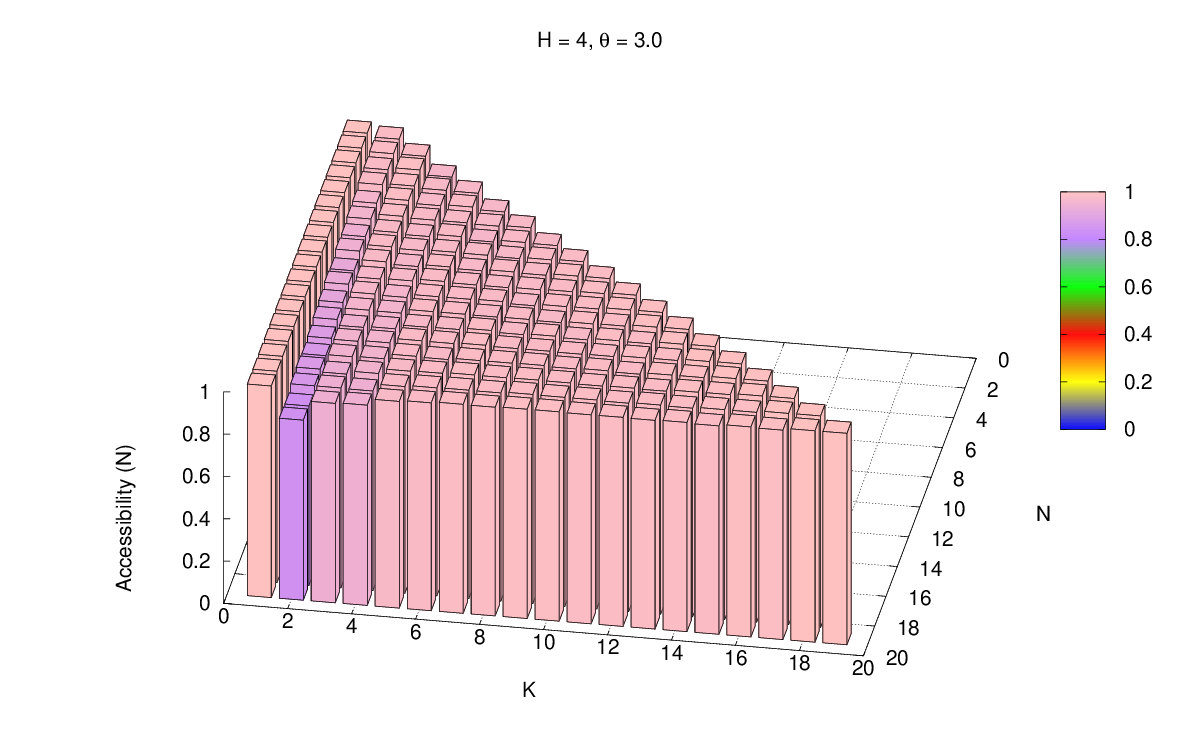}
        \end{tabular}
    \end{center}
    \caption{Accessibility plots across $H=1,2,3,4$, $\theta=0,1,2,3$, and $N,K$ from 1 to 20 for $H\theta NK$ landscapes}
    \label{4x4}
\end{figure}

We also investigated the influence of $\theta$ being applied to the $N$-segment $NK$ model to see if its effect dominated $H$ or not. The graphs in Figure \ref{4x4} show accessibility versus $N$ and $K$ for $\theta=0,1,2,3$ and $H=1,2,3,4$ on blocked neighborhood landscapes.

We observed that $H$ dominates $\theta$, since $\theta$ tends to influence the $N=K$ ridge more and $H$ influences the $K < N$ cases more (and can influence them directly, versus for blocked neighborhood landscapes where said region of these graphs is just the product of values on the ridge). However, as previously described, $HNK$ reverts to $NK$ on the $N=K$ ridge since each $H$-segment locus must mask the entire $N$-segment in that case, so the behaviors discussed in the preceding section regarding the $N=K$ ridge under the influence of $\theta$ still apply. 

\section{Future Directions}
There are many ways that this investigation could be furthered.

\subsection{Flexible $HNK$}
One crucial area for potential expansion of this work is the modification of the $HNK$ model so that $|G_h| \neq K$. This would free the effect of $H$ to be studied fully independently from the influence of $K$, and would eliminate some of the more unintuitive behaviors (like the reversion to effectively $H=0$ when $N=K$). The original justification for modeling $HNK$ with $|G_h|=K$ was that the parameter $K$ gives a sense of how interconnected the loci are to each other, and if there are loci interacting with $K$ others in the $N$-segment, it would be most biologically reasonable to assume that loci in the $H$-segment would also interact with $K$ others. This does not strictly need to be true.

\subsection{$N=K$ Fitting}
There is also considerable opportunity to further the work done fitting curves to the $N=K$ ridge. The fits presented above are less true analytical descriptions of the central tendencies of the underlying functions producing $p_1$ on these landscapes and more attempts to describe the trends and end behavior of the simulations analytically. 

One particularly promising shape that was not explored in this investigation would be fitting gamma distributions, or inverse gamma distributions, to the overall shape of the accessibility on the $N=K$ ridge rather than attempting to separate it into a $\theta=0$ and $\theta>0$ component added together. Besides just being a simpler solution, there is evidence from the close fit found to the gamma distribution in the shape of the even-odd accessibility deviations for large $\theta$ that gamma distributions may be closely related to the overall shape of these functions.

\subsection{Disruption Over Time}
Removing the assumption that $f$ is constant over time is a novel direction in the study of fitness landscapes of this kind. A method for doing this was considered early in this investigation, but not developed beyond definition and some early simulations to probe feasibility.

Let a greedy adaptive walk be a sequence of genotypes $\sigma, \sigma^1, \sigma^2, ...$ terminating at a local optimum where each pair $\sigma^i, \sigma^{i+1}$ satisfies $\sigma^{i+1} = \text{arg}\max\limits_{d(\sigma^i, \sigma^j)}(f(\sigma^j))$. That is, that each step in the walk is the most improving step possible from the current genotype.

Then, suppose we define the rank order of a fitness landscape to be the ordered list $\sigma^1, \sigma^2, \dots, $ $\sigma^{2^N}$ such that $f(\sigma^1) < f(\sigma^2) < ... < f(\sigma^{2^N})$. We define $g$, the rank function, to be the position of a genotype in the rank order -- that is, using the superscripting style shown previously, $g(\sigma^k) = k$. We note that $g$ can be interpreted as a fitness function, and that for purposes of adaptive walks, both greedy and not, the landscape induced by fitness function $g$ is identical to the landscape induced by fitness function $f$.

We define a disruptable fitness landscape to be one in which the fitness function is the same as the rank order function. 

We define a disruption $S$ to be an algorithm that takes as input a disruptable fitness landscape and gives as output a disruptable fitness landscape.

We comment that there is no inherent restriction on how the algorithm must work; disruptions that generate the same output no matter what landscape is input, disruptions that appear uncorrelated between their input and output, disruptions that reverse their input's rank order, and many other potentially pathological cases are all permitted by the definition.

Where $S$ is a disruption, $F$ a disruptable fitness landscape, and $f$ the fitness function of $F$, we will write $S \circ f$ to mean the fitness function (equivalent to the rank order function) of $S(F)$.

We define the $j$-relative rank of a genotype $\sigma$ given a rank order function $g$ to be $r_j(\sigma, g) = \displaystyle \frac{g(\sigma)}{\displaystyle \left(\max_{d(\sigma, \delta)\leq j} g(\delta)\right)}$. Qualitatively, this is a value between 0 and 1 where 0 indicates $\sigma$ is the worst among its $j$-neighbors and 1 indicates $\sigma$ is the best among its $j$-neighbors.

We define a $(j,t)$-disaster of a disruption $S$ on a landscape $F$ to be a genotype $\sigma$ where $r_j(\sigma)=1$ and $r_j(\sigma, S \circ g)<t$. Qualitatively, a disaster is a genotype that was a $j$-local optimum before the disruptions and is sufficiently worse compared to its neighbors after the shuffle. We restrict $0 < t < 1$.

We define $\Omega_j(F)$ to be the number of $j$-local optima in $F$.

We define the $j$-disaster function $D$ of a disruptions $S$ on a landscape $F$ to be $D_j(F, S, t) = \displaystyle \frac{1}{\Omega(F)}\sum_{\{\sigma : \sigma \text{ is a $(j,t)$ disaster}\}} 1$. In other words, $D_j(F, S, t)$ counts the number of unique $(j,t)$-disasters of $S$ on $F$, normalized to the total number of optima in $F$. When it is not ambiguous what landscape and disruptions are being discussed, we will omit the $F$ and $S$ arguments to discuss only $D_j(t)$. Further, we will usually restrict $j=1$ to discuss only immediate, Hamming distance 1, neighbors when defining locality.

Let a shape function be a continuous function $\Theta(t)$ defined over $t\in [0, 1]$ such that $\Theta(0)=0$, $\Theta(1)=1$, and $\Theta$ is monotonically increasing on its domain.

A good disruption $S$ will:
\begin{enumerate}
    \item Over any natural number $k > 10000$ of fitness landscapes of a chosen structure for which the disruption process is constructed, at each $0 < t < 1$, have no more than $k e^{-(3x)^2}$ disaster functions where $|D_1(t) - \Theta(t)| > x$ for a chosen shape function;
    \item Satisfy $\displaystyle v\Omega(S^k(F)) < \Omega(F) < \frac{1}{v}\Omega(S^k(F))$ for all natural numbers $k$ for some real number $0 < v < 1$; and
    \item Be describable as an algorithm that contains no conditional logic based on any non-uniformly randomly chosen genotype or on the fitness of any non-uniformly randomly chosen genotype.
\end{enumerate}

The intention here is to create a non-pathological disruption function where local optima tend to remain good compared to their peer genotypes after a single disruption and where the overall disaster function roughly follows $\Theta(t)$ to within a degree of tolerance across many landscapes. The number of local optima should not be considerably inflated or deflated by disruption either.

The purpose of these constraints on a good disruption $S$ is to ensure that the landscape only transforms gradually over repeated application of $S$. 

The first major question is whether such a good disruption even exists for shape functions like $\Theta(t)=t^2$ or $\Theta(t) = \displaystyle \frac{1}{1+e^{-3(t-0.7)}}$, or what shape functions, if not these, it might be possible to find good disruptions to fit. 

If such a good disruption could be found, the next question would be how it interacts with $H\theta NK$, and whether repeated greedy walks with disruptions applied in between would increase or decrease accessibility to the global optimum on the whole. 

Some attempts to find a good disruption are detailed in Appendix C.

\printbibliography 

\appendix

\newpage

\section{Sources of Randomness}

When working with random numbers in a computer context, it is inevitably the case that the random number generator used in testing will be at best pseudorandom. While usually not an issue, for an investigation of this kind where the random numbers are often added together in strided patterns (for example, every eighth or sixteenth output is part of a sum), certain kinds of random number algorithms are unacceptably poor.

The most common C-language random number source is the \texttt{rand()} function, but because its implementation is as a simple linear congruential generator, it would produce an unacceptable degree of collision between landscapes meant to be independent.

The Mersenne Twister 19937 algorithm was used instead, as this algorithm is both more sophisticated and generally accepted to have better (``more random") statistical properties \cite{MT19937}. Because the Mersenne Twister algorithm holds internal state information, earlier landscapes generations would still be able to influence later ones and distort the results, but re-seeding the generator across changes in the parameters (so, if $H=2$, $N=15$, $K=3$ had just run, the generator would be seeded again before running $H=3$, $N=15$, $K=3$) prevents this issue.

All random number generation in the final implementation of the simulation was done with deterministic, seeded random number generators. The output of the code is therefore consistent if run multiple times. 

\newpage

\section{Functions to Fit the $N=K$ Ridge}
Since the behavior of accessibility on the $N=K$ ridge of these plots can be used to infer the shape of the remainder of the plot when dealing with blocked neighborhood landscapes, it became a focus in this investigation to fit a function that would describe the shape of the simulated data on the $N=K$ ridge.

A first observation is that the ridge seems to behave fundamentally differently in the case of $\theta=0$ and $\theta>0$, with $\theta=0$ tending toward a nonzero constant and $\theta>0$ tending to 1. Our chosen approach was to view the case of $\theta=0$ as a level of accessibility conferred to the landscape by the $NK$ structure, and to describe it as being additive with some further contribution made by $\theta>0$. This decomposes the outer ridge into two functions that can be summed together. If we take the data from the cases of $\theta=0$, $\theta=1$, $\theta=2$, and $\theta=3$, and plot them both in their original form (left) and with $\theta=0$ subtracted from the others (right):

\begin{figure}[H]
    \begin{center}
        \includegraphics[scale=0.08]{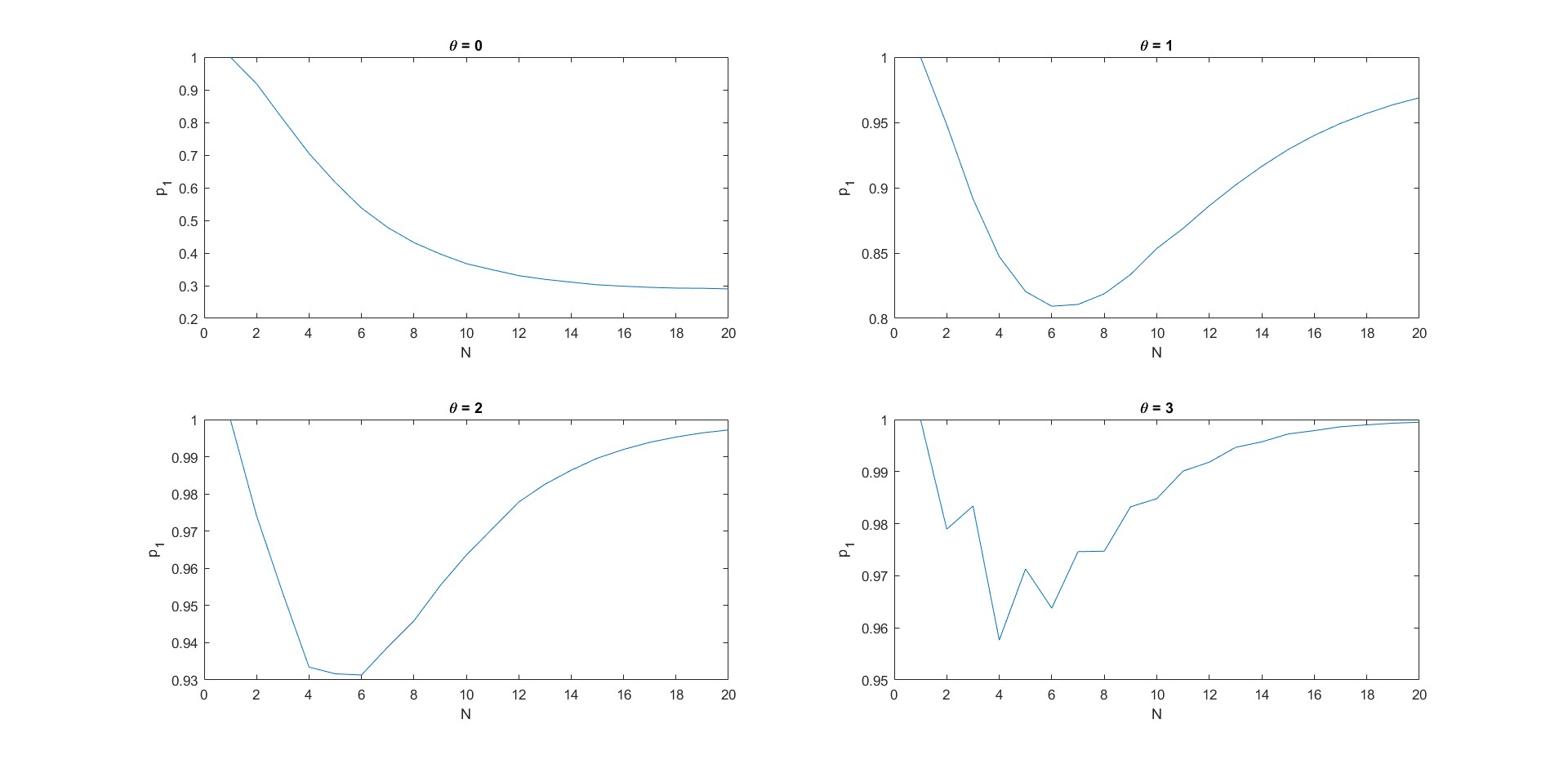}
        \includegraphics[scale=0.08]{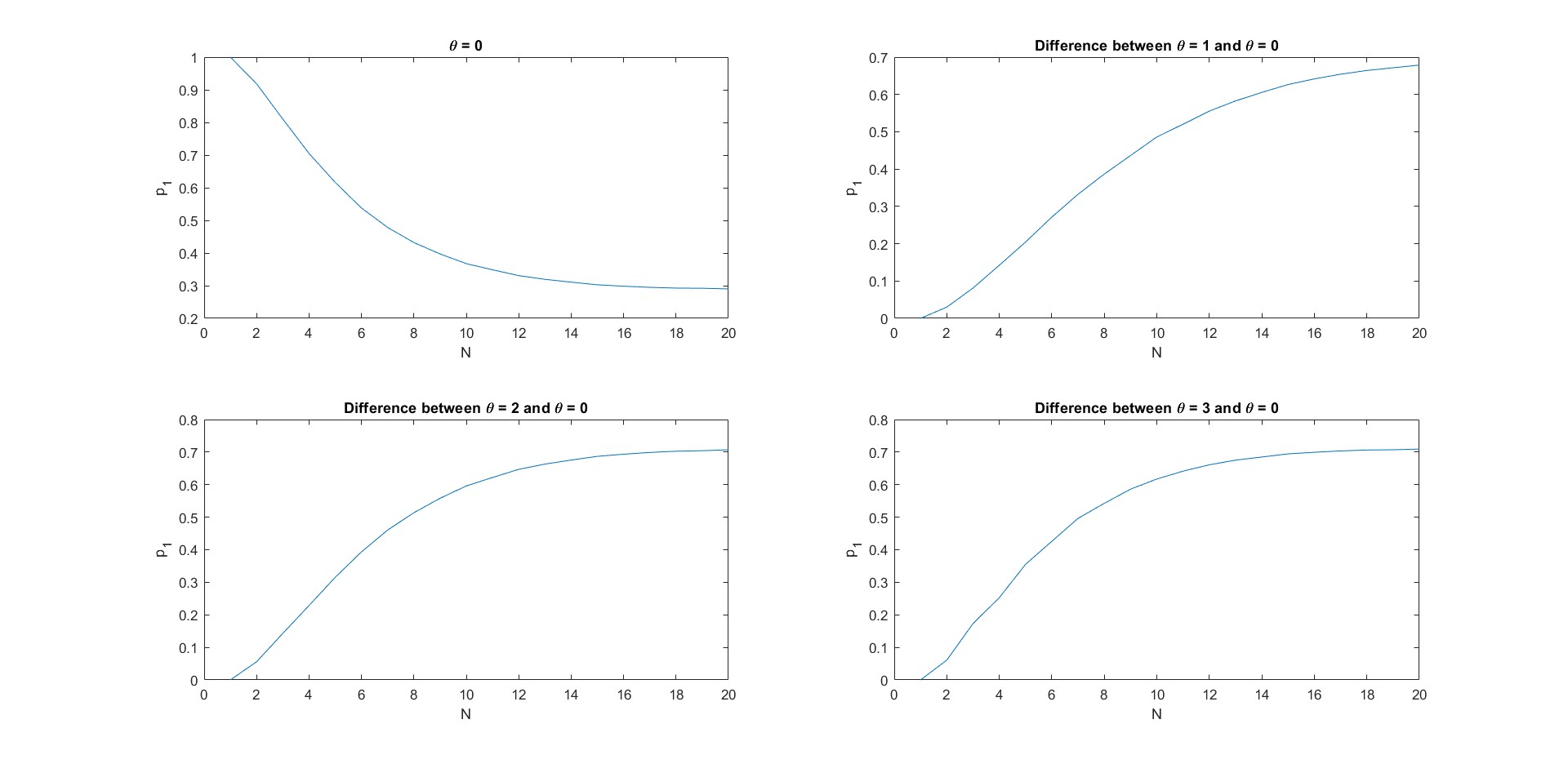}
    \end{center}
    \caption{$N=K$ ridge original accessibilities (above) and differences from $\theta=0$ (below)}
\end{figure}

We see that they appear to separate into exponential decays, or functions that resemble them. In the case of $\theta=0$, if we attempt to fit the mode model $p_{1,\theta=0}(N) = a_1 + a_2 e^{-a_3 (N-1)}$ to the data where $a_1,a_2,a_3$ are parameters, the least square error result returned by MATLAB \texttt{fminsearch()} is $a_1=0.258, a_2=0.7816, a_3=0.2029$. That fit is:

\begin{figure}[H]
    \begin{center}
        \includegraphics[scale=0.15]{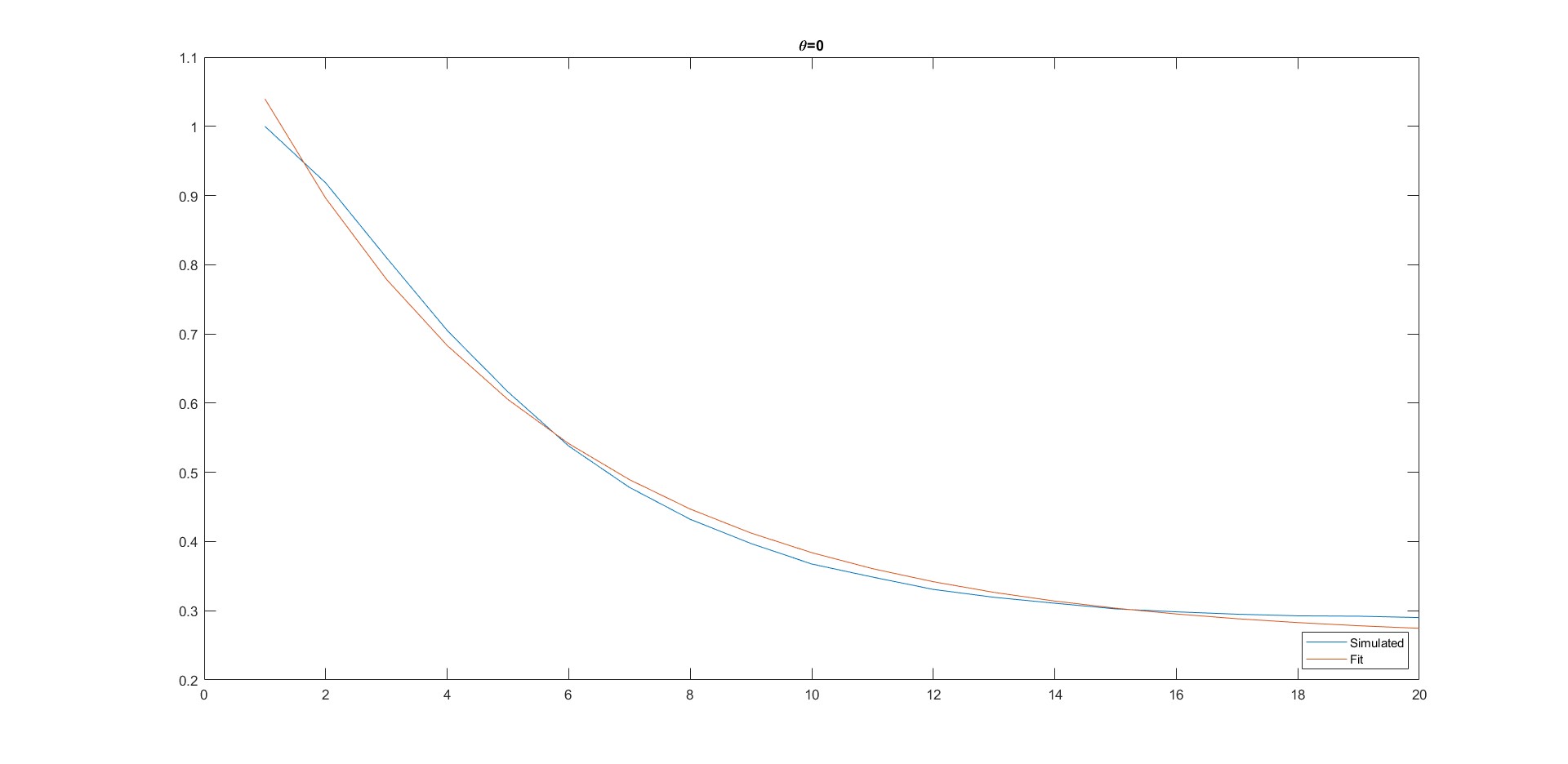}
        \caption{Simple fit to $\theta=0$ on the $N=K$ ridge}
    \end{center}
    \label{z_simple}
\end{figure}

This is a poor result, and since the functional form given was reasonably general, it suggests that the shape was not exponential to begin with. Attempting to fit functions using $\tanh(N-1)$ or the Hill Equation also produced similarly poor results, but the form $p_{1,\theta=0}(N)=a_1 + a_2 e^{-a_3 (N-1)^{a_4}}$ returns:

\begin{figure}[H]
    \begin{center}
        \includegraphics[scale=0.15]{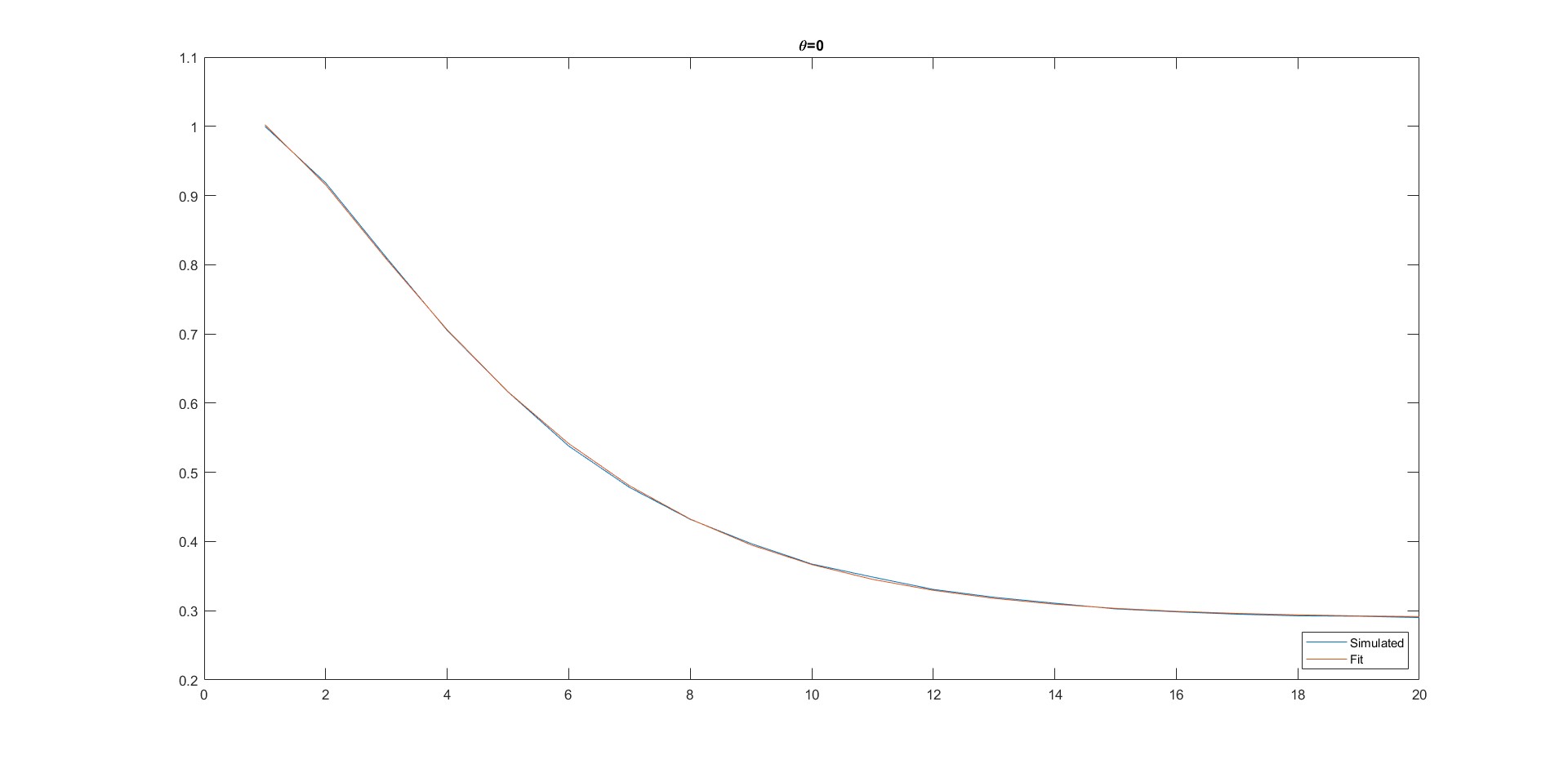}
    \end{center}
    \caption{$\theta=0$ double-exponentiated fit on the $N=K$ ridge}
\end{figure}
with the parameters $a_1=0.2894, a_2=0.713, a_3=0.1301, a_4=1.2917$. This is a superior fit, but it leaves open the question of whether overfitting has also rendered it overly specific to the data at hand. Its end behaviors for $N=1$ and the trend as $N\rightarrow \infty$ appear reasonable, but that is not in itself a guarantee that this fit shape is reasonable.

Similarly, trying to fit the data for $\theta>0$ with a simple exponential form $p_{1,\theta>0}(N, \theta) = a_1 + a_2 e^{-a_3 \theta (N-1)}$ provides a minimum square error with the simulation results that yields:

\begin{figure}[H]
    \begin{center}
        \includegraphics[scale=0.15]{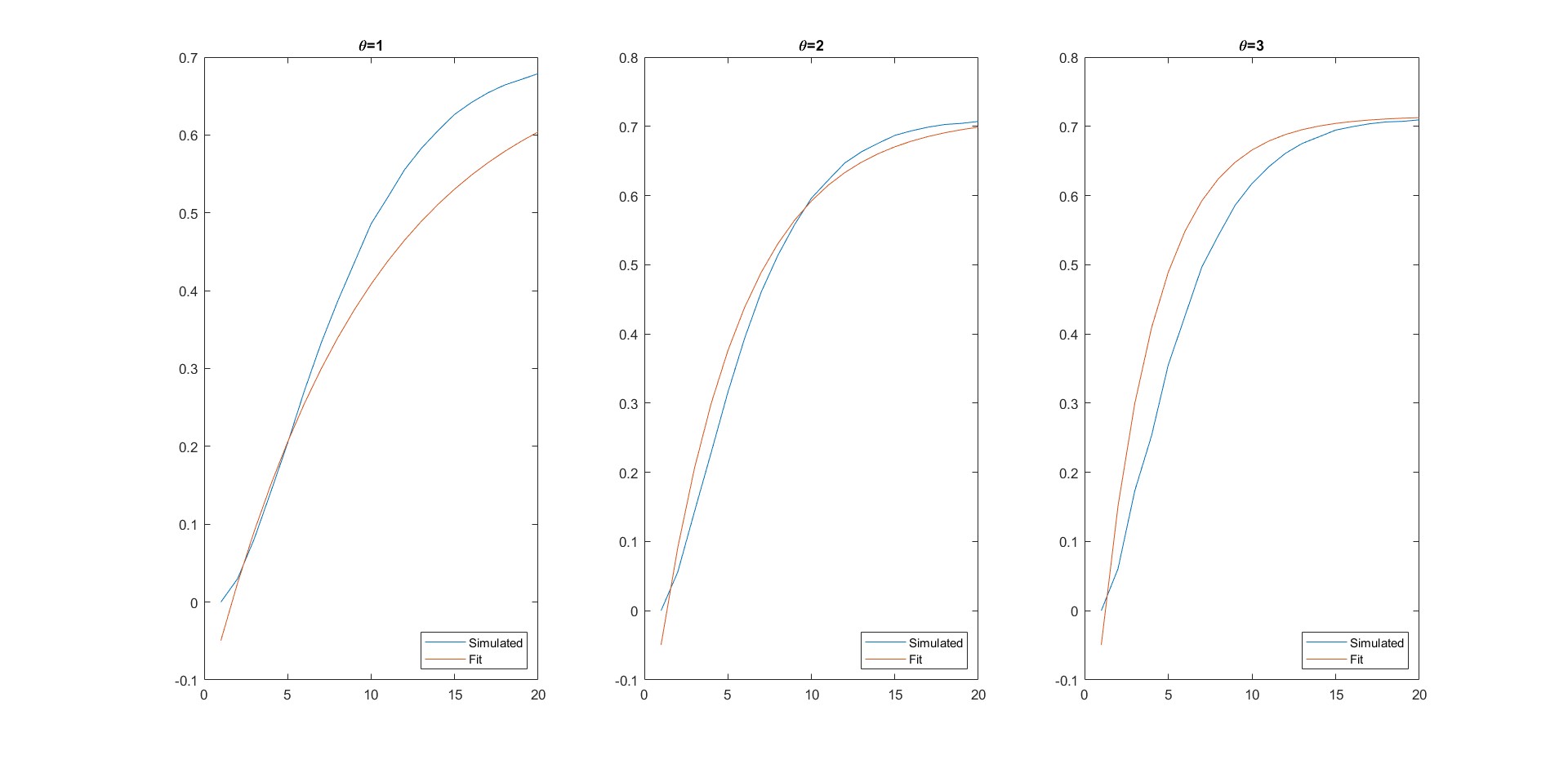}
    \end{center}
    \caption{Simple fit to $\theta>0$ on the $N=K$ ridge}
    \label{gt_simple}
\end{figure}

Like Figure \ref{z_simple}, the result in Figure \ref{gt_simple} improves substantially if we allow an double-exponent on $N-1$, but even then, the fit quality is still underwhelming:

\begin{figure}[H]
    \begin{center}
        \includegraphics[scale=0.15]{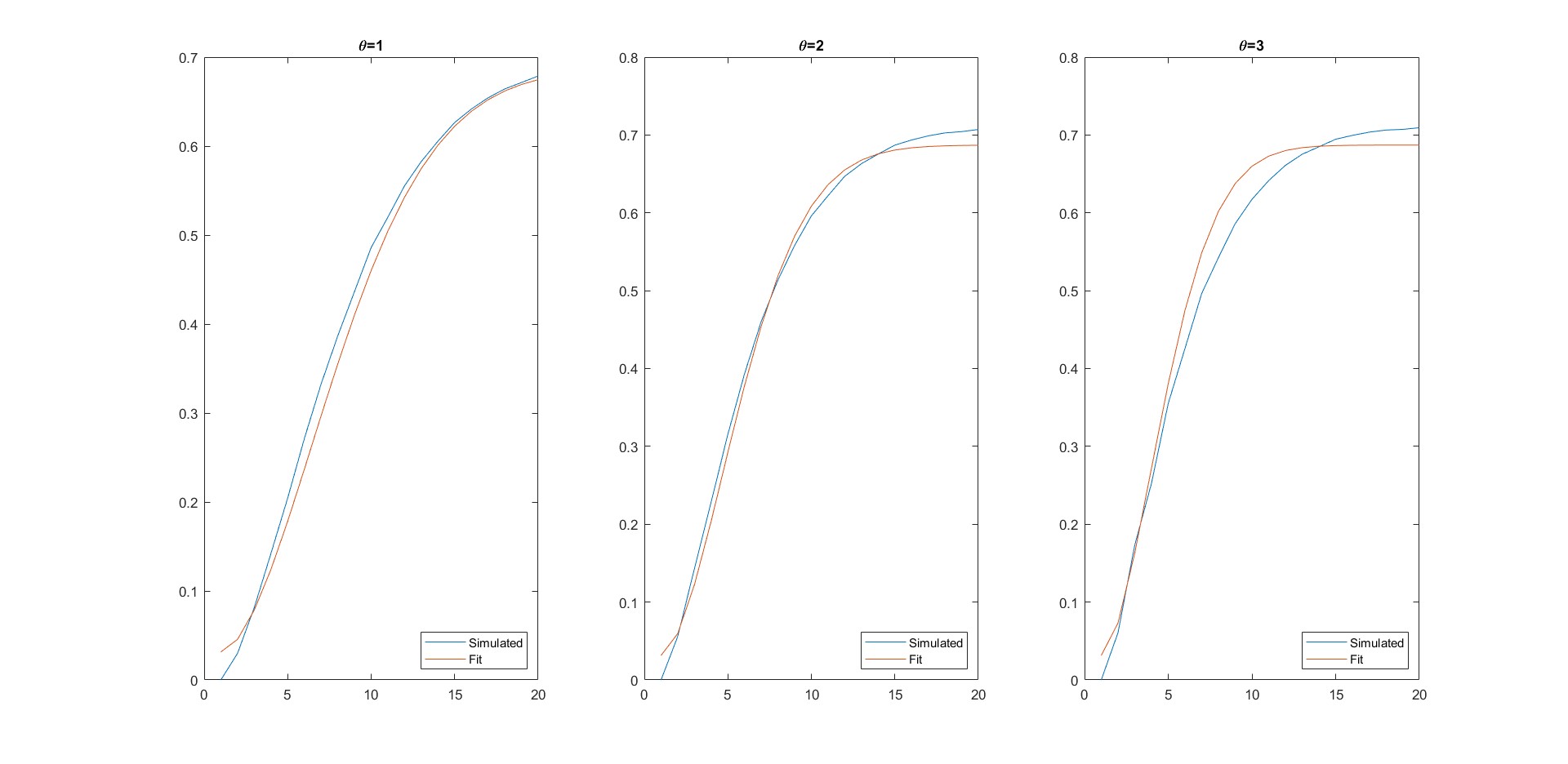}
    \end{center}
    \caption{Double-exponentiated fit to $\theta>0$ on the $N=K$ ridge}
\end{figure}

This points to the need to fit a better function for changing the transitional speed than just to put $\theta$ in the exponent. If we add additional data for $\theta=1.5$, $\theta=5$, and $\theta=10$, and instead of fitting all of them at once instead allow the exponent $a_3$ term to be fit to each $\theta$ individually, we obtain the following shapes:

\begin{figure}[H]
    \begin{center}
        \includegraphics[scale=0.15]{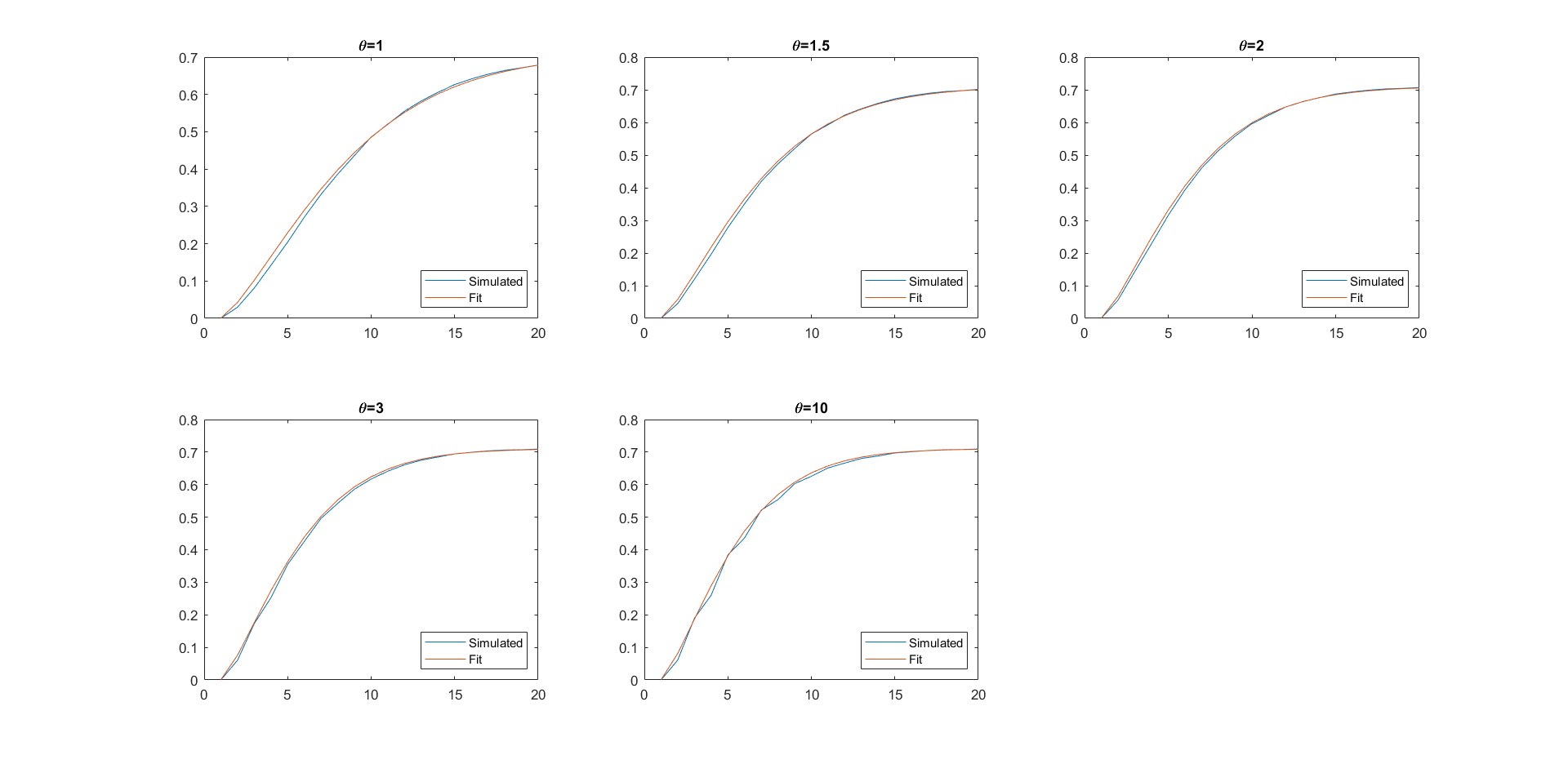}
    \end{center}
    \caption{$\theta>0$ on the $N=K$ ridge with each $\theta$ value having a separately fit exponent}
\end{figure}

The parameters were $a_1 = 0.7089, a_2=-0.7118, a_{3,\theta=1}=0.0534, a_{3,\theta=1.5}=0.0739, a_{3,\theta=2}=0.0866, a_{3,\theta=3}=0.986, a_{3,\theta=5}=0.1047, a_{3,\theta=10}=0.1058, a_4=1.3907$ (with $a_4$ being the exponent on $(N-1)$). This is an improvement, but it shows that at larger $\theta$ we cannot ignore the even-odd bias previously discussed, and it allows for the function to be slightly non-zero even at $N=1$, which we know should be impossible. We will adjust that by requiring $a_2=-a_1$ on subsequent fits. 

For the even-odd bias, we will do the following: for $\theta=10$ we can presume that every odd $N$ should roughly have $p_1 = 1$ overall. If we look at how much smaller the even-$N$ accessibilities are than this, we obtain the top plot:

\begin{figure}[H]
    \begin{center}
        \includegraphics[scale=0.15]{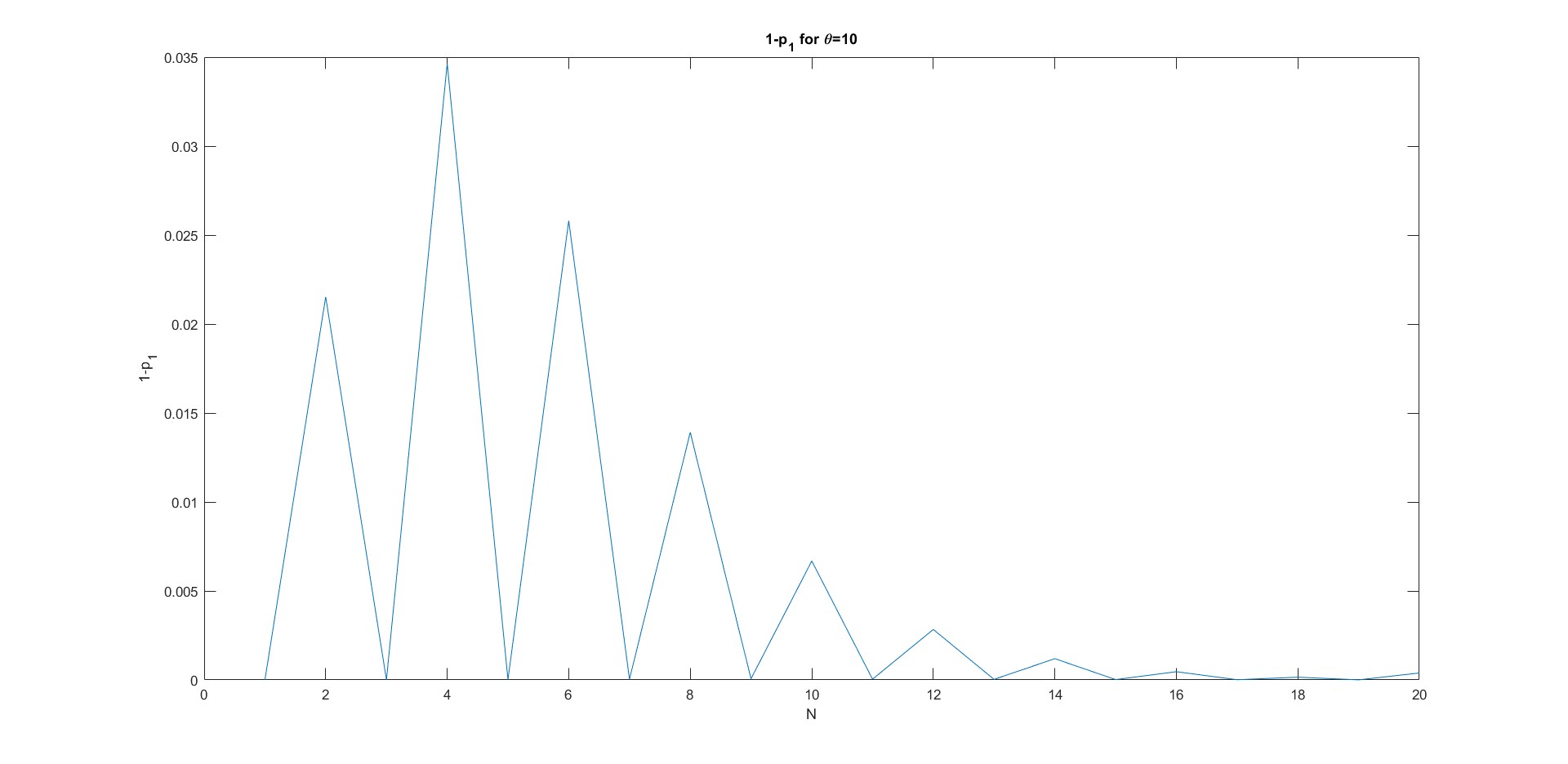}
    \end{center}
    \caption{Difference between accessibility of $\theta=10$ and $\theta=0$ on the $N=K$ ridge}
\end{figure}

We can select for these points with $1-(N \mod 2)$ as a coefficient on any function that fits to them, and seeing that it appears they follow a classic right-skewed distribution, we can assume that a gamma distribution would be appropriate. So, fitting the form $1-p_{1, \theta=10}= \displaystyle a_3 \left(\frac{N^{a_1} e^{-a_2 N} a_2^{a_1}}{\Gamma(a_1)}\right) \left( 1 - (N \mod 2) \right)$ in MATLAB, we obtain the lower graph, an essentially perfect match, with $a_1=3.6475,a_2=0.6822,a_3=0.2134$.

\begin{figure}[H]
    \begin{center}
        \includegraphics[scale=0.15]{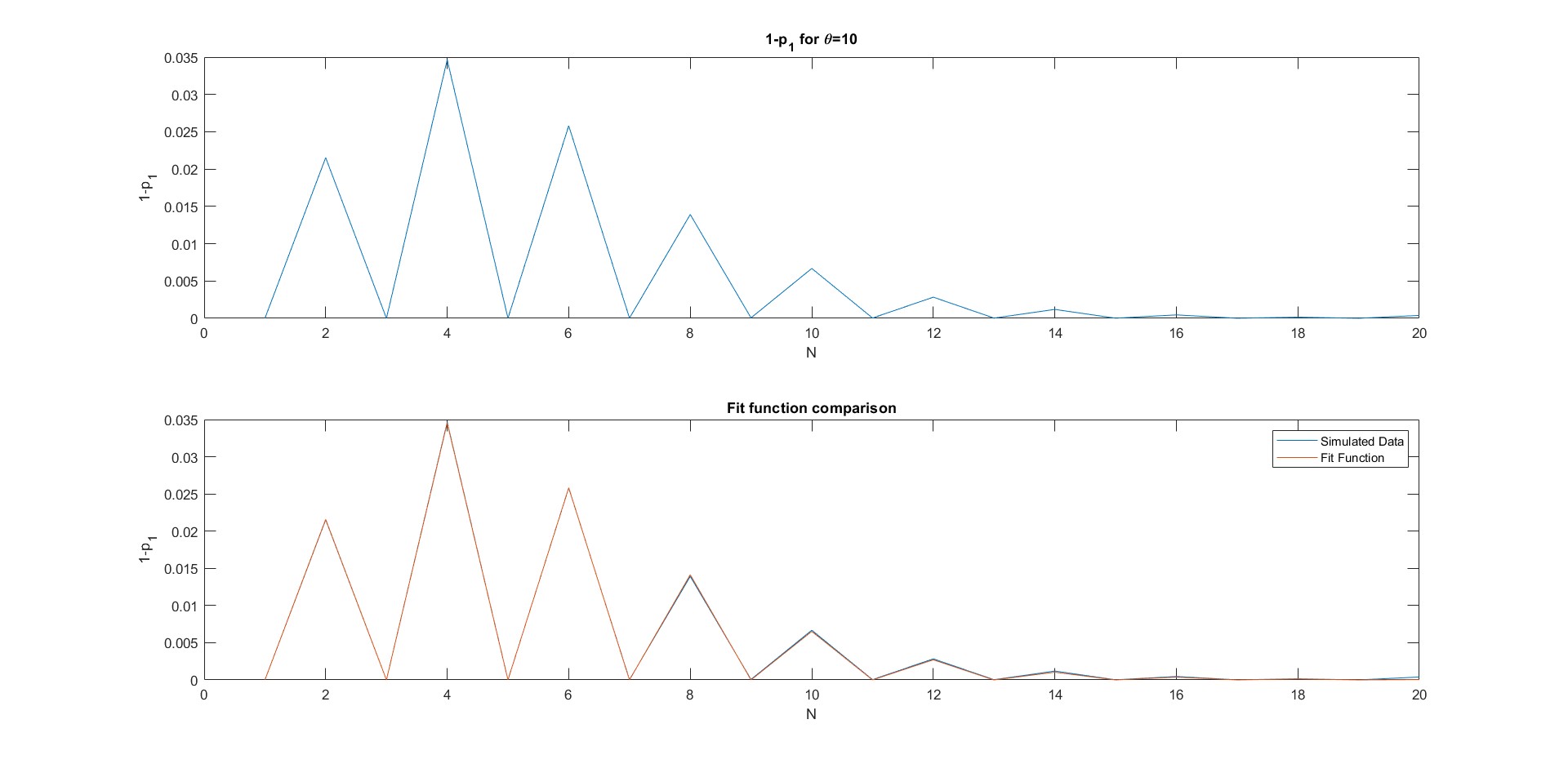}
    \end{center}
    \caption{Fit to the $\theta=10$ differences from $\theta=0$ on the $N=K$ ridge}
\end{figure}

Going back to the exponent values fit to each $\theta$, we can see that their trend appears to quickly hit a maximum value as $\theta$ rises, which produces the troubling suggestion that if we were to fit a function to this, it would require at least further exponentiation. The risk that this overfits the simulation data is virtually guaranteed, but as with the $\theta=0$ case, the $\tanh$ and Hill Equation forms do not perform any better on this data, and the fact that the $\theta=0,\theta>0$ separation produces monotonic functions suggests it is not a step in the wrong direction.

If we plot the exponent value ($a_3$) versus $\theta$:

\begin{figure}[H]
    \begin{center}
        \includegraphics[scale=0.15]{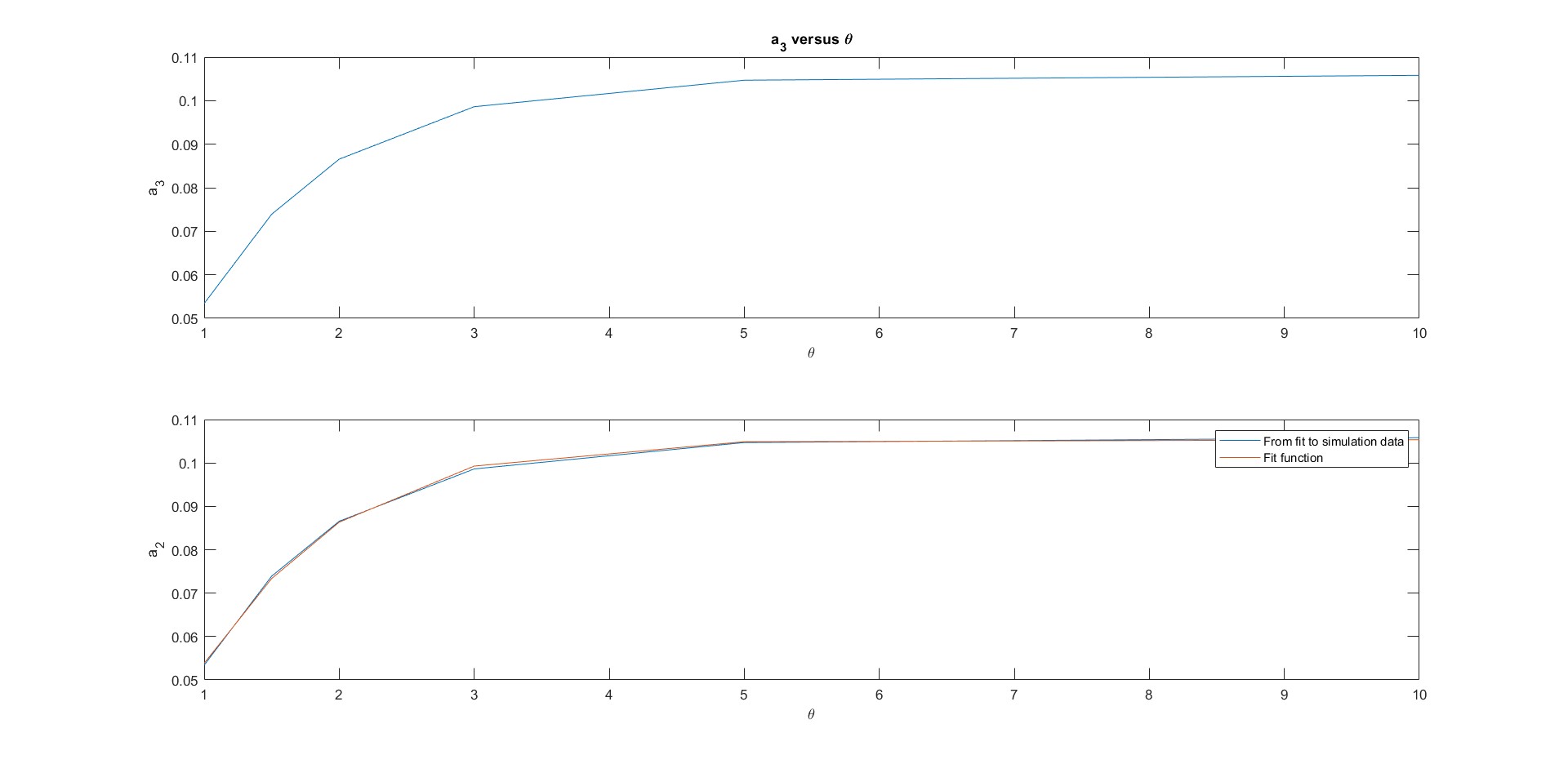}
    \end{center}
    \caption{Fit to the exponential bandwidth terms seen in the $\theta>0$ fits on the $N=K$ ridge by an exponential transition function}
    \label{exp_transition}
\end{figure}

This appears to have the form of an exponential transition function, roughly, but fitting one directly does not perform well if we condition that it must be 0 at $\theta=0$ (a sensible restriction to have, since we want the entire $\theta>0$ contribution to revert to 0 at $\theta=0$), we don't get a good fit. A better result can be found fitting the form $a_3(\theta) = b_1-b_1e^{-b_2 \theta^{b_3}}$ with $b_1=0.1053$, $b_2=0.7155$, and $b_3=1.2594$. This is compared to the results from the original plot in the lower graph of the above figure.

When this is combined with the rest of the functional form, it gives a best-fit function for the $\theta>0$ contribution to accessibility of $\displaystyle a_1-a_1e^{-(b_1-b_1 e^{-b_2 \theta^{b_3}}) (N-1)^{a_2}}$, if we renumber for eliminated variables. This is an unwieldy function, but the performance is a good match to the data, and the end behaviors are sensible for both increasing $\theta$ and increasing $N$, limiting the risk normally posed by overfitting (since even using this form for points not in the data taken seems to imply a reasonable shape). 

Combined with $p_{1,\theta=0}(N)$, we obtain a final overall form:

\begin{equation}\label{overall}
    p_1 = a_1 + (1-a_1) e^{-a_2 (N-1)^{a_3}} +
          b_1-b_1e^{-(c_1-c_1 e^{-c_2 \theta^{c_3}}) (N-1)^{b_2}}
\end{equation}

Despite having been able to fit the even-odd variation, Eq. \eqref{overall} does not combine well with efforts to fit the non-even-odd behavior, introducing excessive error compared to just fitting without it. If we fit Eq. \eqref{overall} all at once for its complete parameters against the data from $\theta=0,1,1.5,2,3,5,10$, we obtain:

\begin{align*}
    a_1 &= 0.2942\\
    a_2 &= 0.1271\\
    a_3 &= 1.3219\\
    b_1 &= 0.7034\\
    b_2 &= 1.4158\\
    c_1 &= 0.1037\\
    c_2 &= 0.6909\\
    c_3 &= 1.3102
\end{align*}

In conjunction with the multiplicative decomposition of blocked neighborhood landscape fitness, Eq. \eqref{overall} can be used to predict the shape of fitness landscapes not simulated, like $\theta=0.25$ out to $N=K=50$:

\begin{figure}[htp]
\makebox[\textwidth][c]{\includegraphics[width=0.55\linewidth]{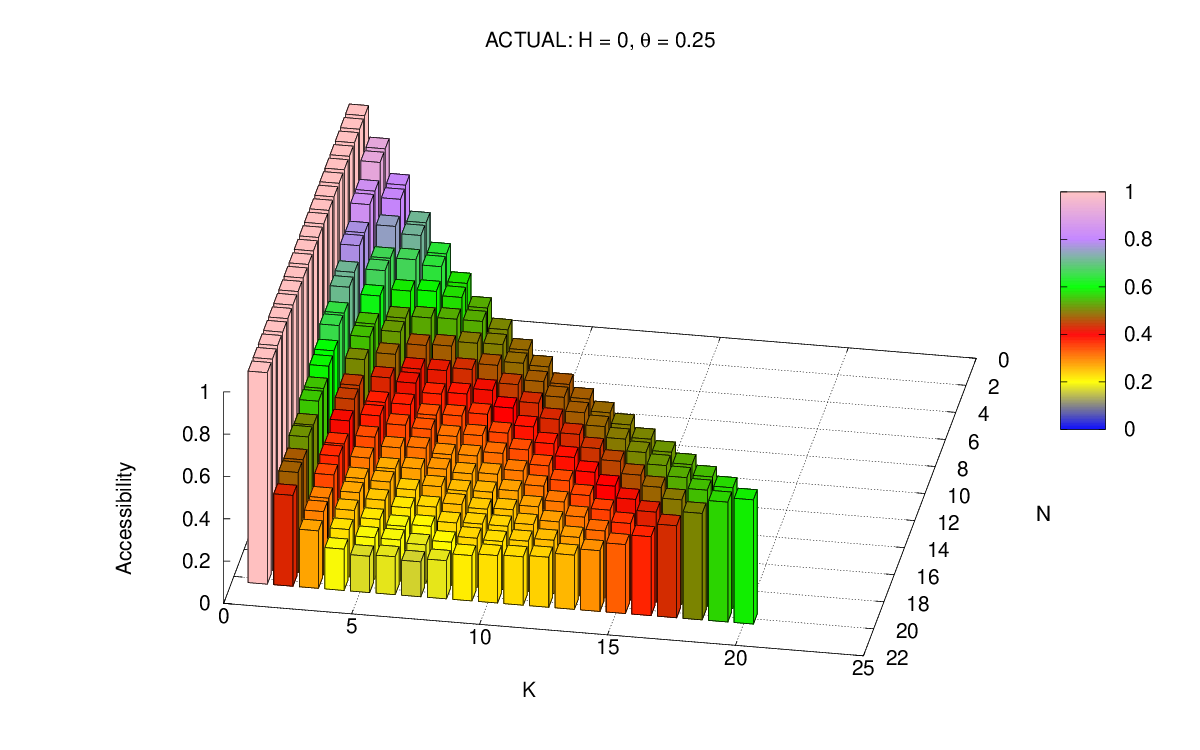}\includegraphics[width=0.55\linewidth]{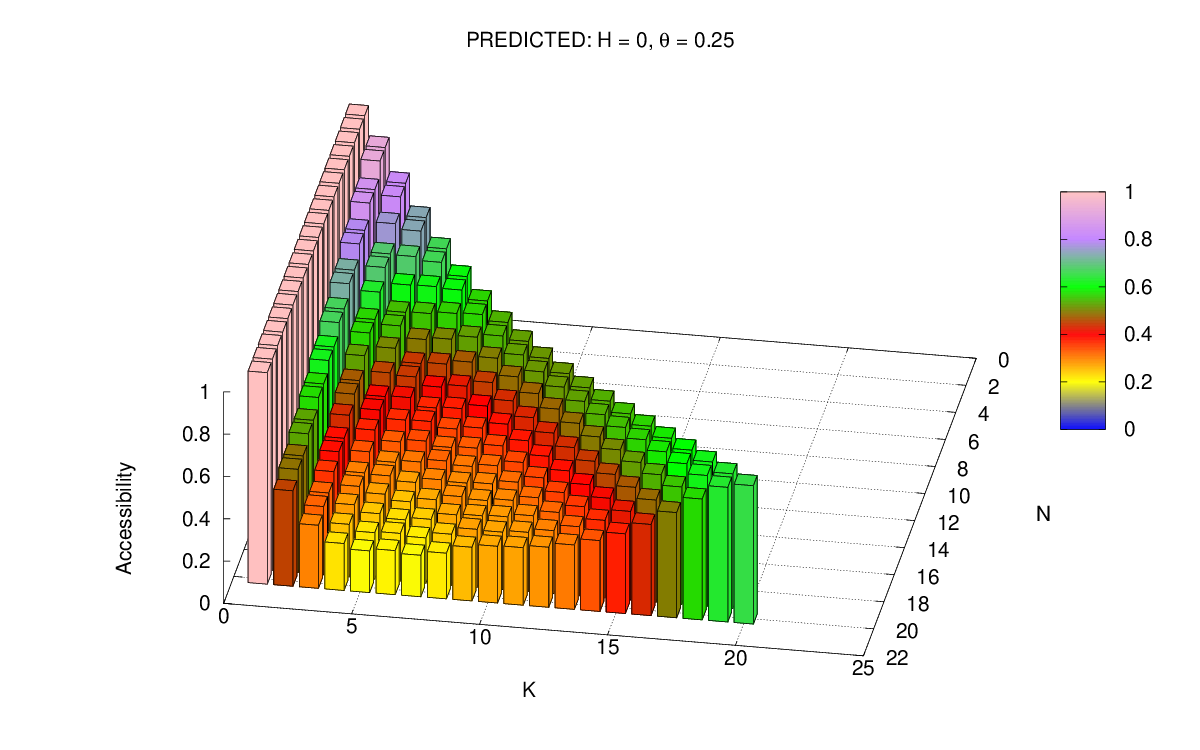}}
\caption{Comparison between direct simulation of $\theta=0.25$ (left) and predicted shape (right) using fit function and blocked landscape multiplicative decomposition}
\label{comparison}
\end{figure}

The computational time needed to verify points on Figure \ref{comparison} outside the $N<20,K<20$ region is considerable, but the simulation up to those values is shown in the top plot above, and the prediction generated based on the decomposition of a blocked landscape extrapolated outward using Eq. \eqref{overall} is shown in the preceding figure in the lower plot. This shows that they are similar in value and structure, but not exact -- the $N=K$ ridge dips slightly lower in simulation compared to the predicted fit, but does show qualitatively the right behavior trending back upward after reaching its lowest point.

Nevertheless, Eq. \eqref{overall} guarantees many of the most important end behaviors (such as reversion to decay in the $\theta=0$ case and tendency to $p_1=1$ as $N\rightarrow \infty$ in all $\theta>0$ cases) and is, at least, a good interpolant for cases both simulated and not. 

For a broader comparison of ridge function shapes across the main datapoints:

\begin{figure}[H]
\makebox[\textwidth][c]{\includegraphics[width=1.1\linewidth]{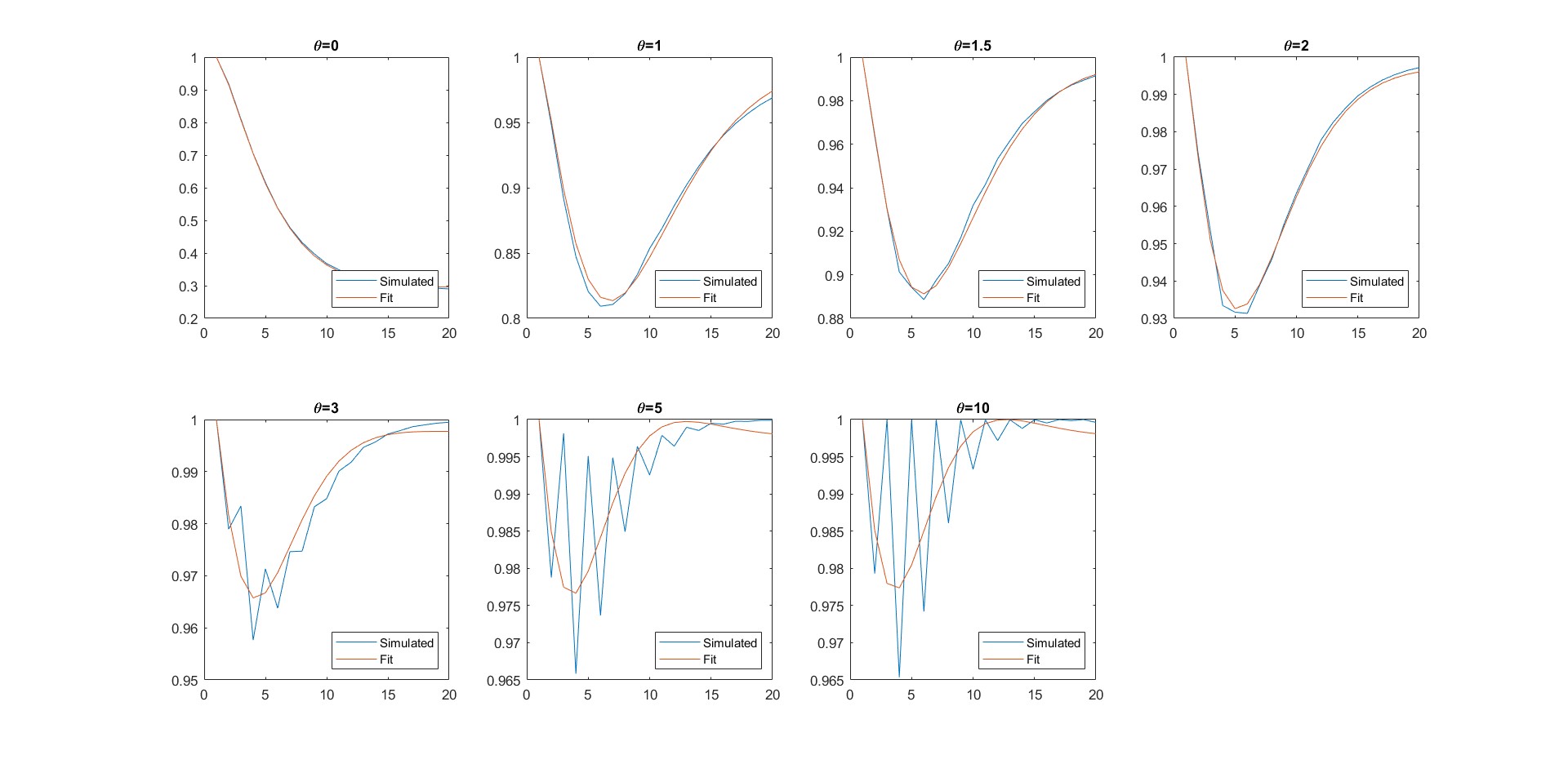}}
\caption{Fit functions in comparison to actual accessibilities on the $N=K$ ridge across a variety of $\theta$ values}
\end{figure}

The most significant shortcoming here is that the model is not able to integrate the even-odd behavior of large $\theta$ into the smooth curves being fit. Combining the previous result that fit the even-odd spikes with a Gamma distribution function naively (that is, adding it into the final approximate form of $p_1$ and re-fitting) did not match the simulation, as the Gamma function was fit against the assumption that $p_1=1$ at $\theta=10$, which the existing interpolant does not quite achieve.

It is recommended for work going forward that a different approach be taken to derive a more general result. Specifically, due to the repeated exponentiations needed to achieve a good fit after splitting the $\theta=0$ and $\theta>0$ cases, it is suggested that it may be more convenient to fit a function \emph{without} this division.

\newpage

\section{Candidates for Disruption Processes}

Some effort has already been made toward finding potential good disruptions, but significant limitations have been found regarding the shape of the disaster function. For example, if the disruption mechanism is taken to be adding a normally distributed random value into the contribution to fitness of a single subgenotype within the $NK$ model (which is in effect the smallest possible unit of change in an $NK$ landscape), the disaster function across many instances of this change will vary with the standard deviation of the random value:

\begin{figure}[H]
    \begin{center}
        \includegraphics[width=0.3\linewidth]{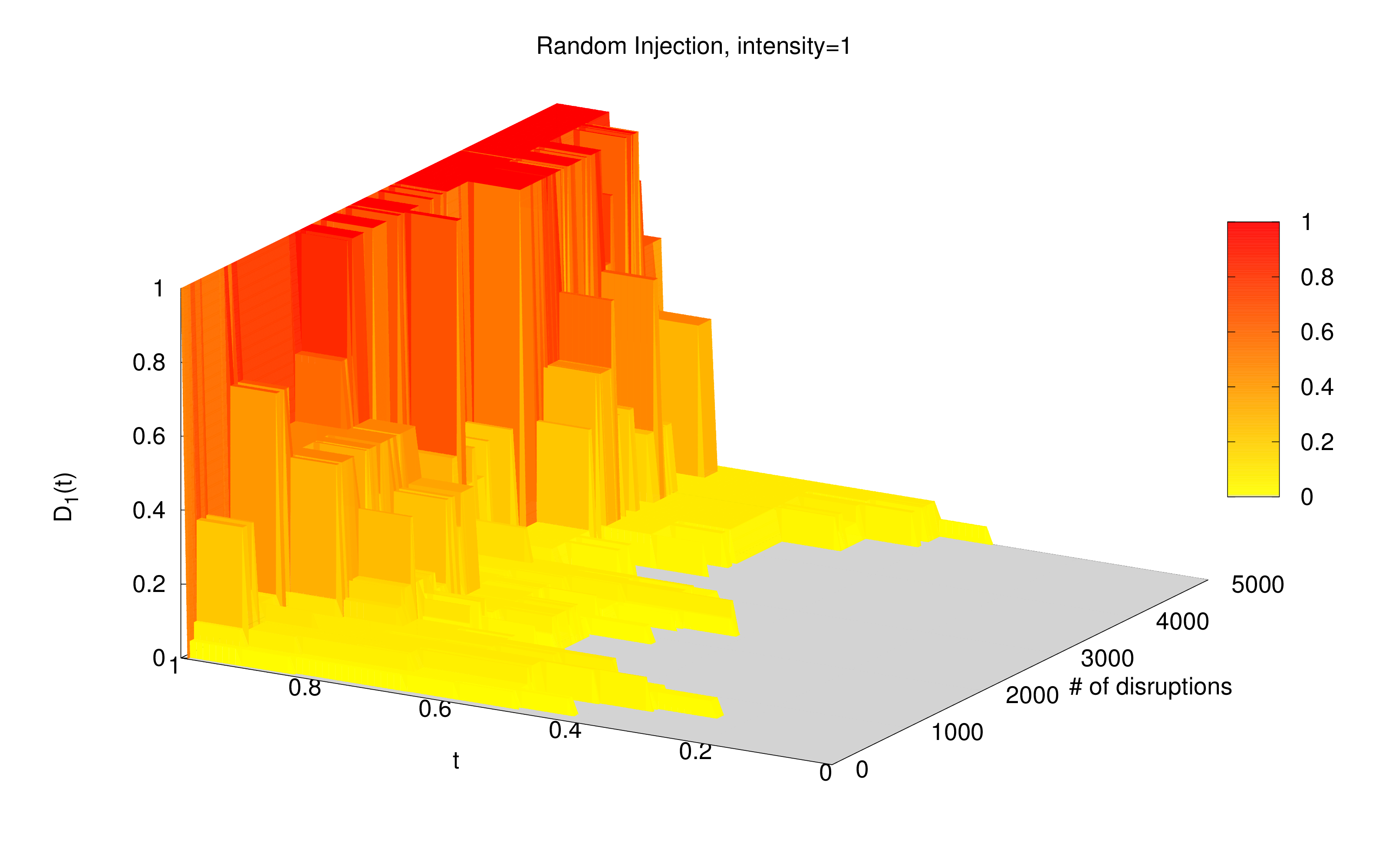}
        \includegraphics[width=0.3\linewidth]{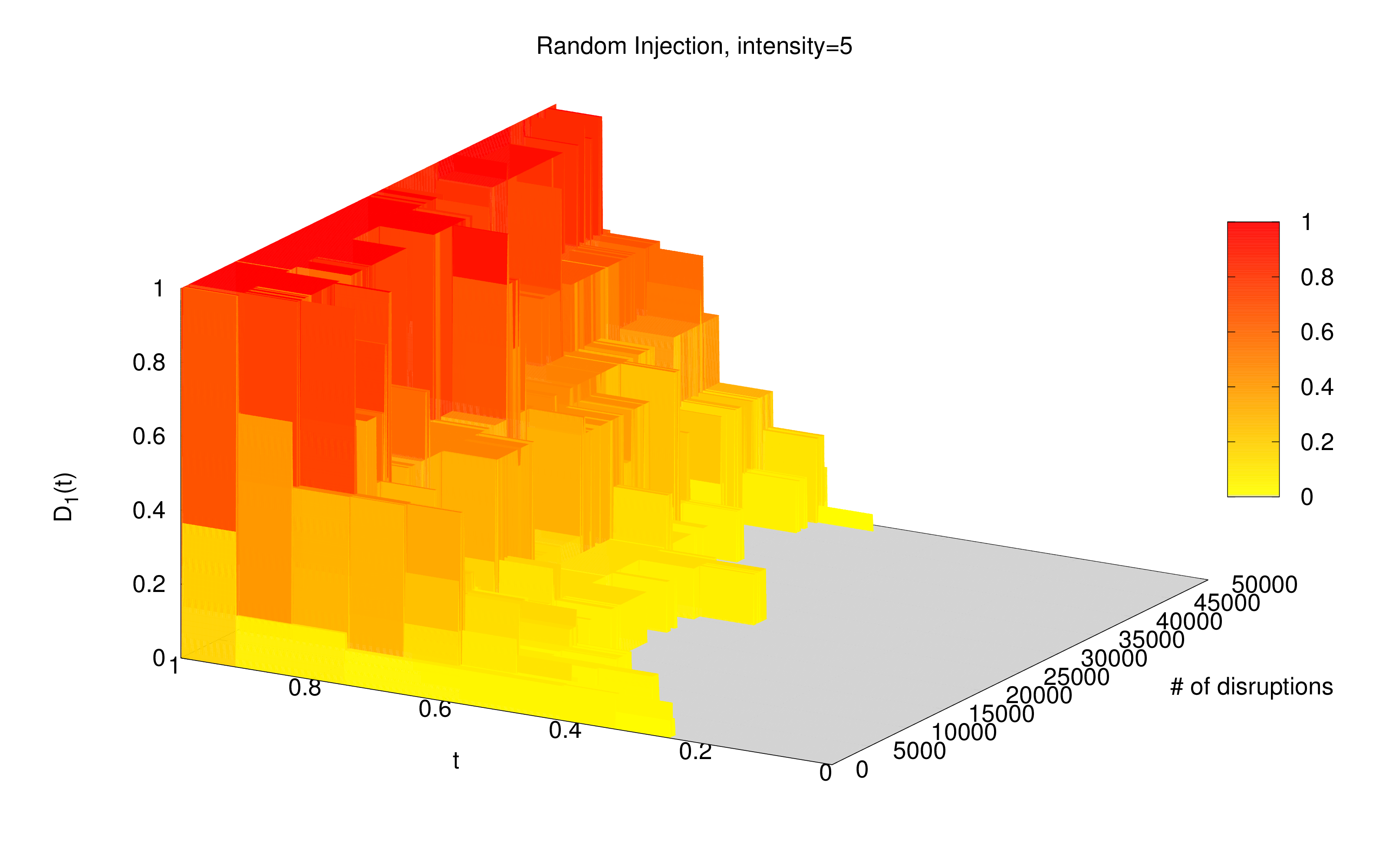}
        \includegraphics[width=0.3\linewidth]{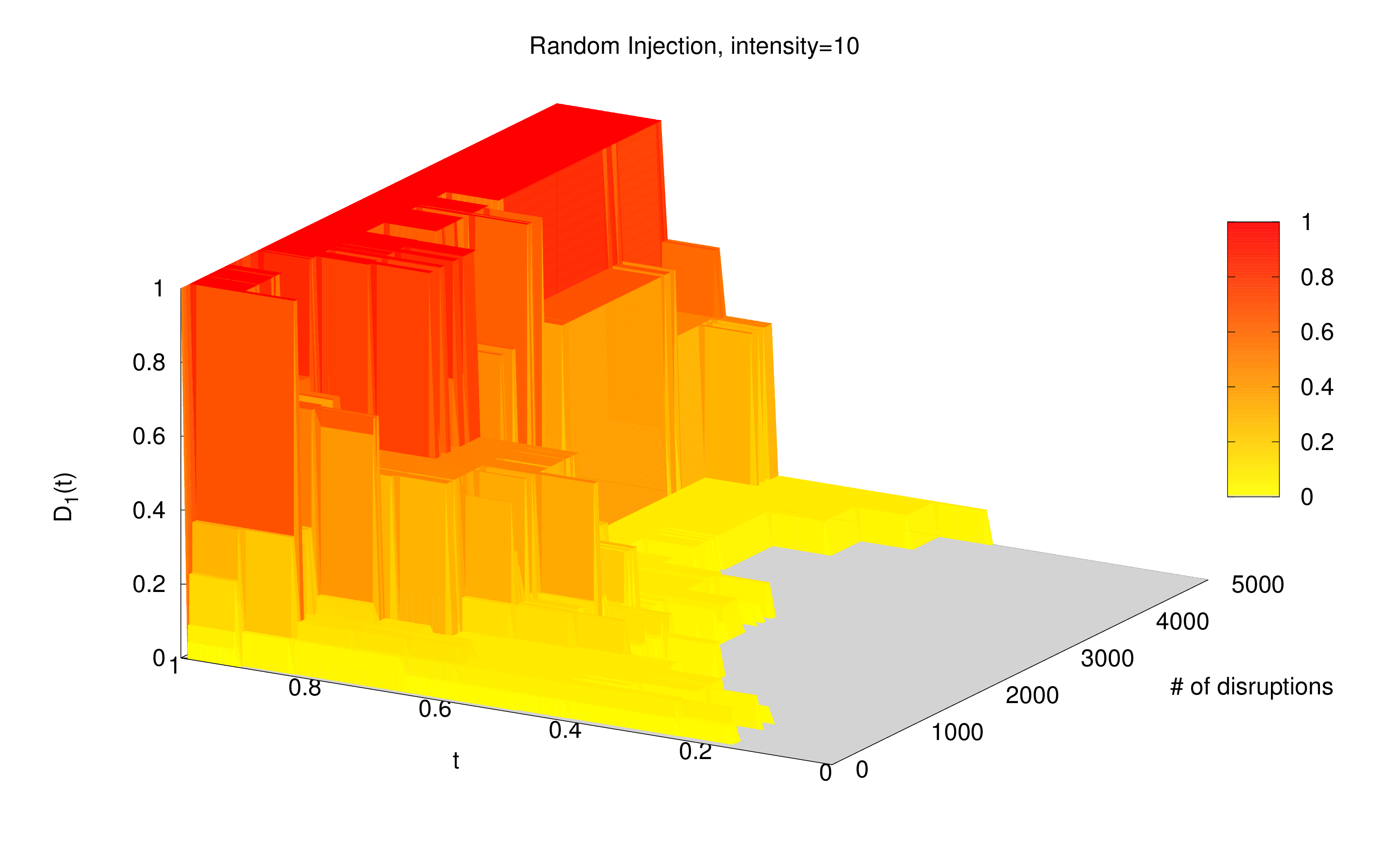}\\
    \end{center}
    \caption{Single-additive-value disruption function graphs for $\sigma=1,5,10$}
\end{figure}

All of these were done in classic $NK$ with blocked neighborhoods with $N=12$, $K=4$. The critical feature here is that the disaster function is often quite rough and step-like and rarely has any non-zero values below $t=0.5$. This is not particularly surprising because non-zero values below $t=0.5$ would indicate the optima of the pre-disruption landscape having fallen below the median of their neighbors, which would likely only happen in any significant number by way of a disruption that systematically punished optima. 

Similarly, if the disruption were to completely re-generate the landscape such that the old and new landscapes were completely uncorrelated, we suggest that the disaster function would likely look like the cumulative distribution function of a binomial distribution, crossing 0.5 at $t=0.5$, since the relative rank of the former optimum after a complete overhaul of the landscape would follow a binomial distribution (there is only 1 way for it to be the worst of all its neighbors, but $2^N-1$ ways for it to be second-worst, and $\begin{pmatrix}2^N-1\\2\end{pmatrix}$ ways for it to be third-worst, etc.). 

If we were instead to try to make disruptions in a correlated manner, we could potentially change the landscape more drastically, but we would be at risk of deflating the number of optima. For example, suppose that the disaster function generates a random vector $v$ of length $N$ where $v_i = N(0, \epsilon^2)$. Then, for each subgenotype $\sigma_B$ in the landscape, its contribution to fitness is altered by adding $v_i$ when locus $i$ is in $B$ and ${\sigma_B}_i=1$, and subtracting $v_i$ when locus $i$ is in $B$ and ${\sigma_B}_i=0$. 

The disaster function and optima count graphs are:

\begin{figure}[H]
    \begin{center}
        \includegraphics[width=0.4\linewidth]{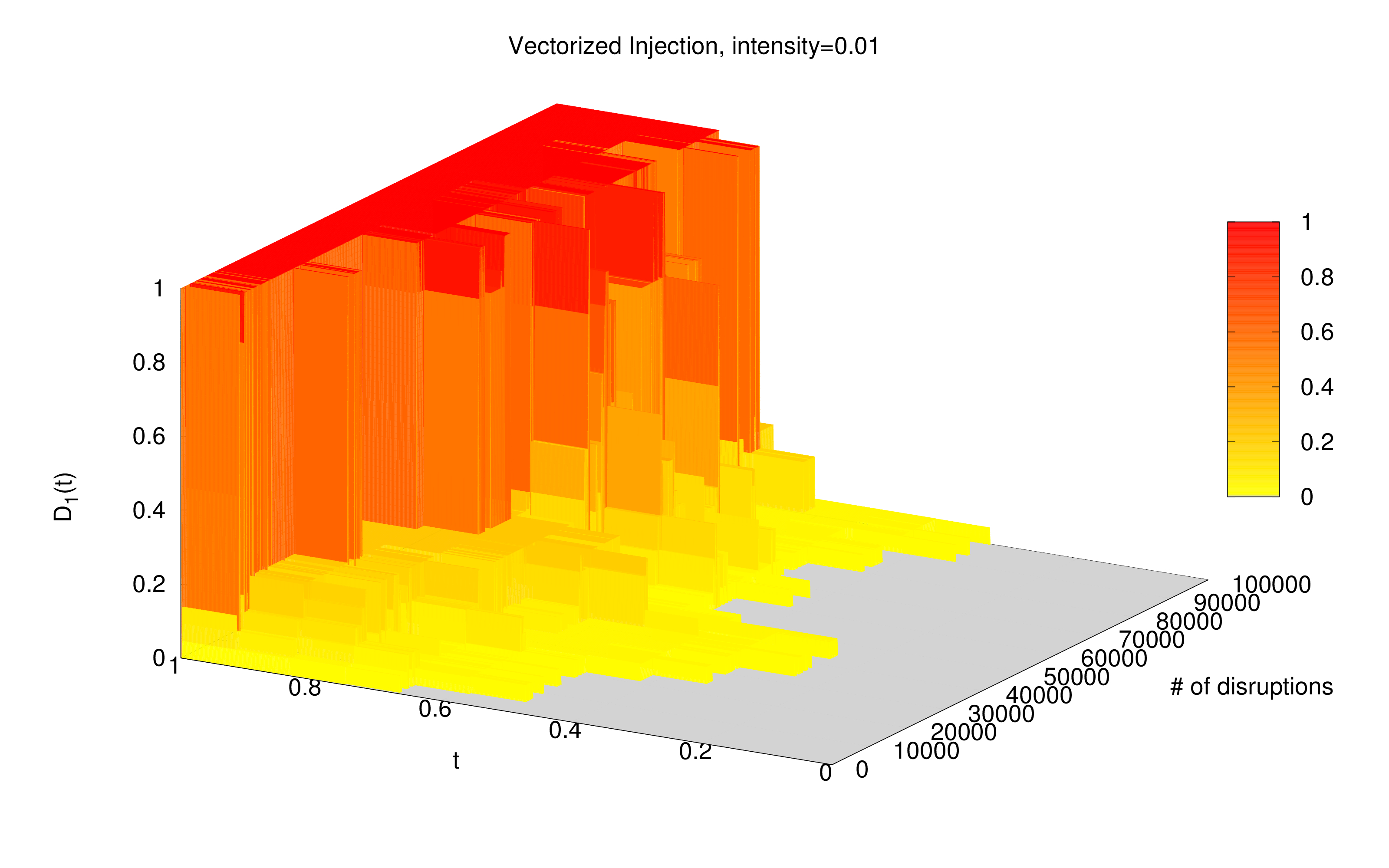}
        \includegraphics[width=0.4\linewidth]{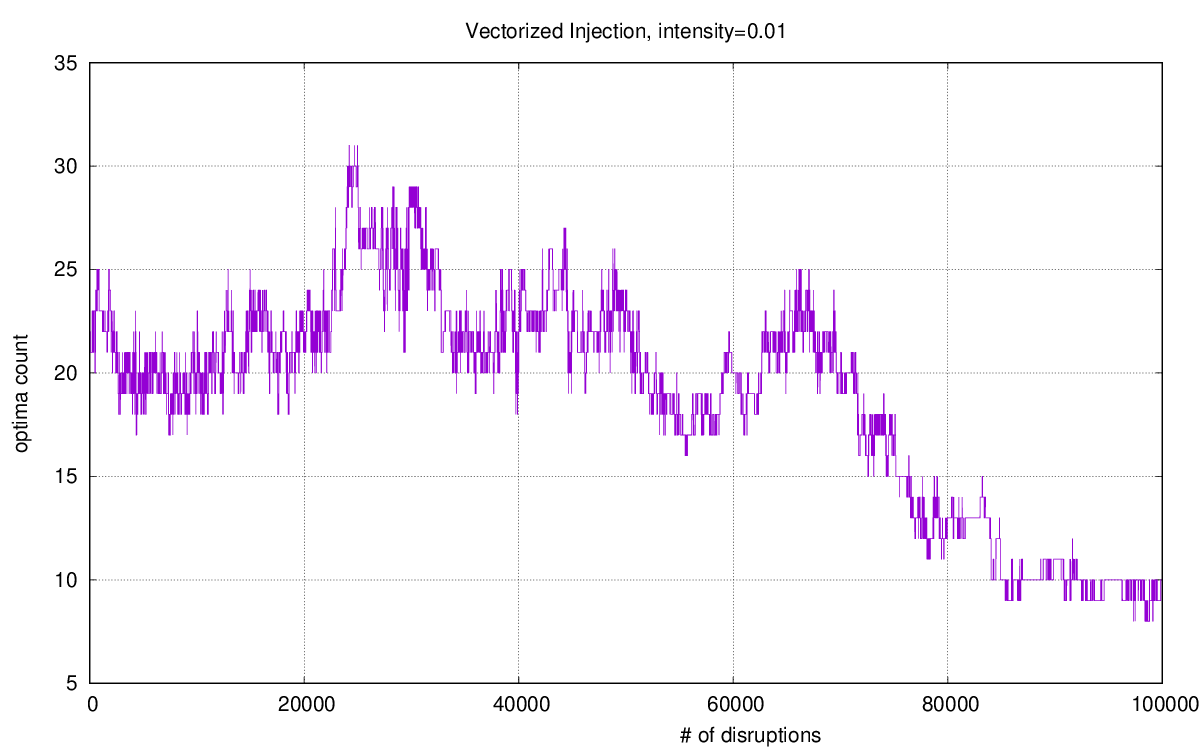}
        \caption{Disaster function and number of optima versus number of times the disruption has been applied with $\epsilon=0.01$}
    \end{center}
\end{figure}

This introduces too much order, concentrating the landscape toward a single global optimum. Finding a disruption process that is capable of achieving significant change from the pre-existing landscape while also minimizing inflation or deflation of the optima count is not trivial, and will likely take considerable effort to specially construct so as to be both biologically sensible, mathematically useful, and computationally tractable.

We conclude with the suggestion that an approach based on swapping genotypes in the rank order, and using the rank order itself as fitness rather than recalculating scores after disruptions to the contributions to fitness in the $NK$ model, could generate a viable approach under the right conditions, but would risk inflating the optimum count (if starting from an $NK$ landscape) if done naively, as scrambling the rank order would be entropy-increasing.

\end{document}